\pdfoutput=1
\documentclass{article}
\usepackage{soul}
\usepackage{framed}

\usepackage{amssymb,cancel}
\usepackage{wrapfig}
\usepackage{mathrsfs}
\usepackage{amsbsy}
\usepackage{color}
\usepackage{graphicx}
\usepackage[colorlinks=true, pdfstartview=FitV, linkcolor=blue, citecolor=blue, urlcolor=blue]{hyperref}
\definecolor{shadecolor}{rgb}{0.95, 0.95, 0.86}
\def\P{\mathbf P}
\def\pole{ v_p}

\def\insrt#1{\blue{#1}}

\def\blue#1{\textcolor[rgb]{0,0,1}{#1}}

\def\O{\Omega}

\usepackage{verbatim}
\newcommand{\eqref}{(\ref)}
\def\wt{\widetilde}

\def\wh{\widehat}

\def\ds{\displaystyle}

\def\res{\mathop{\mathrm{res}}\limits_}

\makeatletter
\@addtoreset{equation}{section}
\makeatother

\def\le{\left}
\def\ri{\right}
\def\QED{{\bf Q.E.D.}\par\vskip 5pt}

\def\bc{\begin{corollary}}
\def\ec{\end{corollary}}
\def\&{&{\hskip -20pt}}
\def\m{\mathop}

\def \s{\mathfrak s}
\def\ov{\overline}
\def\br{\begin{remark}\rm\small}
\def\1{{\bf 1}}
\def\er{\end{remark}}
\def\bt{\begin{theorem}}
\def\et{\end{theorem}}

\def\bx{\begin{example}}
\def\ex{\end{example}}
\def\bi{\begin{itemize}}
\def\ei{\end{itemize}}
\def\bd{\begin{definition}}
\def\ed{\end{definition}}
\def\bp{\begin{proposition}\rm}
\def\bl{\begin{lemma}\em}
\def\el{\end{lemma}}
\def\ep{\end{proposition}}
\def\bea{\begin{eqnarray}}
\def\eea{\end{eqnarray}}
\def \pa{\partial}
\def\C{{\mathbb C}}
\def\R{{\mathbb R}}
\def\N{{\mathbb N}}
\def\Z{{\mathbb Z}}

\newenvironment{bmatrix}
{\le[\begin{array}{cc}} {\end{array}\ri]}
\renewenvironment{pmatrix}
{\le(\begin{array}{cc}} {\end{array}\ri)}
\newtheorem{problem}{Problem}[section]

\newtheorem{theorem}{Theorem}[section]
\newtheorem{example}{Example}[section]
\newtheorem{coroll}{Corollary}[section]
\newtheorem{examps}{Examples}[section]

\newtheorem{lemma}{Lemma}[section]
\newtheorem{remark}{Remark}[section]
\newtheorem{remarks}[remark]{Remarks}
\newtheorem{proposition}{Proposition}[section] 
\newtheorem{definition}{Definition}[section]
\def\br{\begin{remark}}
\def\er{\end{remark}}
\def\bt{\begin{theorem}}
\def\et{\end{theorem}}
\def\bc{\begin{coroll}}
\def\ec{\end{coroll}}
\def\brs{\begin{remarks} \rm\
\begin{enumerate}}
\def\ers{\end{enumerate}\end{remarks}}
\def\bl{\begin{lemma}}
\def\el{\end{lemma}}
\def\bxs{\begin{examps}. \rm\begin{enumerate}}
\def\exs{\end{enumerate}\end{examps}}
\def\bd{\begin{definition}}
\def\ed{\end{definition}}
\def\bp{\begin{proposition}}
\def\ep{\end{proposition}}
\def\be{\begin{equation}}
\def\ee{\end{equation}}

\def\bea{\begin{eqnarray}}
\def\eea{\end{eqnarray}}
\def\beas{\begin{eqnarray*}}
\def\eeas{\end{eqnarray*}}
\def\gt{\hat\gamma}
\def \hf{\frac{1}{2}}
\def\part{\partial}
\def \qt{\frac{1}{4}}
\def \pa{\partial}

\def \ra{\rightarrow}

\def\C{{\mathbb C}}

\def\a{\alpha}
\def\b{\beta}
\def\d{\delta}
\def\g{\gamma}

\def\l{\lambda}
\def\m{\mu}
\def\o{\omega}

\def\s{\sigma}

\def\t{\tau}
\def\x{\xi}
\def\e{\varepsilon}
\def\z{\zeta}

\def\L{\Lambda}
\def\R{{\mathbb R}}
\def\N{{\mathbb N}}
\def\h{{\mathbf h}}
\def\Z{{\mathbb Z}}

\date{}
\begin{document}

\baselineskip 16pt plus 1pt minus 1pt
\begin{titlepage}
\begin{flushright}
\end{flushright}
\vspace{0.2cm}
\begin{center}
\begin{Large}
\textbf{\large Asymptotics of orthogonal polynomials with complex varying  quartic weight: }
\textbf{\large global structure, critical point behaviour   and the first Painlev\'e\  equation}\\
\end{Large}
\bigskip
M. Bertola$^{\dagger\ddagger}$\footnote{Work supported in part by the Natural
  Sciences and Engineering Research Council of Canada (NSERC)}\footnote{bertola@crm.umontreal.ca},  
A. Tovbis$^{\sharp}$ 
\\
\bigskip
\begin{small}
$^{\dagger}$ {\em Centre de recherches math\'ematiques,
Universit\'e de Montr\'eal\\ C.~P.~6128, succ. centre ville, Montr\'eal,
Qu\'ebec, Canada H3C 3J7} \\
\smallskip
$^{\ddagger}$ {\em  Department of Mathematics and
Statistics, Concordia University\\ 1455 de Maisonneuve W., Montr\'eal, Qu\'ebec,
Canada H3G 1M8} \\
\smallskip
$^{\sharp}$ {\em  University of Central Florida
	Department of Mathematics\\
	4000 Central Florida Blvd.
	P.O. Box 161364
	Orlando, FL 32816-1364
} \\
\end{small}
\end{center}
\bigskip
\begin{center}{\bf Abstract}\\
\end{center}

We study the  asymptotics of recurrence coefficients for monic orthogonal polynomials $\pi_n(z)$ with the 
quartic exponential  weight $\exp\le[-N\le(\hf z^2 + \qt t z^4\ri)\ri]$, where 
$t\in \C$ and $N\in\N$, $N\ra\infty$. Our goals are: A) to describe the regions of different asymptotic behaviour
(different genera) globally in $t\in\C$; B) to identify all the critical points,  and; C) to study in details the asymptotics 
in a full neighborhood near of critical points (double scaling limit), including at and near the poles of   Painlev\'e\ I
solutions $y(v)$ that are known to provide the leading correction term in this limit. Our results are:
A) We found global in $t\in\C$ asymptotic of recurrence coefficients and  of ``square-norms''
for the orthogonal polynomials $\pi_n$  for  different configurations of the contours of integration.
Special code was developed to analyze all possible cases. 
B) In addition to the known critical point $t_0=-\frac 1{12}$,
we found new critical points $t_1=\frac{1}{15}$ 
and $t_2=\frac 14$.
C) We 
derived the leading order behavior of the recurrence coefficients  (together with the error estimates) at and around the poles of 
 $y(v)$ near the critical points $t_0,t_1$ in what we called the triple scaling limit. We proved  
that the recurrence coefficients 
 have unbounded $\mathcal O(N^{-1})$-size (in $t$) ``spikes'' near the poles of $y(v)$
and calculated the ``universal''  shape of these spikes for different cases
(depending on the critical point $t_{0,1}$ and on the configuration of the contours of integration). 
The nonlinear steepest descent method for Riemann-Hilbert Problem (RHP) is the main technique used in the paper. 
We note that the RHP near the critical points is very similar to the RHP describing the semiclassical limit
of the focusing NLS near the point of gradient catastrophe that the authors solved
in \cite{BT2}. Our approach is based on the technique developed in \cite{BT2}.

\medskip
\bigskip
\bigskip
\bigskip
\bigskip

\end{titlepage}
\tableofcontents
\section{Introduction and main results}\label{sectintro}

In this paper we consider monic polynomials $\pi_n(z)$, orthogonal with respect to the quartic exponential weight  
$e^{-N f(z,t)}$, where $f(z,t)=\hf z^2 + \qt t z^4$, 
$t\in \C$ and $N\in\N$.
As $z\ra \infty$, 
the weight function  is exponentially decaying in four sectors $S_j$ of the opening $\pi/4$, centered around the rays 
$\O_j=\{z:~\arg z=-\frac {\arg t}4+ \frac {\pi (j-1)} 2\}$, $j=1,2,3,4$. We consider the most general case when the polynomials $\pi_n(z)$ are 
integrated on the ``cross'' formed by the rays $\O_j$, where the rays $\O_{1,2}$ are oriented outwards (away from the origin)
and the rays $\O_{3,4}$ - inwards.
The corresponding bilinear form is
\be\label{orthog} 
\langle p,q\rangle_{\varrho_1,\varrho_2,\varrho_3,\varrho_4}=
\sum_{j=1}^4\varrho_j\int_{\O_j}p(z)q(z)e^{-N f(z,t)}dz,\ \ \ f(z,t):=\hf z^2 + \qt t z^4
\ee
where $\varrho_j$ are  fixed complex numbers chosen to satisfy $\varrho_1+\varrho_2=\varrho_3+\varrho_4$. Moreover, since multiplying all the $\varrho_j$'s by a common nonzero constant does not affect the families orthogonal polynomials, these parameters are only defined modulo the action of the group $\C^*$ and hence the orthogonal polynomials are naturally parametrized by points in $\C \mathbb P^2$.
%

\begin{wrapfigure}{r}{0.4\textwidth}
\resizebox{0.4\textwidth}{!}{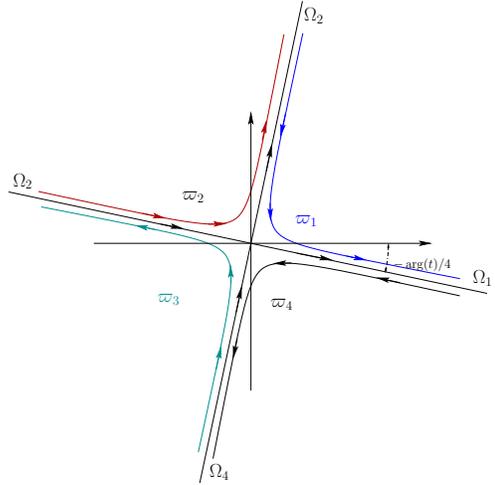}
\label{Figcontours}
\caption{The contours of integration and the asymptotic directions. The contour $\varpi_4$ is homologically equal to $-\varpi_1-\varpi_2-\varpi_3$, and it is unnecessary for the definition of the pairing (\ref{orthogA}).}
\end{wrapfigure}

Alternatively, the bilinear form (\ref{orthog}) can be represented as
\bea\label{orthogA} 
\langle p,q\rangle_{\vec\nu}=
\sum_{k=1}^3 \nu_k \int_{\varpi_k} p(z)q(z)e^{-N f(z,t)}dz,\\
\nu_1 = -\varrho_1\ ,\qquad \nu_2 = -\varrho_2 - \varrho_1\ ,\qquad \nu_3 = -\varrho_4
\eea
where $\varpi_j$, $j=1,2,3$, are  simple contours emanating from $\infty$ along $\O_j$ and returning to  $\infty$ along $\O_{j+1}$ \cite{BertolaMo} (Fig. \ref{Figcontours}).

Then in the case a)  we have $\nu_2\neq 0$ {(and we therefore can, and will,  normalize it to be $\nu_2=1$)}  and the following three cases are possible: 
\begin{enumerate}
 \item The "generic case": $\nu_{1}\nu_3\neq 0$, $\nu_1\neq 1\neq \nu_3$,  so that   
there are three contours $\varpi_j$  in (\ref{orthogA});
 \item  The "consecutive wedges": either $\nu_1$ or $\nu_3$ (but not both) is zero  so that there are two  
adjacent contours $\varpi_j$ in (\ref{orthogA});
\item  The "real axis": $\nu_3=0$  and $\nu_1=1$.
\end{enumerate}
The remaining
case b): $\varrho_1+\varrho_2=0$, corresponds to $\nu_2=0$, so that the following three cases are possible:
\begin{enumerate}
\item The "single wedge": $\nu_1=0$, so that there is only one contour $\varpi_3$ in (\ref{orthogA}), 
{for which we can, and will, set $\nu_3=1$};
\item The "opposite wedges, generic": $\nu_1\nu_3\neq 0$ and $\nu_1\neq \nu_3$;
\item The "opposite wedges, symmetric": $\nu_1=\nu_3 \neq 0$.
\end{enumerate}

The orthogonality condition for the monic polynomials $\pi_n(z)$  can now be written as
\be\label{orthog1} 
\langle \pi_n, z^k\rangle_{\vec \nu}=\h_n \delta_{nk},~~~~~~~k=0,1,2,\cdots, n,~~~~\vec\nu=(\nu_1,\nu_2,\nu_3),
\ee
where the coefficient $\h_n$ can also be written as $\h_n=\langle \pi_n,\pi_n\rangle_{\vec \nu}$ and hence is the 
equivalent of the ``square norm'' of $\pi_n$ (but it is in general a complex number).
The existence of orthogonal polynomials $\pi_n(z)$ is not {\em a priori}  clear. However, if three 
consecutive monic polynomials exists,
they are related by a three-term recurrence relation  
\be
\pi_{n+1}=(z-\b_n)\pi_n(z)-\a_n\pi_{n-1}(z),
\ee
where $\a_n=\a_n(t,N)$, $\b_n=\b_n(t,N)$ are called recurrence coefficients. 

If the bilinear pairing is invariant under the map $z\mapsto -z$ then it follows immediately that the orthogonal polynomials are even or odd  according to their degree and thus $\b_n=0\,, \forall n \in \N$ (for example in case (a1) with coefficients $\nu_1=\nu_3$, and $\nu_2=0$).
 Then
the remaining recurrence coefficients $\a_n$ satisfy 
\be
\a_n[1+t(\a_{n+1}+\a_n+\a_{n-1})]=\frac{n}{N},\label{m0}
\ee
which is known in literature as the string equation or the Freud equation \cite{Freud}. 
We are interested in the asymptotic limit of $\a_n,\b_n$ as $N\ra\infty$
and $\frac{n}{N}=x>0$ is fixed and finite, so we will use notations $\a_n=\a_n(x,t),~\b_n=\b_n(x,t)$ instead of 
$\a_n=\a_n(t,N)$, $\b_n=\b_n(t,N)$. 

In the case   of $\nu_2=-1$ and $x=1$ (that is, $n=N$) and a fixed $t\in (-\frac{1}{12 },0)$, the asymptotics of $\a_n,~\b_n$ 
was obtained  in \cite{ArnoDu} as
\be\label{alpharegass}
\a_n(1,t)=\frac{\sqrt{1+12t}-1}{6t}+O(n^{-1}),
\ee
and $\b_n$ decaying exponentially  as $n\ra\infty$. (To be more precise, Theorem 1.1 from  \cite{ArnoDu} states that there exists some $n_0=n_0(t)$,
such that $\a_n(1,t),\b_n(1,t)$ exists for all $n\geq n_0$ and have the above mentioned asymptotics.)  
In the non symmetrical case, however, the recurrence coefficients $\beta_n$ are, generically, different from 
zero.  Then, instead of (\ref{m0}), we have   the general Freud system (we indicate how to derive it in Section. \ref{ssect-string}):
\bea
0 &\&= \b_n + t\le[(2\b_n + \b_{n+1}) \a_{n+1} + ( \b_{n}^2 +2 \a_n(1-\delta_{n0}))\b_n + \a_n \b_{n-1}(1-\delta_{n0})\ri],\label{m1}\\
\frac n N &\& = \a_n + t\le[
\a_n\a_{n-1} + \a_n^2 (1-\delta_{n0}) + \a_{n+1} \a_n + \b_n^2 \a_n + \a_n \b_{n-1}(\b_n + \b_{n-1})
\ri].\label{m2}
\eea
where $\delta_{ij}$ denotes the Kronecker's delta.
Assuming that $\b_n \to \b$ and $\a_n\to \a$ as $n\ra\infty$, we obtain two leading order  algebraic equations 
\be\label{leorFr}
\b(1 + 6t\a + t\b^2)=0\ ,\qquad \a\le( 1+ 3t \a + 3 t \b^2\ri) = x,
\ee
which have two solutions 
\bea\label{firstsol}
&& \b=0,\qquad~~~~~~~~~~~~ \a=\frac {\sqrt{1+12xt}-1}{6t}, \\
&&\rm{~and}\cr
\label{secsol}
&& \b^2 =- 6\a- \frac 1 t\ ,\qquad \a = \frac {\sqrt{1-15 xt}-1} {15t}.
\eea
The dependence on $x$ is rather fictitious: indeed, looking at the pairing (\ref{orthog}) one sees that if $0<x \neq 1$ then 
we can rescale $\wt z= \frac z{\sqrt{x}}$, $\wt t= x\,t$ and obtain the case $N=n$ ($x=1$) without loss of 
any generality; we shall assume this done throughout the paper, but still distinguish $n$ and $N$ because they 
play a slightly different role.

We point out that the asymptotics of the recurrence coefficients for orthogonal polynomials with integration on the 
real axis (and analytic continuation thereof in the complex $t$--plane) satisfies identically $\beta_n=0$, and hence only 
the first  solution (\ref{firstsol}) is relevant. 
However, in view of applications to combinatorics of maps, it is not clear what role, if any, 
the second solution (\ref{secsol}) play. It could be,  perhaps,  of some
relevance that the critical point $t_1=\frac{1}{15}$ 
of the second solution is actually {\em closer} to the origin than the  critical point $t_0=-\frac{1}{12}$ of the ``standard'' 
(first) solution. 

The  recurrence coefficients problem  was studied in \cite{ArnoDu}, 
following earlier work \cite{FIK}, for negative values of $t\in (-\frac 1{12},0)$. 
For definiteness we shall assign $\arg(t)=\pi$, so that (refer to Fig. \ref{Figcontours}) the contour $\varpi_1$ 
consists of the two rays $\arg(z) = \pm \frac \pi 4$, $\varpi_2$ of $\arg(z) = \frac \pi 4, \frac {3\pi}4$ and 
$\varpi_3$ of $\arg(z) = \frac {3\pi}4,  \frac {5\pi}4$ (other determinations of $\arg(t^\frac 14)$ would  
only reshuffle the contours around). To describe the results of \cite{ArnoDu} and to set 
the stage for our results, we need to recall that
the general solution of the Painlev\'e\ I (P1) equation 
\be
y''(v) = 6 y^2(v)-v\ \label{P1ODE}
\ee
is parametrized by two parameters in terms of a certain Riemann--Hilbert problem described in Section \ref{P1sec}. 
It is known that {\em any solution} to P1 has infinitely many poles  with a Laurent expansion of the form 
\be
y(v) = \frac 1{(v-\pole)^2} + \frac \pole {10} (v-\pole)^2 + \frac 1 6 (v-\pole)^3 + \beta  (v-\pole)^4 + \frac {\pole^2}{300} (v-\pole)^6 + \mathcal O((v-\pole)^7).
\label{TaylorP1}
\ee
The Painlev\'e\ property asserts that the only singularities that can occur are of this form, that is, 
the {\em position} of these poles depends on the chosen solution, and it is largely unknown, except for some 
asymptotic localization of the remote poles, see, for example, \cite{KapaevP1}. 
In the following theorem and henceforth we use notations $\a_n(t)=\a_n(1,t),~ \b_n(t)=\b_n(1,t)$.
\begin{theorem} (\cite{ArnoDu}) \label{theo-ArnoDu}
Let $y^{(0)}(v):= y(v;1-\nu_3), y^{(1)}(v):= y(v;\nu_1)$ (see  Def. \ref{defyrl} below),  be the two solution of P1(\ref{P1ODE})
\footnote{The parameters that were 
indicated with $\alpha, \beta$ in \cite{ArnoDu}  in our notations are: $\alpha = 1-\nu_1$, $\beta = \nu_3$.
}.
Let $K\subset \C$ be a compact set that does not contain any of the poles of $y^{(0)}, y^{(1)}$.
Let $t\in\C$ approach the critical value  $t_0= -\frac{1}{12}$ in such a way that
\be\label{tlim}
N^\frac 45\le(t+\frac{1}{12}\ri)=-\frac{v}{2^\frac 95 3^\frac 65},
\ee
where  $v\in K $. 
Then, for  large enough $n$, the recurrence coefficients $\a_n(t),\b_n(t)$ have asymptotics
\bea\label{anbn}
\a_n(t)=2-2^\frac 35 3^\frac 25(y^{(1)}(v)+y^{(0)}(v))N^{-\frac 25} +\mathcal O(N^{-\frac 35}), \cr
\b_n(t)=2^\frac{1}{10} 3^\frac 25(y^{(0)}(v)-y^{(1)}(v))N^{-\frac 25} +\mathcal O(N^{-\frac 35})
\eea
as $n=N\ra +\infty$, which is valid uniformly in $K$.
Moreover, the $\mathcal O$ terms can be expanded into a full asymptotic expansion in powers of $n^{-\frac 15}$.
\end{theorem}
The statement of Theorem \ref{theo-ArnoDu} is an example of what is known as the double scaling limit  near a critical point  and it is obtained using the steepest descent analysis and a special "Painlev\'e\ I parametrix" that was first introduced in \cite{FIK}:
it does not address, however, the asymptotics 
of the recurrence coefficients when $v$ is at or close to (in a  ``triple'' scaling sense to be specified later)  a pole of either $y^{(1)} (v)$ or $y^{(0)}(v)$. Also, no information is available
for the case  $t<-1/12$, as well as  for general complex values of $t$. Thus,
{\bf the main results of this paper are:}
\begin{enumerate}
 \item finding  the  global (in $t\in\C$) leading order behavior of the recurrence coefficients $\a_n(t), \b_n(t)$ 
and of the "square-norms" $\h_n$ for the orthogonal polynomials $\pi_n$  for the cases to different configurations of the 
contours $\varpi_j$ as listed above;

\item  deriving new critical points $t_1=\frac{1}{15}$ in the case $\nu_2=0$ (this case  was not considered in \cite{ArnoDu}) 
and $t_2=\frac 14$; note that $t_1$ is closer to the origin $t=0$ than $t_0$;

\item deriving the leading order behavior of $\a_n(t), \b_n(t)$ (together with the error estimates) at and around the poles of 
the P1 solutions $y^{(0),(1)}(v)$ near the critical points $t_0,t_1$; we will see that  $\a_n(t),\b_n(t)~ {\rm and}~ \h_n$ 
have unbounded ``spikes'' near the poles of $y^{(0),(1)}(v)$
and study the shape of these spikes in certain cases.
\end{enumerate}
In regard to the point $t_1 =\frac 1{15}$ (which,  to the best of our knowledge, was not considered  in the literature), 
we also find  an asymptotics that is related to the same Painlev\'e\ I equation in the theorem below.
\begin{shaded}
\bt
\label{t1away}
Let $y(v):=y^{(1)}(v)=y(v;\nu_1)$ (see Def. \ref{defyrl}) be the solution of P1 (\ref{P1ODE}); let $K\subset \C$ be a compact set not containing any poles of $y^{(1)}$.  Let $t$ depend on $N$ so that
\be
N^\frac 4 5 \d t := N^{\frac 4  5} \le(t - \frac 1{15}\ri) = {\rm e}^{-\frac {3i\pi }5} \frac v {3^\frac 6 5 2^\frac 1 5  5}, \label{t1lim}
\ee
where $ v\in K$ (see \ref{v(t)2}). Then, uniformly for $v\in K$,  we have 
\bea
\a_n &\&=-1  + \frac{i 6^{\frac 25} {\rm e}^{-\frac {3i\pi}{10}}}{N^{\frac 25} } y(v) + \mathcal O(N^{-\frac 35})\ ,\qquad
\b_n=
 -3i -  \frac {6^\frac 25 {\rm e}^{-\frac {3i\pi}{10}}}{N^{\frac 25}} y(v) + \mathcal O(N^{-\frac 35}),
\\
\h_n&\& =    2i\pi  (-1)^N\le(
1 -\frac{3^\frac 25}{2 ^{\frac 35} }{\rm e}^{-\frac {4}{5}i\pi} \frac {y(v) }{N^{\frac 25 }}\ri)\exp\le[
\frac {9N}4  - 
 \frac {195N }4 \d t
  + {\rm e}^{-\frac 25 i\pi} \frac{6^\frac 15}{N^{\frac 1 5}} H_I 
\ri] (1 + \mathcal O(N^{-\frac 35})),
\eea
where $H_I = \frac 1 2 (y')^2 + y v  - 2y^3$ is the Hamiltonian of P1 (\ref{P1expansion}) and $H_1'(v) = y(v)$.
\et
\end{shaded}
Theorem \ref{t1away} is, of course, of the same nature as Theorem \ref{theo-ArnoDu}.
It is clear, though, that in order to study the full neighborhood of a critical point $t_j$, $j=0,1$,
which, by analogy with the zero dispersion limit of the focusing Nonlinear  
Schr\"odinger equation (NLS) will be called a point of gradient catastrophe,  one must separate the asymptotic analysis in two distinct regimes:
\begin{itemize}
\item {\bf Away from the poles}: the variable $v$ is chosen within a fixed compact set that does not include any pole of the relevant solutions to P1;
\item {\bf Near the poles}: the variable $v$ undergoes its own scaling limit and approaches a given pole at a certain rate.
\end{itemize}
Theorems \ref{theo-ArnoDu}, \ref{t1away} are examples of the regime ``away from the poles". To investigate the regime ``near the poles" we must use a novel modification that we could call {\bf triple scaling}.
\begin{shaded}
\begin{theorem}\label{theor-nonsym}
Consider the setups as in Thm. \ref{theo-ArnoDu} with $\nu_1 \neq 1-\nu_3$ or  Thm. \ref{t1away}, with the same notation for $y^{(0),(1)}$ in 
the former case and $y^{(1)}$ in the latter. Let $v_p$ denote any chosen pole of $y^{(1)}$ (which is, if in the first setup, not a pole of $y^{(0)}$). 
Let $t$ approach $t_0=-\frac 1{12}$ or $t_1=\frac 1 {15}$   in such a way that 
it  satisfies respectively
 \bea
 t+\frac 1 {12} =  -\frac{  v_p}{N^{\frac 45}3^{\frac 6 5 } 2^{\frac 9 5}} -\frac {s}{3\sqrt 2 N}\ \qquad {\rm or} \qquad 
t-\frac 1{15} = -\frac {v_p {\rm e}^{-\frac {3i\pi}5}} {3^\frac 6 5 2^\frac 1 4  5 N^\frac 45} -i \frac s {2N},
 \eea
 Then
\bea
\a_n(t)&\& =\frac{b_0^2}{4} - \frac{1}{4s^2} + \mathcal O\le ({N^{-\frac 1 5 }} s^{-1}\ri),   \label{assnon}
\\
\b_n(t)&\& = a_0+\frac 1{2s(1-b_0s) 
+  \mathcal O(N^{-\frac 1 5})
 },  \label{betnon}
\eea
where $a_0,b_0$  (the limiting values of $a(t), b(t)$  from Table \ref{expansions}) are given by 
$a_0 = 0$, $b_0 = \sqrt{8}$ and $a_0 = -3i\ ,\ \ b_0 = 2i$ in the  cases $t\sim t_0$ and $t\sim t_1$ respectively. 
The numbers $\h_n$ satisfy:
\bea
\h_n &\&=\pi
2^N \exp \le[-\frac {3 N}2   + \frac{  N^{\frac 1 5} v_p}{3^{\frac 1 5 } 2^{\frac 4 5}} +\sqrt{2} \,{s}     \ri]
\le(
\sqrt{8} - \frac 1 s   + \mathcal O(N^{-\frac 1 5}s^{-1})\ri)\ ,\ \ \ t\sim -\frac 1{12},  \\
\h_n &\& = \pi 
 (-1)^N \exp\le[\frac {9N}4  -   \frac {13 }4\frac {N^\frac 1 5v_p {\rm e}^{-\frac {3i\pi}5}} 
{3^\frac 1 5 2^\frac 1 {5}    }  +  \frac {13}{2}  is 
\ri]\le(2i - \frac 1 s   + \mathcal O(N^{-\frac 1 5}s^{-1})\ri) \ ,\ \ \ t\sim \frac 1 {15}.
\eea
These formul\ae\ hold uniformly for bounded values of $s$ as long as the indicated error terms remain infinitesimal. In particular $s$ can approach $s=0$ or $s = \frac 1 {b_0}$ at any chosen rate $\mathcal O(N^{\rho})$ with $\rho \in [0,\frac 1 5)$ (the case $\rho=0$ allowing any given fixed value $s \neq 0, \frac 1{b_0}$). 
\end{theorem}
\end{shaded}
As the reader notices, the asymptotics has a dramatically changed form and does not involve now any transcendental function. 
Note that the scale of the phenomenon in this case is $\mathcal O(N^{-1})$ around the location of 
the image of the pole $v_p$ (see (\ref{v(t)}) or (\ref{v(t)2})respectively)
in the $t$-plane, whereas the scale at which the transcendental nature of the asymptotic is shown is $N^{-\frac 45}$. 
To study this new phenomenon, it is convenient to set a {\bf triple scaling} of the form 
\be
t= t_0 + \frac{c_1}{N^\frac 45} + \frac {c_2}N,
\ee
where the value of parameter $c_1$ corresponds to a particular pole of the  Painlev\'e\ I transcendent. Note that, according to 
(\ref{assnon}), (\ref{betnon}), the values of $\a_n,\b_n$ are unbounded as $s\ra 0$ and $s\ra b_0$ (the latter is valid only for $\b_n$).
A quite different phenomenon occurs instead if we are in the setup of Theorem \ref{theo-ArnoDu} with the same triple scaling 
limit {\em but with additional symmetry} $\nu_1=1-\nu_3$ (the case excluded from Theorem \ref{theor-nonsym}). 
In this case the two functions $y^{(0)}$ and $y^{(1)}$ are the {\em same} solution to P1 (\ref{P1ODE}) and a sort of cancellation  in 
(\ref{assnon}), (\ref{betnon}) occurs. 

\begin{shaded}
\begin{theorem}\label{theor-sym}
Consider the setup of Thm. \ref{theo-ArnoDu} with $t$ approaching $t_0$ and $\nu_1=
1-\nu_3$. Let  $v_p$ be a 
pole of  $y(v):= y^{(1)}(v)=y^{(0)}(v)$ and let t vary so that 
\be\label{tsym}
t =t_p(s) =  - \frac 1 {12} - \frac {v_p}{2^{\frac 9 5} 3^{\frac 6 5} N^{\frac 4 5}}  + \frac {s}{2^3 3 N},
\label{tps}
\ee 
where $s=\mathcal O(N^{-\rho})$ with an arbitrary $\rho\in [0,\frac 15)$.
Then the following holds:
\bea\label{asssym}
  \a_n&\&=\frac{b^2}{4}\frac{9-s^2+ \mathcal O(N^{-\frac 1 5})}{1- s^2 +\mathcal O(N^{-\frac 1 5})}\ ,\qquad
\b_n=0 
,\\
 \h_n  &\& = \pi  \sqrt{8}\,2^N
 \exp\le[-\frac {3N}2  +
    \frac {N^\frac 1 5 \,v_p}{ 3^\frac 15 2^\frac 45} - \frac {s}{4 } 
   \ri]\le(
  \frac {3-s }{1+s }+  \mathcal O(N^{-\frac 1 5}(s^2-1)^{-1}) 
 \ri).
\eea
The variable $s$ may approach the points $s = \pm 1$ 
at some rate (a {\em quadruple scaling}) as long as the corresponding error indicated in the formul\ae\ above  terms are infinitesimal.
\end{theorem}
\end{shaded}

\br
Note that the values of $\a_n$ {is unbounded in the vicinity of $s = \pm 1$ and $\h_n$ is  unbounded in the vicinity of $s=- 1$ (there is no information for $\h_n$ in the vicinity of $s = +1$). }
Let us denote the Hankel determinants of the moments  by $\Delta_n(t,N)$ (see Remark \ref{rem-det}) and use $t_p(s)$ as in (\ref{tps}): since   $\a_n = \frac {\Delta_{n-1} \Delta_{n+1}}{\Delta_n^2}$ we deduce that $\Delta_n(t(s),n)$ vanishes at $s = \pm 1$ (within our error estimates), while $\Delta_{n\pm 1}(t(s), n)$ vanish at  $s\in \{1, 3\}$ and $s \in \{ -3, -1\}$ respectively.
\er
\section{The Riemann--Hilbert problem for  Painlev\'e\ I}

\begin{wrapfigure}{r}{0.5\textwidth}
\resizebox{0.45\textwidth}{!}{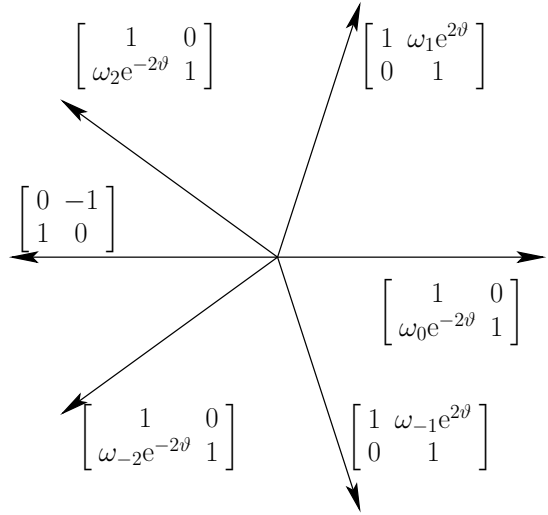}
\caption{The jump matrices for the Painlev\'e\ 1 RHP: here $\vartheta:= \vartheta(\xi;v):= \frac 45 \xi^{\frac 52} - v \xi^{\frac 12}$.}
\label{RHPP1}
\end{wrapfigure}

\label{P1sec}
Let the   invertible matrix-function $\P=\P(\xi,v)$ be analytic in each sector of the complex $\xi$-plane
shown on Fig. \ref{RHPP1} and satisfy the  multiplicative jump conditions along the oriented boundary of each 
sector with jump matrices shown on Fig. \ref{RHPP1}.

The entries of the jump matrices satisfy 
\be\label{jumpsym}
\begin{array}{l}
1+\o_0\o_1= - \o_{-2},\\
1+\o_0\o_{-1}= - \o_{2},\\
1+\o_{-2}\o_{-1}=\o_{1},
\end{array}
\label{betaconditions}
\ee
so that the jump matrices in Fig. \ref{RHPP1} depend, in fact, only on $2$ complex parameters
(that uniquely define a solution to P1).
The matrix function $\P(\x,v)$ is uniquely defined by the following RHP.

\begin{problem}[Painlev\'e\ 1 RHP \cite{KapaevP1}]
\label{P1RHP}
The matrix $\P(\xi;v;\vec \omega)$  is locally bounded, admits boundary values on the rays shown in Fig. \ref{RHPP1} and satisfies
\bea
&\& \P_+ = \P_- M,\\
&\& \P(\xi;v;\vec \omega)=\frac{\xi^{\s_3/4}}{\sqrt{2}}
\begin{bmatrix} 1 & -i \cr 1 & i \cr \end{bmatrix}
\left(I + O(\xi^{-\hf}) \right), \label{asymppsi}
\eea
where the jump matrices $M=M(\xi;v,\vec \omega)$ are the matrices indicated on the corresponding ray in 
Fig. \ref{RHPP1}, with $\vec \omega:=(\omega_{-2},\omega_{-1},\omega_0,\omega_1,\omega_2)$ satisfying (\ref{betaconditions}).
\end{problem}

For any fixed values of the parameters $\o_k$, Problem \ref{P1RHP} admits a unique solution for generic values of $v$;
 there are isolated points in the $v$--plane where the solvability of the problem fails as stated.
The piecewise analytic function  
\be\label{PsiPain}
\Psi(\xi,v;\vec \omega) = \P(\xi,v;\vec \omega) {\rm e}^{\vartheta \sigma_3}
\ee
solves a slightly different RHP with {\em constant} jumps on the same rays and thus solves an ODE.
Direct computations using the ODE
 and formal algebraic manipulations of series along the lines of \cite{JMU1, JMU2, JMU3} show that $\Psi$ admits the following formal  expansion
\bea
\Psi(\xi,v;\vec \omega) =&\& \frac{\xi^{\s_3/4}}{\sqrt{2}}
\begin{bmatrix} 1 & -i \cr 1 & i \cr \end{bmatrix}\times \cr
&\& \times
\left(\1  -  \frac {H_I \sigma_3}{\sqrt {\xi}} + \frac { H_I^2\1 + y \sigma_2}{ 2\xi} + 
\frac {( v^2 \, -\, 4 H^3_I - 2y')}{24 \xi^{\frac 32 }}\sigma_3 + \frac {iy' - 2 iH_I y}{4\xi^{\frac 32 }}\sigma_1 + \mathcal O(\xi^{-2})
\right) {\rm e}^{\vartheta \sigma_3},\cr
H_I &\& := \frac 12 (y')^2 + y v - 2 y^3,\qquad \vartheta:= \vartheta(\xi;v)= \frac 45 \xi^{\frac 52} - v \xi^{\frac 12},~~
{\rm as}~~ \xi\ra\infty.
\label{P1expansion}
\eea
where
 $y = y(v)$ solves the Painlev\'e\ I  equation (\ref{P1ODE}).
The matrix $\Psi(\x,v ;\vec \omega )$ uniquely defines a solution $y(v;\vec \omega)$ of P1 (\ref{P1ODE}), and viceversa. 
The family of solution we shall use consists of the choice $\omega_0=0$ in (\ref{jumpsym}). 
Then the constant jump matrix for $\Psi$ depends  only on one free parameter: we shall choose it to be  $\o_1$, with $\o_{\pm 2}=-1$ and $\o_{1}=1-\o_{-1}$.
 
\bd
\label{defyrl}
The functions  $y (v;\varkappa)$ are the solutions to Painlev\'e\ 1, which are defined 
via $\Psi\le (\x,v, (-1,\varkappa, 0,1-\varkappa, -1)\ri)$ in Problem \ref{P1RHP}.
We shall abbreviate this notation by $\Psi\le(\x,v;\varkappa\ri)$.
\ed

\section{The RHP for recurrence coefficients}\label{RHPsect}

It is well known (\cite{Deift}) that the existence of the above-mentioned orthogonal polynomials $\pi_n(z)$ is equivalent to the existence of the solution
to the following RHP (\ref{RHPY}).
%
More precisely, relation between the RHP (\ref{RHPY}) and the orthogonal polynomials $\pi_n(z)$ is given by the following
proposition (\cite{ArnoDu}), which has the standard proof (see \cite{Deift}).

\bp\label{proparno1}
Let  $\O:=\bigcup_{j=1}^3 \varpi_j$, and define  $\nu :\Omega \to \C$ by $\nu(z)=\nu_j$ when $ z\in \varpi_j$.
Then the solution of the following RHP problem 
\be
\le\{\begin{array}{cc} 
& Y(z) ~~{\rm is~ analytic~in~} \C\setminus \O ~~~~~ ~~~~~~~~~~~~~~\\
&Y_+(z)=Y_-(z) \left(
\begin{array}{cc}
1 & \nu(z){ e}^{-N f(z,t)}\\
0& 1
\end{array}
\right),    z\in\O, \\
%
&Y(z)=(\1+O(z^{-1}))
\left(
\begin{array}{cc}
z^n & 0\\
0& z^{-n}
\end{array}
\right),  z\ra\infty. \\
\end{array}
\ri.
\label{RHPY}
\ee
exists (and it is unique) if and only if there exist a monic polynomial $p(z)$ of degree $n$ and a polynomial
$q(z)$ of degree $\leq n-1$ such that
\bea\label{pqorth}
&\langle p(z), z^k\rangle_{\nu_1,\nu_2,\nu_3}=0,~~&{\rm for~all}~~k=0,1,2,\cdots, n-1, \\
&\langle q(z), z^k\rangle_{\nu_1,\nu_2,\nu_3}=0,~~&{\rm for~all}~~k=0,1,2,\cdots, n-2,~~~{\rm and}~~\langle q(z), 
z^{n-1}\rangle_{\nu_1,\nu_2,\nu_3}=-2\pi i.
\eea 
In that case the solution to the RHP (\ref{RHPY}) is given by 
\be\label{YandCauchy}
Y(z)=\left(
\begin{array}{cc}
p(z) & C_\O[p(z)\nu(z){ e}^{-N f(z,t)}] \\
q(z) & C_\O[q(z)\nu(z){ e}^{-N f(z,t)}]
\end{array}
\right),~~~ z\in \C\setminus \O,~~{\rm where}~~C_\O[\phi]=\frac{1}{2\pi i}\int_\O\frac{\phi(\z) d\z}{\z-z}\ 
\ee
is the Cauchy transform of $\phi(z)$.
\ep
\br\label{rem-det}
It follows immediately that the polynomials $p,q$ in Proposition \ref{proparno1} 
coincide with 
\bea
\pi_n(z) =\frac 1{\Delta_n} \det \le[\begin {array}{ccccc}
\mu_0&\mu_1 & \dots  \mu_{n-1} & \mu_n\\
\mu_1 &&&\mu_{n+1}\\
\vdots &&&\vdots\\
\mu_{n-1} & \dots &\mu_{2n-2}& \mu_{2n-1}\\
1&z&\dots &z^n
\end{array}\ri]\qquad
q(z) =\frac {-2i\pi} {\Delta_n} \det \le[\begin {array}{ccccc}
\mu_0&\mu_1 & \dots  \mu_{n-2} & \mu_{n-1}\\
\mu_1 &\dots &&\mu_{n}\\
\vdots &&&\vdots\\
\mu_{n-2} & \dots &\mu_{2n-4}& \mu_{2n-3}\\
1&z&\dots &z^{n-1}
\end{array}\ri]
\eea
respectively, where $\mu_j:= \langle z^j,1\rangle_{\nu_1,\nu_2,\nu_3}$ are the moments and $\Delta_n:= \det \le[\mu_{i+j}\ri]_{0\leq i,j\leq n-1}$.
It {is clear} from these expressions but it  is also a well known fact \cite{Chihara} 
that the condition of existence \insrt{of the $n$-th orthogonal polynomial $p_n$ is that $\Delta_n\neq 0$; on the other hand it is known from \cite{FIK}  that existence of $p_n$ is equivalent to the solvability of the RHP and hence the existence } of the solution for the RHP problem (\ref{RHPY}) is equivalent to $\Delta_n\neq 0$. This determinant is sometimes referred to as the ``tau function'' of the problem \cite{Bertola:MomentTau}.
Note also that the "square-norms" of the polynomials are ratios of Hankel determinants
\be\label{hndet}
\h_n:= \le\langle \pi_n(z) , \pi_n(z) \ri \rangle_{\vec \nu} =\le\langle \pi_n(z) , z^n \ri \rangle_{\vec \nu}  = \frac {\Delta_{n+1}}{\Delta_n}
\ee
\er
If  $\pi_{n+1}(z)$, $\pi_{n}(z)$ , $\pi_{n-1}(z)$ are monic orthogonal polynomials then they satisfy
\be\label{3termrecrel}
\pi_{n+1}(z)=(z-\b_n)\pi_{n}(z)-\a_n \pi_{n-1}(z)
\ee
for certain recurrence coefficients $\a_n,\b_n$ \cite{Szego, Chihara}. The following well known statements (see, for example, \cite{FIK}, 
\cite{Deift}, \cite{ArnoDu}) show the connection 
between the RHP (\ref{RHPY}), the orthogonal polynomials $\pi_n(z)$ and their recurrence coefficients.

\bp\label{proparno3}  
Let $Y^{(n)}(z)$ denote the solution of the RHP (\ref{RHPY}). If we write 
\be\label{assYz}
Y^{(n)}(z)=\left(\1+\frac{Y^{(n)}_1}{z} + \frac{Y^{(n)}_2}{z^2} +O(z^{-3})\right)
\left(\begin{array}{cc}
z^n & 0\\
0& z^{-n}
\end{array}
\right), ~~~ z\ra\infty,
\ee
then
\be\label{expressalpbet}
\h_n = - {2i\pi}  (Y_1^{(n)})_{12}\ ,\ \ 
\a_n=\left(Y^{(n)}_1\right)_{12}\cdot \left(Y^{(n)}_1\right)_{21},~~~~\b_n=\frac{\left(Y^{(n)}_2\right)_{12}}{\left(Y^{(n)}_1\right)_{12}}-\left(Y^{(n)}_1\right)_{22}.
\ee
\ep

\bp\label{proparno4} 
Suppose the RHP (\ref{RHPY}) has solution  $Y^{(n)}(z)$. Let (\ref{assYz}) be its expansion at $\infty$ and let $\a_n,\b_n$ be given by
(\ref{expressalpbet}). If $\a_n\neq 0$, then the monic orthogonal polynomials $\pi_{n+1}(z)$, $\pi_{n}(z)$ , $\pi_{n-1}(z)$
exist and satisfy the three term recurrence relation (\ref{3termrecrel}).
\ep

\subsection{String equation for $\a_n(x,t),~\b_n(x,t)$}\label{ssect-string}
The string equations, or Freud's equations, for the recurrence coefficients $\a_n,\b_n$ are nonlinear 
difference equations. Assuming that  the corresponding orthogonal polynomials exist, they can be obtained as follows. 
On one hand we have 
\be
z\pi_n(z) = \pi_{n+1}(z)  + \b_n\pi_n(z) + \a_n\pi_{n-1}(z). \label{zrec}
\ee
One can iterate (\ref{zrec}) to find $z^k\pi_n(z)$ for any $k\in \N$. On the other hand we have 
\bea
0\equiv \sum_{j=1}^3 \nu_j \int_{\varpi_j} \pa_z \le( \pi_n\pi_m {\rm e}^{-N f} \right)  {\rm d} z  =  \sum_{j=1}^3 \nu_j \int_{\varpi_j} \le(\pi_n'\pi_m  + \pi_n\pi_m'  - N \pi_n\pi_m f'(z)\ri){\rm e}^{-N f} {\rm d} z = \\
= \le\langle \pi'_n,\pi_m \ri\rangle_{\nu} +  \le\langle \pi_n,\pi'_m \ri\rangle_{\nu} -   N\le\langle \pi_n,f'(z)\pi_m \ri\rangle_{\nu}
\label{master}
\eea
Since $f'(z)$ is a polynomial, the last term above can be written as a polynomial in the recurrence coefficients using repeatedly (\ref{zrec}).  For $n=m$ the first two terms are the same and vanish because $\pi'_n$ is  a polynomial of degree $n-1$ and $\pi_n$ is orthogonal to any polynomial of lower degree. Then (\ref{master})$_{nn}$ yields a recurrence relation. For $m=n-1$ we have 
\be
\le\langle \pi'_n,\pi_{n-1} \ri\rangle_{\nu} +  \le\langle \pi_n,\pi'_{n-1} \ri\rangle_{\nu} -   N\le\langle \pi_n,f'(z)\pi_{n-1} \ri\rangle_{\nu} = n \le\langle \pi_{n-1},\pi_{n-1} \ri\rangle_{\nu} -  N\le\langle \pi_n,f'(z)\pi_{n-1} \ri\rangle_{\nu}\label{master2}.
\ee 
Equation (\ref{master2}) yields (\ref{m2}) while (\ref{master}) for $n=m$ yields (\ref{m1}).  
If  the orthogonality pairing is symmetric under $z\mapsto -z$, that is, if  
\be
\le\langle p(z), q(z) \ri\rangle_\nu =\le\langle p(-z), q(-z) \ri\rangle_\nu 
\ee
then it follows easily that $\b_n\equiv 0$ and then (\ref{m1}, \ref{m2}) reduce simply to (\ref{m0}).

\section{Steepest descent analysis of the RHP (\ref{RHPY})} \label{solRHPsect}

The steepest descent analysis in general terms for these kind of orthogonal polynomials with a polynomial external field was 
investigated in \cite{BertolaMo} and so we refer the reader there for details. 
The schematic of the approach is outlined here; as customary, the problem undergoes a sequence of modifications into equivalent 
RHPs until it can be effectively  solved in approximate form while keeping the error terms  under control.

$\bullet$ \ One starts with the problem for $Y$ (\ref{RHPY}) and seeks an auxiliary scalar function $g(z)$, called the $g$--function, 
which is analytic except for a collection $\Sigma$ of appropriate contours to be described subsequently and behaves like 
$\ln z + \mathcal O(z^{-1})$ near $z=\infty$: the contour $\O$ of the RHP (\ref{RHPY}) can be deformed because the  RHP (\ref{RHPY}) has 
an analytic jump matrix. {\em The final configuration of $\Omega$ must contain all the contours where $g$ is not analytic}. 

$\bullet$ Then we introduce a new matrix 
\be
T(z):={\rm e}^{-N\ell\frac {\s_3} 2}Y(z){\rm e}^{-N (g(z,t)-\frac \ell2){\s_3} }.\label {defT}
\ee
As a result, $T$ solves a new RHP
\be
\le\{\begin{array}{lc} 
 T(z) ~~{\rm is~ analytic~in~} \C\setminus \O ~~~~~ ~~~~~~~~~~~~~~\\
T_+(z)=T_-(z) \left(
\begin{array}{cc}
{\rm e}^{-\frac N2 \le(h_+-h_-\ri)} & \nu(z){ e}^{\frac N2 \le(h_+ + h_-\ri)}\\
0& {\rm e}^{\frac N2 \le(h_+-h_-\ri)}
\end{array}
\right),    z\in\O,\qquad {\rm where}~~ h(z,t):= 2g(z,t)-f(z,t) - \ell. \\
%
T(z)=(\1+O(z^{-1})), \qquad z\ra\infty. \\
\end{array}
\ri.
\label{RHPT}
\ee 
$\bullet$ At this point the Deift--Zhou method can proceed provided that the function $g(z)$, the constant $\ell$ and 
the collection of contours $\Omega$ into  which we have deformed the problem fulfill a rather long collection of equalities 
and --most importantly-- {\em inequalities} that we set out to briefly describe \cite{TVZ1, BertolaMo}: we say here that if all 
these requirements are fulfilled the full asymptotic for the problem can be obtained in terms of Riemann Theta functions on a suitable (hyper)elliptic 
Riemann surface of a positive genus  (with the case of zero genus not requiring any special function).

\subsection{Requirements on the $g$--function}
The (deformed) contour $\Omega$ can be partitioned into two disjoint subsets of oriented
arcs that we shall denote by  $\mathfrak M$ and term {\bf main arcs}, 
and $\mathfrak C$ or {\bf complementary arcs}; this partitioning is subordinated to a list of requirements for $g$ and $h$. 

\subsubsection{Equality requirements for $g$}
\label{eqreqs}
\begin{enumerate}
\item $g(z)$ (to shorten notations, we drop the $t$  variable in this subsection) is analytic in $\C\setminus \le(\mathfrak M\cup\mathfrak C\ri) $ 
and has the asymptotic behaviour 
\be\label{assg}
g(z) = \ln z + \mathcal O(z^{-1})\ ,\  \ z\ra\infty;
\ee
\item $g$ is analytic along all the {\em unbounded} complementary arcs  except for exactly one {\em unbounded} complementary arc which we will denote 
by $\gamma_0$, where 
\be
g_+(z)-g_-(z) = 2i\pi \,\ \ z\in \gamma_0	
\ee 
(note that the function ${\rm e}^{N g(z)}$ is analytic across all the {unbounded}  complementary arcs since $N=n\in \N$ by definition);
\item on the  {\em bounded} complementary arcs $\gamma_c$, the function $g$ has a  jump
\be
 g_+(z) - g_-(z) = 2\pi i \eta_c\ ,\ \ \ \eta_c\in \R\ ,\ \ z\in \gamma_c,
\ee
where $\eta_c$ is a constant on each connected component of the complementary arcs;
\item across each main arc (which are all bounded by assumption) we have the jump
\be\label{jumpgmain}
 g_+(z) + g_-(z)= f(z) + \ell,\ \ \ z\in \mathfrak M .
\ee
We stress that the constant $\ell$ is the same for all the main arcs.
\end{enumerate}
Assuming that the contours $\mathfrak M$, $\mathfrak C$ are known, the function $g(z)$ can be considered as the
solution of the scalar  RHP,  defined by conditions 1-4. Similarly, the function $h=2g-f$  can be considered as the
solution of the scalar  RHP with the jumps
\be\label{jumph}
\frac 12 \le(h_+(z)-h_-(z)\ri)= 2\pi i \eta_c,~~ z\in\mathfrak M,~~ ~~       \frac 12\le(h_+(z) + h_-(z)\ri) =0,~~ z\in\mathfrak C, ~~~
h_+(z)-h_-(z) = 4i\pi,\ \ z\in \gamma_0,	
\ee
and the asymptotic behavior 
\be\label{asshgen}
h(z) = -f(z) +2\ln z + \mathcal O(z^{-1})\ ,\  \ z\ra\infty.
\ee
It follows immediately from (\ref{jumph}) that  $\Re h(z)$ is continuous across the complementary arcs $\mathfrak C$.
\subsubsection{Inequality (sign) requirements (or sign distribution requirements) for $h$ and the modulation equation}
\label{signreqs}
\begin{enumerate}
\item along each complementary arc $\g_c$ we have $\Re h(z)\leq 0$;with the equality holding at most at a 
finite number of points. In the generic situation these would be only the endpoints 
(we shall call this case {\bf regular}, with the same connotation as in \cite{DKMVZ});
\item on both sides in close proximity of each main arc $\gamma_m\subset \mathfrak M$ we have $\Re h(z)> 0$
\end{enumerate}
The sign distribution requirement for the main arcs implies that {\bf $\Re h(z)$ is continuous 
 everywhere in $\C$} and the main arcs belong necessarily 
to its zero level set. The main arcs $\g_m$ can be considered as the branch-cuts of a hyperelliptic Riemann surface 
 $\mathfrak R(t)$, associated with $g$ and $h$. The number of main arc (the genus of $\mathfrak R(t)$ plus one) 
needs to be chosen in such a way that the above sign conditions will be satisfied.
The location of  the endpoints $\lambda$ of each main arc 
(which are the branch-points of $\mathfrak R(t)$) is governed
by the requirement 
\be
\Re h(z) = \mathcal O (z-\lambda)^\frac 32 )\ ,\ \ z\to\lambda,\label{modeq}
\ee
known as the {\em modulation equations}.
Since the jumps on the complementary arcs are constants, the above requirement can also be stated as 
\be
h'(z) = \mathcal O(\sqrt{z-\lambda})\ ,\ \ z\to\lambda,\label{modeq'}
\ee
where the discontinuity is placed on the main arc. The logic behind all the above requirements and the modulation 
equations will be briefly discussed in Subsections  \ref{sect-exist}, \ref{sect-stdesc}.
Note that the modulation equation (\ref{modeq}) implies that there are three zero level curves of $\Re h$
emanating from each branch-point $\l$. 
\subsubsection{The $g$-function and the modulation equations in the genus zero case}\label{modeqsubsec}
Due to the modulation equations \ref{modeq}, solutions of the RHPs for $g$ and for $h$ commute with differentiation.
Thus, the (scalar) RHP for $g'$ is:
\begin{enumerate}
\item  $g'(z)$ is
analytic (in $z$) in $\bar\C\setminus 
\mathfrak M$ and 
\be\label{assg'}
g'(z) = \frac 1 z +O(z^{-2})~~~~{\rm as }~~~~z\ra\infty;
\ee
\item $g'(z)$ satisfies the jump condition
\be\label{rhpg'}
g'_+ + g'_-=f'~~~~ {\rm on}~~\mathfrak M.
\ee
\end{enumerate}
%
Let us consider the case of a single main arc $\g_m$ with the endpoints $\l_0,\l_1$.
Using the analyticity of $f'$, the solution of the latter RHP is given by 
\be\label{g'forman}
g'(z)={{R(z)}\over{4\pi i}} \int_{\gt_m}{{f'(\z)}\over{(\z-z)R(\z)_+}}d\z~,\qquad R(z):=\sqrt{(z-\l_1)(z-\l_0)} ~,
\ee
where the contour $\gt_m$ encircles the contour $\g_m$ and has counterclockwise orientation ($z$ is outside $\gt_m$).
It is  known (\cite{ArnoDu}) that  the case $t\in(-\frac 1{12},0)$ is the genus zero case with real  branch-points 
(we will derive the same result shortly).
Using (\ref{g'forman}), the asymptotics (\ref{assg'}) yields two equations
\be\label{mom}
\int_{\gt_m}{{f'(\z)}\over{R(\z)_+}}d\z=0~~~~~{\rm and}~~~~~~\int_{\gt_m}{{\z f'(\z)}\over{R(\z)_+}}d\z=-4i\pi ,
\ee
called moment conditions, which are equivalent to the endpoint condition (modulation equation) (\ref{modeq}).
{\it We use the moment conditions (\ref{mom}) to define the location of the endpoints $\l_{0,1}$,} where we put $\l_0<\l_1$.

In the case of a  polynomial $f(z,t)$, equations (\ref{mom}) can be solved  using the residue theorem.
 Setting $\l_0=a-b,~\l_1=a+b$ and using the residue theorem on equations (\ref{mom}) we obtain
\be\label{modalg}
a+ta(a^2+\frac 32 b^2)=0~~~~~{\rm and}~~~~a^2+\hf b^2 +t (a^4+ 3a^2b^2+\frac 38 b^4)=2.
\ee
There are two possibilities: $a=0$ and $a\neq 0$. In the first case we obtain solutions to the system (\ref{modalg}) as 
\bea \label{branchpeq}
a=0\,\qquad b^2=-\frac{2}{3t}(1\mp \sqrt{1+ 12 t}),\\
\label{lambdasym}
\l_{0,1}=\mp b=\mp \sqrt{-\frac{2}{3t}(1 - \sqrt{1+ 12 t})}
\eea
The choice of the negative sign in (\ref{branchpeq}) comes from the requirement that $b$ is bounded as $t\ra 0$.
Observe that for
$t>t_0=-\frac 1 {12}$ the values $\pm b$ coincide with the 
branch-points, derived in \cite{ArnoDu}. At the critical point
\be\label{tcrit}
t=t_0=-\frac{1}{12}
\ee
the two pairs of roots (\ref{branchpeq}) coincide, creating five zero level curves of $\Re h(z)$ emanating from the endpoints $\pm b_0$, where
$ b_0=\sqrt{8}$. 
The second pair of roots $b=\pm \sqrt{-\frac{2}{3t}(1 + \sqrt{1+ 12 t})}$ 
are sliding along the real axis from $\pm\infty$ to $\pm b_0$ as real $t$ varies from $0^-$ to $t_0$,
and sliding along the imaginary axis from $\pm i\infty$ to $0$ as real $t$ varies from $0^+$ to $+\infty$.

In the second case  $a\neq 0$, the  system of modulation equations (\ref{modalg}) becomes
\be \label{remeq1}
\le\{
\begin{array}{cc} 
t(a^2+\frac 32 b^2)=-1,\\
tb^2(a^2-\frac 38 b^2)= 2,
\end{array}\right.
\ee
which yields
\be\label{othersols}
b^2=-\frac{4}{15t}(1\pm\sqrt{1-15t})~~~~~{\rm and}~~~~~~a^2=-\frac{1}{5t}(3\mp 2\sqrt{1-15t}).
\ee

\subsubsection{Explicit computation of $g$ and $h$.}\label{ghcalc}
%
Once the values of branch-points (endpoints) $\l_{0,1}$ are determined, one can calculate explicitly $g(z)$ and 
 $h(z)=h(z;t)$, where $h(z)=2g(z) - f(z)- \ell$. 
The expression 
\be\label{h'forman}
h'(z)=
{{R(z)}\over{2\pi i}} \int_{\gt_m}{{f'(\z)}\over{(\z-z)R(\z)_+}}d\z~,
\ee
for $h'(z)$ is readily available from (\ref{g'forman}) by placing
 $z$  inside the loop $\gt_m$. However, it seems easier to calculate $h'(z)$ explicitly by solving the scalar  RHP that $h'(z)$ satisfies: 
\begin{enumerate}
\item  $h'(z)$ is
analytic (in $z$) in $\g_m$
and 
\be\label{assh'}
h'(z) = -z-tz^3 +
2z^{-1} +O(z^{-2})~~~~{\rm as }~~~~z\ra\infty;
\ee
\item $h'(z)$ satisfies the jump condition
\be\label{rhph'}
h'_+ + h'_-=0~~~~ {\rm on}~~\g_{m},
\ee
\end{enumerate}
which can be easily obtained from the RHP (\ref{rhpg'}) for $g'(z)$. There are two cases, {\em symmetric} and {\em nonsymmetric} depending on the value $a=0$ or $a\neq 0$.
\paragraph{Symmetric case: $a=0$.}
The RHP for $h'$ has a unique solution (with  $h'_\pm\in L^2(\g_m)$)
that is given by  $h'(z)=-(k+tz^2)\sqrt{z^2-b^2}$, where the endpoints $\pm b$ of $\g_m$ are known and the constant
  $k$ is to be determined.
Assuming that $b^2$ is given by (\ref{branchpeq}), we obtain
$k=1+\hf tb^2$, so that 
\be\label{h'}
h'(z) =- \le[tz^2 +1+\frac{tb^2}{2}\ri](z^2-b^2)^\frac 12 = -\le[t z^2  + \frac {\sqrt{1+12  t}}3 + \frac 2 3\ri] \le(z^2 - b^2\ri)^{\frac 1 2}
\ee
Since the branch-cut of the radical is $[-b,b]$ we conclude that $h'(z)$ is an odd function. Direct calculation yield
\be\label{h}
h(z) = 2 \ln\frac{z+\sqrt{z^2-b^2}}{b} - \frac z8 (2tz^2 +tb^2 +4) (z^2-b^2)^\frac 12.
\ee
It is clear that $h(b)=0$. There is the oriented branch-cut of $h(z)$ along the ray $(-\infty,-b)$, 
where $h_+(z)-h_-(z)=4\pi i$. 
Combined with (\ref{h'}), that implies 
\be\label{modh}
h(z)=O(z-b)^\frac 32
\ee
(or a higher power of $(z-b)$).  
At the point of gradient catastrophe
$t_0=-\frac{1}{12}$,  the order $O(z-b)^\frac 32$ in (\ref{modh}) should be replaced by $O(z-b)^\frac 52$.
Because of the $4\pi i $ jump along $(-\infty, -b)$,  the function $h(z)$ does not have $O(z+b)^\frac 32$
behavior near $z=-b$; however, $\Re h(z)$ does have $O(z+b)^\frac 32$ behavior near $z=-b$.

\paragraph{Non-symmetric case: $a\neq 0$.}
Following the same lines (the algebraic computation being a bit more involved) one obtains 
\be
h'(z)  = - t\le(z^2 + az - b^2\ri) \sqrt{ (z- a)^2-b^2}   \label{nsh'}
\ee
where $a,b$ are given by (\ref{othersols}).
A direct computation yields in $w = z-a$ as (using $tb^2 a^2 - \frac 38 tb^4=2$)
\be
h(z) = \int_{a-b}^z h'(\zeta)\d\zeta = 
- \sqrt{w^2-b^2} \le[ \frac t 4 (w^2-b^2)(w+4a) + \frac {2} {b^2} w\ri] +  2\ln \le(\frac {w+\sqrt{w^2-b^2}}b\ri)
\label{nsh}\ .
\ee

\br\label{rhphremark}
One can 
verify directly that $h(z)$ satisfies the following RHP:
\begin{enumerate}
\item  $h(z)$ is
analytic (in $z$) in $\C\setminus\{ 
\g_m\cup (-\infty,\l_0)\}$
and 
\be\label{assh}
h(z) =-\qt tz^4 - \hf z^2+2\ln z -\ell +O(z^{-1})~~~~{\rm as }~~~~z\ra\infty,
\ee
where 
\bea
\label{l}
\ell&\&=\ln\frac{b^2}{4}-\frac{b^2}{8}-\hf ;\qquad ~~~\hbox{(symmetric case, with $b$ given by (\ref{branchpeq}).)}\\
\label{l2}
\ell&\&=  \ln \frac {b^2} 4 - \frac {2a^2+ b^2}{8} - \frac 1 2\ ,\hbox{(nonsymmetric case,  with $a,b$ given by (\ref{othersols})}
\eea
\item $h(z)$ satisfies the jump condition
\be\label{rhph}
h_+ + h_-=0~~~~ {\rm on}~~\g_{m}~~~ ~{\rm and}~~~~~h_+-h_-= 4\pi i ~~~~ {\rm on}~~(-\infty,\l_0),
\ee
\end{enumerate}
\er

\br\label{rhphtxremark}
As it was mentioned above, the solution to the scalar RHP for $h$ commutes with differentiation
in $z$;  on the same basis, it commutes with differentiation
in $t$ as well.
Thus, we obtain the following
RHP for $h_t$ (symmetrical case):
\begin{enumerate}
\item  $h_t(z)$ is
analytic (in $z$) in $\bar\C\setminus 
\g_m$ and 
\be\label{assht}
h_t(z) = -\qt z^4 - \ell_t +O(z^{-1})~~~~{\rm as }~~~~z\ra\infty,
\ee
where 
\be\label{lt}
\ell_t=\frac{3}{32}b^4=\frac{2+12t-2\sqrt{1+12t}}{24t^2};
\ee
\item $h_t(z)$ satisfies the jump condition
\be\label{rhpht}
h_{t+} + h_{t-}=0~~~~ {\rm on}~~\g_{m}.
\ee
\end{enumerate}
These RHP has the unique solution
\be\label{hthx}
h_t(z)=- \frac{z}{8}(2z^2+b^2)\sqrt{z^2-b^2}
\ee
that can be verified directly. In the non-symmetrical case,
$h_t$ can be calculated in a similar way.
\er

\br\label{eqmea}
The $g$-function $g_t(z)$ was defined in \cite{ArnoDu},  eq. (3.2), as 
\be\label{gAD}
g_t(z)=\int_{-b}^b\ln(z-\x) d\mu_t (\x),
\ee
where $\mu_t$ is the equilibrium measure  in the external field
$f(z,t)$ and $x=n/N=1$. In the case $t\ge 0$, the equilibrium measure $\m_t$ is the unique Borel probability measure  on $\R$  that
minimizes the functional
\be\label{funct}                          
I_f (\mu) =  \int\int \ln\frac{1}{|x-y|}  d\mu(x)d\mu(y) + \int f(x,t))d\mu(x)
\ee                           
among all Borel probability measures on $\R$. (Here the subindex $t$ does not mean differentiation.) 
The measure $\mu_t$ can be calculated
explicitly. It turns out to be  supported on the interval $I=[-b , b]$, where $b$ is defined by (\ref{branchpeq}),
and it has a density given by
\be\label{dmu}            
\frac{d\mu}{dx}=\frac{t}{2\pi}\left(x^2+\frac{\sqrt{1+12t}+2}{3t}\right)\sqrt{b^2-x^2},~~~~~~~x\in[-b,b].
\ee
(In fact,  the case $t\in (t_0,0)$, the equilibrium measure $\m_t$                                                                            
minimizes (\ref{funct}) among all Borel probability measures with the support on $[-b , b]$.)
The function $g_t(z)$ satisfies the requirements of Section \ref{eqreqs} because of (\ref{gAD}).
Since the RHP  for $g$ has a unique solution, we have $g_t(z)=g(z)$.

\er

\subsection{Discussion about existence of $h$}\label{sect-exist}

To the reader it could be a little bit of a mystery as  to  why there exists any function $h(z,t)$ satisfying the above long list of conditions.
However,  this result was proven in a general setting, that is, for any polynomial  $f(z,t)$ and for any $t\in\C$ in \cite{BertoBoutroux}. 
The idea of the proof is quite simple. Suppose that we have our contours $\Omega$ and we want to find the $h(z,t)$ 
function for a specific value $t\in \C$. Assume,
on the other hand, that for a certain value of the parameter $t$ (for example, for  $t_*\in(- \frac 1{12},0)$,) 
one can somehow find $h(z,t_*)$,  satisfying all of the above requirements (for example,
$h$ can be calculated directly using the residue theory, as above, or by use of the potential theory). 
Then one chooses a path in the parameter space ($t$-plane) that connects $t_*$ to $t$ and shows that 
the requirements can be maintained throughout the path; we shall call this the {\bf continuation principle in the parameter space}. 
This idea (implemented in slightly different form) was at the basis of the discussion of \cite{TVZ1} and \cite{BertoBoutroux}. 
In a general situation with $f(z)$ being an arbitrary polynomial, the existence of a suitable 
$h(z,t_*)$ 
was established in \cite{BertoBoutroux}, but in the present paper we will prove all the inequalities for $h(z,t_*)$,  $t_*\in(- \frac 1{12},0)$,
directly. In fact, the  continuation principle is not limited to the polynomial or even rational potentials $f$. For example, 
in the context of the semiclassical limit of the focusing NLS,
the continuation principle for a large class of  analytic  $f$,
 was stated and proven in  \cite{TV1}

To indicate the obstacles that make the continuation principle nontrivial we point out that, as we follow our path in $t$--plane,
 it may happen that the regions where $-\Re h<0$ (the {\bf sea}) moves in such a way to either pinch off one 
of the complementary arcs or to ``expose'' one of the main arcs (or {\bf causeways}); 
in that case  we can use  local analysis  to guarantee that a new main arc (causeway) or complementary arc respectively
can be ``sewn in'' in order to adjust the situation; such an adjustments increases the genus of the solution. 
In fact it is quite a daunting task to try and describe in words this process; we invite the reader to have a close look at 
the pictures of the ``phase diagrams'' Figs. \ref{Generic}, \ref{RealAxis}, \ref{Wedge}, \ref{Tri}, \ref{BiWedge}, \ref{BiWedgeSym}. 
The reader should try and imagine how the main arcs and complementary arcs (which are not marked in the pictures) deform 
as we cross the phase-transition curves, also known as breaking curves, indicated there.
In fact an interactive exploration tool  was designed in Matlab and it is available upon request.
\subsection{Schematic conclusion of the steepest descent analysis}\label{sect-stdesc}
The final steps in the steepest descent analysis involve adding additional contours, the {\bf lenses}, 
which enclose each main arc and lie entirely within the $-\Re h<0$ region (the sea). One then re-defines $T(z)$ within 
the regions between the main arc and its corresponding lens by using the factorization
\begin{equation}\label{mainfact}
\begin{pmatrix} a&d\\0&a^{-1}\end{pmatrix}=
\begin{pmatrix} 1&0\\a^{-1}d^{-1}&1\end{pmatrix}
\begin{pmatrix} 0&d\\-d^{-1}&0\end{pmatrix}
\begin{pmatrix} 1&0\\ad^{-1}&1\end{pmatrix}
\end{equation}
of the jump matrices of $T$ so that 
\be
T_+(z)= T_-(z) 
\le[
\begin{array}{cc}
1 & 0\\
\nu_m^{-1}{\rm e}^{-Nh_-} & 1
\end{array}
\ri]
\le[
\begin{array}{cc}
 0& \nu_m\\
-\nu_m^{-1} & 0
\end{array}
\ri]
\le[
\begin{array}{cc}
1 & 0\\
\nu_m^{-1}{\rm e}^{-Nh_+} & 1
\end{array}
\ri],
\ee
where $\nu_m$ is the (constant!) value of $\nu(z)$ on the main arc under consideration.
Therefore, defining $\wh T(z)$ as $T$ outside of the lenses and  by 
\be
\wh T(z):=  T(z) 
\le[
\begin{array}{cc}
1 & 0\\
\mp \nu_m^{-1}{\rm e}^{-Nh} & 1
\end{array}
\ri]
\label{hatT}
\ee
in the regions within the lenses and  adjacent to the  $\pm$ sides of $\gamma_m$ one achieves a 
new problem with jumps that are constant on the main arcs and exponentially close to the identity or constant jumps on the lenses and complementary arcs.

We spend a few more words for the ``genus zero'' case, namely, when there is a single main arc $\gamma_m$ connecting two 
endpoints $\lambda_0, \lambda_1$, since this is the situation mostly relevant to the analysis here; the case with several arcs, 
for the case of real potentials on the real line was fully treated in \cite{DKMVZ} and in the complex plane in \cite{BertolaMo}; 
while not being conceptually more difficult, it requires the introduction and use of special functions called {\em Theta functions}. 

\subsubsection{The ``genus zero" case}
This is the case when there is a single main arc $\gamma_m$ that connects two endpoints $\lambda_0,\lambda_1$; 
since the coefficients $\nu_j$ are defined up to multiplicative constant, we can and will assume without loss of 
generality that they have been normalized so that the $\nu_m$ on the main arc satisfies $\nu_m=1$. 
Then the RHP for $\wh T$ is 
\be
\le\{\begin{array}{ccc} 
\wh T_+(z)=\wh T_-(z) \left(
\begin{array}{cc}
1 &0\\
 e^{-Nh}& 1
\end{array}
\right)~~~&{\rm on~the~upper~and~lower~ lips~respectively,}  \\
%
\wh T_+(z)=\wh T_-(z) \left(
\begin{array}{cc}
0 &1\\
-1& 0
\end{array}
\right)=\wh T_-(z)i\s_2~~~&{\rm on}~ \gamma_m. \\
\wh T_+(z) = \wh T_-(z)
\left(
\begin{array}{cc}
1 & e^{Nh}\\
0& 1
\end{array}
\right) & \hbox{on} \ \ \Omega \setminus \gamma_m
\end{array}
\ri.
\label{RHPlips}
\ee
Due to the sign requirements, the off--diagonal entries of the jumps on the lenses and complementary axis tend to zero exponentially fast in any $L^p$-space of the respective arcs, $p\geq 1$, but not in  $L^\infty$ because at the endpoints $\lambda_0,\lambda_1$ we necessarily have $\Re h=0$. Near these points one has to construct explicit local solutions of the RHP called {\em parametrices} \cite{DKMVZ}. The type of local RHP depends on the behavior of $h(z)$ near the endpoints.

In a generic situation one has ${\rm e}^{N h(z)}  = {\rm e}^{N C_0^{(j)} (t) ((z-\lambda_j)^{\frac 32} (1+ \mathcal O(z-\lambda_j))},\ j=0,1$ with $C_0^{(j)}(t)$ some nonzero constant. 
The critical case (or "gradient catastrophe" case) correspond to those special case whereby $C_0^{(j)} (t)=0$ at one or the other or both endpoints,  and thus 
\be
{\rm e}^{N h(z)}  = {\rm e}^{N C_1^{(j)} (t_c) ((z-\lambda_j)^{\frac 52} (1+ \mathcal O(z-\lambda_j))}
\ee
where $C_1^{(j)} (t_c)$ is now nonzero (nondegenerate gradient catastrophe). 
In the former case the local parametrix can be constructed in terms of Airy functions and its construction is very well known 
since  \cite{DKMVZ} (see also \cite{ArnoDu}, \cite{BertolaMo}). The latter case requires the solution of a special RHP which can 
be reduced to an instance of the RHP for the Painlev\'e\ I Problem \ref{P1RHP}. This was done in \cite{ArnoDu} and will not be repeated here.
We point out that one of the main distinctive features is that 

\begin{figure}[t]
 \resizebox{0.35\textwidth}{!}{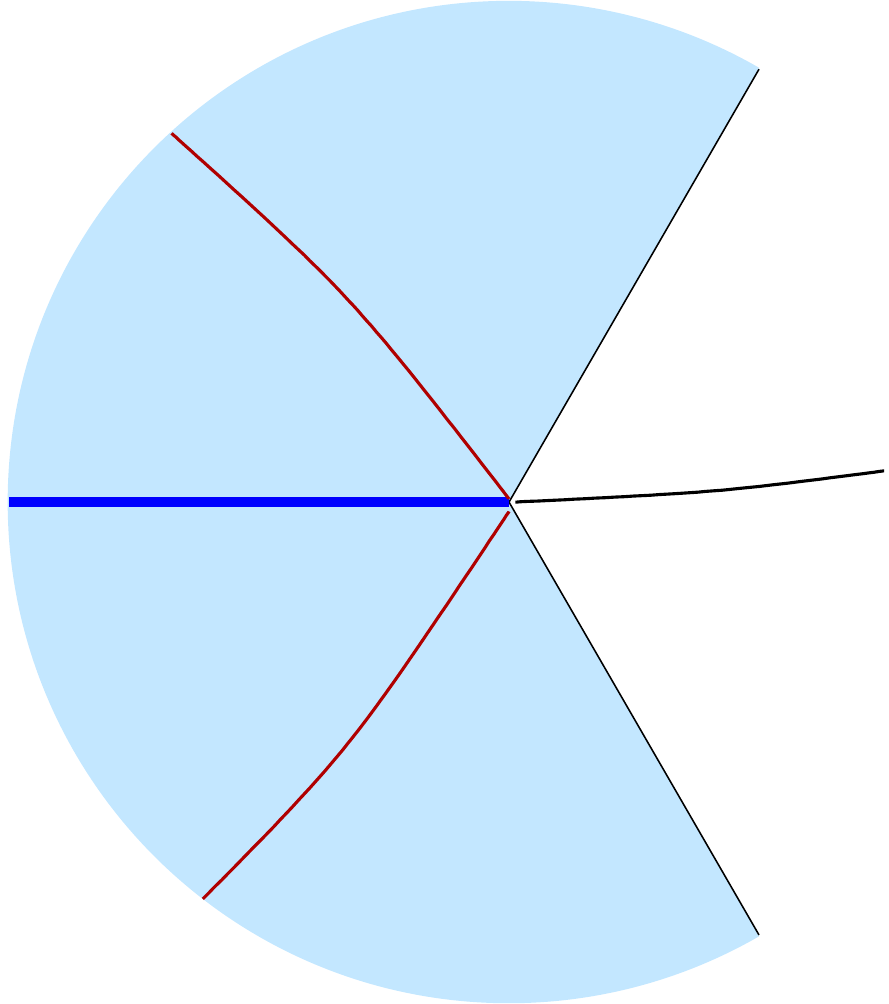} \hspace{0.1\textwidth}  \resizebox{0.4\textwidth}{!}{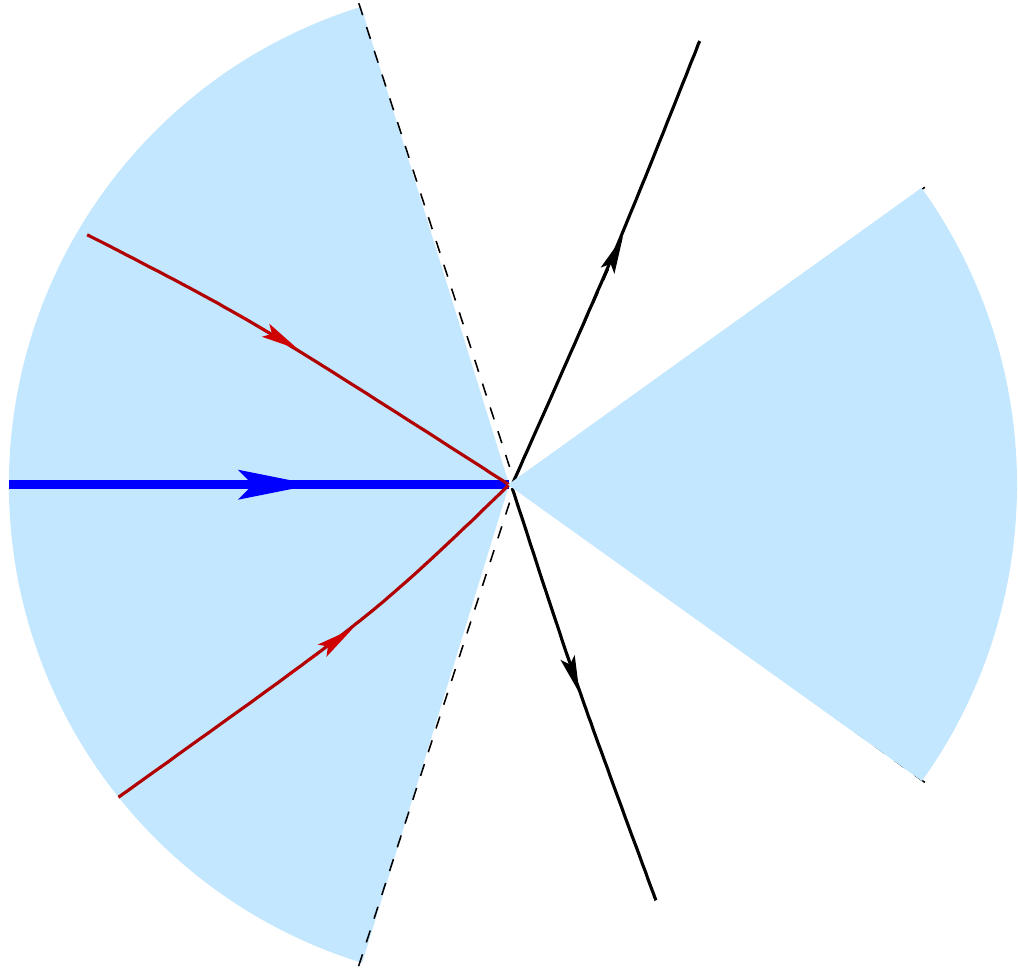}
 \caption{The two typical configurations of level curves and sign distributions near the endpoint in the generic case (left) and critical case (right). 
Indicated are the complementary arcs $\gamma_c$ (there might be only one complementary arc in the critical case, 
depending on the function $\nu(z)$) and the lenses. The blue (darker) color corresponds to the region where $-\Re h<0$ (the sea).
The parameter $\varkappa$ equals $\nu_1$ or $1 -\nu_3$ depending on the endpoint under consideration (see Fig. \ref{YRHP} and parameters therein).}
 \label{levelcurves}
 \end{figure}
 
\begin{itemize}
\item
 in the generic case there are {\bf three} level curves $\Re h=0$ that emanate from the corresponding endpoint $\lambda_j$ (one of them being the main arc), see Fig. \ref{levelcurves};
\item 
in the critical case there are {\bf five} level curves $\Re h=0$, one of them being the main arc (see Fig. \ref{levelcurves}) 
\end{itemize}

The final steps in the approximation mandates that we fix two disks $\mathbb D_0,\mathbb D_1$ (small enough not to enclose any other endpoint) around  the endpoints $\lambda_0,\lambda_1$ and define a suitable approximate solution 
\be
\Phi(z):= \le\{
\begin{array}{cc}
\Phi_{ext}(z) & \hbox{ for $z$ outside $\mathbb D_{0,1}$}\\
\Phi_{ext}(z) \mathcal P_0(z) & \hbox {inside $\mathbb D_0$}\\
\Phi_{ext}(z) \mathcal P_1(z) & \hbox {inside $\mathbb D_1$}
\end{array}
\ri.
\ee
such that the error matrix $\mathcal E(z):= \wh T(z) \Phi^{-1}(z)$ solves a small--norm Riemann--Hilbert problem (as $N\to \infty$) 
and thus can be -in principle- be completely solved in Neumann series. Here by $\mathcal P_{0,1}(z)$ we denote the 
parametrices near the endpoints $\l_{0,1}$ respectively.

In all situations the matrix $\Phi_{ext}$ (``model solution'' or ``exterior parametrix'') solves a RHP
of the form (model problem) 
\be
\Phi_{ext}(z)_+ = \Phi_{ext} (z)_- i\s_2\ ,\ \ \ z\in \gamma_m=[\lambda_0,\lambda_1],\ ,\qquad
\Phi_{ext}(z) = \1 + \mathcal O(z^{-1})\ ,\ \ \ \ z\to\infty
\ee
with some particular growth behavior near the endpoints which depend on the scaling limit under consideration. 
In the  usual case it satisfies 
\be
\Phi_{ext}(z) = \mathcal O(z-\lambda_j)^{-\frac 1 4}\ ,\ \ z\to \lambda_j,
\ee
but in special cases the behavior needs to be modified.

At any rate, once we have achieved a suitable approximation for $\wh T(z)$, 
the recurrence coefficients for the orthogonal polynomials can and will be recovered via the formulae
\be
\h_n = -{2i\pi}{\rm e}^{N\ell} \le(T_1\ri)_{12} \ ,\qquad\a_n  = \le(T_1\ri)_{12} \le(T_1\ri)_{21}\ ,\qquad
\b_n = \frac {\le(T_2\ri)_{12}}{\le(T_1\ri)_{12}} - \le(T_1\ri)_{22}\label{Talphabeta}
\ee
where $\wh T(z)$  near $\infty$ equals $T(z)$ (since we are in the exterior region) and has expansion 
\be
\wh T(z) = T(z) = \1 + \frac {T_1}z + \frac {T_2} {z^2} + \dots\ ,\ \ \ z\to\infty\ .
\ee
The latter coefficient matrices can be obtained from the corresponding expansion of $\Phi_{ext}(z)$ near infinity, to within the error determined by $\mathcal E$; in the generic ({\em regular}) case  ($t\neq t_0,t_1, t_2$ and not on the breaking curves) the  parametrices $\mathcal P_0,\ \mathcal P_1$ are the well--known Airy parametrices and the standard error analysis (which we do not report here) shows that $\mathcal E$ introduces an error of order $\mathcal O(N^{-1})$. 

In this case the {\bf exterior parametrix (model solution)} $\Phi_{ext}$ in the genus $0$ region
is the ``standard''  solution  (that we shall denote by $\Psi_0$)  to the following ``model RHP'':
\be
\le\{\begin{array}{ccc} 
& \Psi_0(z)  & {\rm is~ analytic~in~} \C\setminus [\lambda_0,\lambda_1], ~~~~~ ~~~~~~~~~~~~~~\\
 &\Psi_{0+}(z)=\Psi_{0-}(z)i\s_2~~~&{\rm on}~~[\lambda_0,\lambda_1],  \\
%
&
\Psi_0 (z)=\1+O(z^{-1})~~&{\rm as}~~ z\ra\infty, \\
&\Psi_0(z) = \mathcal O(z-\lambda_{0,1})^{-\frac 14}\ ,\& \ z\to\lambda_{0,1}.
\end{array}
\ri.
\label{RHPPsi_0}
\ee
The solution to the RHP (\ref{RHPPsi_0}) is given by
\be\label{Psi0}
\Psi_0(z)=
\frac{(\s_3+\s_2)}2 \left(\frac{z-\lambda_1}{z-\lambda_0}  \right)^{\frac{\s_3}{4}}\!\!\!\!\!(\s_3+\s_2)=
\left(\frac{z-\lambda_1}{z-\lambda_0}  \right)^{\frac{\s_2}{4}},
\ee
which has expansion (recall our notation $\l_1 =a+b$, $\l_0 = a-b$)
\be\label{Psi0ass}
\Psi_0(z)=\1-\frac{b}{2z}\s_2+\frac{b^2}{8z^2}\1 - \frac {ab \s_2}{2 z^2} +O(z^{-3})\ ,\qquad z\to\infty.
\ee 
Thus, near $z=\infty$, one finds
\be
T(z) = \le(\1 + \frac 1 z\mathcal O (N^{-1}) \ri) \Psi_0(z)  \ \Rightarrow\ \ \
T_j = (\1 + \mathcal O(N^{-1})) \Psi_{0,j},\ \label{T0app}
\ee
where $\Psi_{0,j}$ denote the Taylor coefficients of $\Psi_0(z)$ at infinity.
\subsubsection{Recurrence coefficients in the genus $0$ cases}
As explained in Section \ref{modeqsubsec} there are two types of genus zero solutions and hence the final formul\ae\ are different.
Using (\ref{Talphabeta}), the approximation (\ref{T0app}), the explicit form of $\Psi_0$ (\ref{Psi0}) and the explicitly 
calculated expressions for $\l_1, \l_2$, one finds the results summarized in Table \ref{genus0asympt}.
\begin{table}[h]
\begin{center}\begin{tabular}{c|c}
Symmetric genus 0 ($\d t:= t+\frac 1 {12}$) & Non symmetric genus 0 ($\d t:= t-\frac 1 {15}$)\\[10pt]
\hline\\
$ \ds
\begin{array}{c}
\ds \h_n =
 2\pi \le( \frac {\sqrt{12\d t} - 1}{6t}\ri)^{\frac 1 2 +N}\hspace{-20pt} \exp\le[ N\frac{4-  (\sqrt{12\d t} +1)^2  }{24t} \ri] \\[10pt]
\hline\\
\ds \a_n = \frac{b^2}{4}=\frac{\sqrt{12\d t}-1}{6t}   \\[10pt]
\hline\\
\ds \b_n = a= 0\hspace{-80pt}\phantom{ -  i\le(\frac 1 {5t} \le(3+2ii\sqrt{15\d t} \ri)\ri)^\frac 12
}
\end{array}
$
&
$ \ds
\begin{array}{c}
\ds \h_n = 
2\pi \le(\frac { i\sqrt{15 \d t} - 1}{15 t} \ri)^{\frac 12 +N}
\hspace{-20pt}\exp\le[N\frac { 9 + 4 i \sqrt{15 \d t} -30 \d t  }{60 t}\ri]
\\[10pt]
\hline\\
\ds
\a_n = \frac{b^2}{4}=  \frac {i \sqrt{ 15\d t} - 1}{15 t} \\[10pt]
\hline\\
\ds
\b_n = a= -  i\le(\frac 1 {5t} \le(3+2i\sqrt{15 \d t} \ri)\ri)^{\frac 12}
\end{array}
$
\end{tabular}
\end{center}
\caption{The leading order approximations of the "square-norms" and recurrence coefficients in the two genus-zero cases: 
all expressions are understood to within an error term of $\mathcal O(N^{-1})$. T
he expressions for $\h_n = \pi b {\rm e}^{N\ell}$ and $\ell$'s are in (\ref{l}, \ref{l2}).
In fact a more careful analysis shows that  $\b_n$  in the symmetric case is exponentially small \cite{ArnoDu}. 
The reason is that the RHP can be seen to be close exponentially to a RHP with a symmetry $z\mapsto -z$, for which 
the expression for $\b_n$ automatically yields zero.
Note that there are two choices of signs for $a$ in (\ref{othersols}) (the choice of signs for $b$  amounts only in 
exchanging the labels of the branch-points) that lead to different (but quite similar)  formul\ae\ and results;
we will formulate all the results for this particular choice whereby $a\simeq -3i$.}
\label{genus0asympt}
\end{table}
\subsubsection{The regions of higher genera}
From the global analysis reported in Figures \ref{Generic}, \ref{RealAxis}, \ref{Wedge}, \ref{Tri}, \ref{BiWedge}, \ref{BiWedgeSym}, the reader can see that there are regions where the hyperelliptic surface of $h'(z)$ has genus $1$ or $2$. In this case, while the general scheme of the steepest descent analysis remains intact, the solution of the relevant model problem for $\Psi_0$ requires Riemann--Theta functions. Formul\ae\ can be found in \cite{DKMVZ, BertolaMo}. The recurrence coefficients also are expressible in terms of Theta functions. In fact the formul\ae\ in \cite{DKMVZ} could be directly applied here, simply by modifying the choice of the $a,b$-cycles (in the standard lore of Riemann surfaces) as described extensively in \cite{BertolaMo}. We shall not write explicit formul\ae\ here since it would require setting up a good deal of additional notation. Suffice it  to say that the nature of the resulting expressions is one of rapidly oscillating functions of $N$ and $t$, with amplitude that depends only on $t$.
\subsection{Contour deformation.}
A general discussion of contour deformations once the appropriate  $g$--function has been found can be read in \cite{BertolaMo} 
and \cite{BertoBoutroux}; we give here a brief sketch. We advise the reader to accompany this part with the pictures that are provided plentiful.

In general, the contours of integration for the pairing $\langle p,q\rangle_{\vec \nu}$ can be deformed by use of 
the Cauchy theorem: any deformation that we shall allow must be such that the deformed contour approaches $\infty$ along the same  
direction $\arg (z) = -\frac 1 4 \arg (t) + \frac k 2 \pi$  of the original contour, so as to preserve integrability 
(we may even mandate that each contour is a straight line outside of a sufficiently large circle). 
Indeed, from  $\Re h(z) = -\Re \le(f(z) -2 g(z) + \ell\ri)$ and from the fact that $g(z)$ is bounded by a logarithm, we see that for $|z|$ 
large enough the sign of $\Re h$ is the same as of $-\Re f$, for which the above directions are the directions of  the steepest descent.

The final deformation  of the contours must fulfill the following requirements, that we describe referring to 
the regions $-\Re h>0$ as {\bf (dry) land},  $-\Re h<0$ the {\bf sea} (or other watery expression) and the main arcs 
(where $\Re h\equiv 0$) as {\bf bridges} or {\bf causeways}:
\begin{itemize}
\item Along each contour $-\Re h$ is always nonnegative, $-\Re h\geq 0$, i.e. each contour does not get wet;
\item if two or more (oriented) contours have been deformed so that they go through the same bridge (main arc), then the {\bf traffic} 
(i.e. the weight of that part of contour) is the (signed) sum of all the traffics. For example if $\varpi_1,\varpi_2$ are deformed so that they pass trough the same main arc, then the weight of that arc shall be $\nu_1-\nu_2$;
\item {\em each bridge} (main arc) must carry a nonzero traffic, or else one needs to find a different $g$--function;
\item the precise form of the deformed contours as they enter/leave a bridge (i.e. the complementary arcs) is largely irrelevant, 
but for definiteness we shall stipulate that they proceed for a short distance along the steepest ascent line or $-\Re h$.
\end{itemize}
In order to offer some rigorous study we consider in more detail the symmetric case of genus $0$.
\begin{lemma}\label{lemma-signs}
In the case $\nu_2=-1$ and $t\in(-\frac{1}{12}, 0)$,  the function $\Re h(z)$, where $h(z)$ is given by (\ref{h}),
satisfies the sign conditions along the contour $\Omega$.
\end{lemma}
{\bf Proof.} 
First, it follows  from (\ref{h}) that $\Re h(z)=0$ on  $[-b,b]$. 
To show that $\Re h(z)>0$ immediately above the main arc  $[-b,b]$,
it is sufficient, by the Cauchy-Riemann equations, to show that $\Im h'(z)<0$ on the upper shore of $[-b,b]$. The latter follows directly 
from (\ref{h'}). We can now use the oddness of  $h'(z)$ to show that $\Re h(z)>0$ also below the main arc.
So, the correct signs around $\g_m$ are proven.
The  correct distribution of signs of $\Re h$ along the complementary arcs follows  from the topology of
zero level curves of $\Re h$. 
Because $\Re h(z)$ is even ( $\left|z+\sqrt{z^2-b^2}\right|$ is even), 
 it is sufficient to consider level  curves only in the right halfpane.
Direct check shows that both terms in (\ref{h}) have positive real part on $i\R^+$.
There are two legs  of zero level curves  of $\Re h$ emanating from $z=b$ and  four legs coming  from  infinity
with asymptotes $\pm \frac \pi 8$, $\pm \frac{3\pi}{8}$, see (\ref{assh}). Denote these legs as $\chi_{\pm j},~j=1,2$ respectively.
Since $\Re h (z)<0$ as real $z\ra b+0$ and $\Re h (z)>0$ as real $z\ra +\infty$, we conclude that $\Re h(z_*)=0$
at some $z_*\in (b,+\infty)$.  So, the only possible topology of the level curves  $\chi_{\pm j}$ is that $\chi_1$ is connected
with $\chi_{-1}$ through $z_*$ and $\chi_2$,  $\chi_{-2}$ are  connected to $b$ (since $\Re h(z)$ is a harmonic function, its level 
curves cannot form bounded loops). Thus, one can choose as complementary arcs any smooth curves ``on the land'' between $\chi_1$ and  $\chi_{2}$
and between $\chi_{-1}$ and  $\chi_{-2}$, i.e., 
in the region where $\Re h(z)<0$  that connect $b$ and $\infty$. Fig. \ref{Generic} shows (in red) main arcs $\g_m$, but not complementary arcs $\g_c$.
However, level curves $\chi_j$ can be visualized in the ``snapshots'' of $z$-plane that correspond to $t\in(-\frac 1{12},0)$.
\QED
\begin{remark}\label{rem-signs}
Similarly to Lemma \ref{lemma-signs}, it is easy to establish the correct sign distribution outside $\g_m$ 
in the case when $h$ is given by (\ref{nsh}), which is valid, for example, when $\nu_1=\nu_2=0$,  $\nu_3\neq 0$
(Single wedge) and $t\in(0,\frac{1}{15})$,
see Fig. \ref{Wedge}. For $t\in(0,\frac{1}{15})$ both $b^2$ and $a^2$ are negative, so that $a,b\in i\C$.
From (\ref{nsh}) it follows that $\Re h(a)=0$  and setting the orientation of $\g_m$ upwards, 
we see that $\Re h'_\pm(a)\lessgtr 0$ respectively. Thus, we have the correct sign distribution outside $\g_m$. 
\end{remark}

\section{Breaking curves and global phase portraits}\label{brcurvesec}
A breaking curve $\L$ in the complex $t$-plane separates the regions of different  genera  in the
asymptotic behavior of the recurrence coefficients, or regions of the same 
genus but with different number of main arcs (see, for example, the breaking curve that joins 
$t= -\frac 1{12}$ to $t= \frac 1{4}$ in Fig. \ref{Generic}, which separates two regions of genus $2$).
It satisfies the system of equations
\be\label{brcurveeqs}
h_k'(z)=0 ~~~~{\rm and}~~~~~\Re h_k(z)=0,\label{saddle}
\ee
which is the system of 3 real equations for two complex variables $z$ and $t$. 
Here the subindex $k$ in $h_k$ indicate the genus of the Riemann surface $\mathfrak R(t)$ where $h_k$ is defined. In our cases the genus 
can be $0,1,2$. 
 To simplify notations,
we will drop the subidnex $k$ whenever the genus of $h$ is obvious.

We will consider the breaking curves in the $t$ plane where the sign requirements fail because a
saddle point $\hat z$ of $\Re h$ (a point satisfying $h_z(\hat z, t)$=0) collided with the contour $\O=\mathfrak M\cup \mathfrak C$.
That means that either a complementary arc is pinched by the rising ``sea'' or a main arc (causeway) is touched by the dry land because of the
receding ``sea''. In any case, equations (\ref{brcurveeqs}) will be satisfied at $z=\hat z$.
There are  {\bf three} cases of  breaking that we consider:
\begin{itemize}
\item genus $0$ symmetric, i.e., $h_0(z)$ is given by (\ref{h});
\item genus $0$ non-symmetric, i.e., $h_0(z)$ is given by (\ref{nsh});
\item genus $1$ (symmetric).
\end{itemize}
The resulting equations when plugging the expression for $h$ into the system (\ref{saddle}) are relatively simple and could be 
analyzed analytically; we find it much more effective and  informative to study and plot them numerically.

\subsection{Genus $0$ symmetric}\label{g0s}
Using (\ref{h'}), we obtain the following equation for the saddle point:
\be\label{z_0}
tz^2=-(1+\hf tb^2)=-\frac 13 \left( 2 + \sqrt{1+12t}\right) ~~~~{\rm or}~~~~z=\pm\sqrt{ \frac{  2 + \sqrt{1+12 t}}{-3t}}.
\ee
Substituting this into $\Re h(z)=0$ and using (\ref{h}) after some algebra yields the following implicit equation for the breaking curve
\be\label{brcurveeq1}
 \Re\left[\ln\frac{1+2 \sqrt{1+12t}+\sqrt{3[(\sqrt{1+12t}+1)^2-1]}}{1-\sqrt{1+12t}}+\frac{\sqrt{3[(\sqrt{1+12t}+1)^2-1]}}{12t}\right]=0
\ee
or 
\be\label{brcurveeq2}
\varphi(u) := \Re\left[\ln\frac{1 + 2u+\sqrt{3u^2-6u}}{1-u}+\frac{\sqrt{3u^2-6u}}{u^2-1}\right]=0,
\ee
where $u=\sqrt{1+12t}$.
Note that according to (\ref{brcurveeq2})  $\L$ is a bounded curve that starts at the point of gradient catastrophe $u=0$ because for large $u$ the expression in (\ref{brcurveeq2}) tends to $\ln (2 + \sqrt 3)$.
To obtain the asymptotics of the breaking curve $\L$ near $t=t_0$, we  use expansion (\ref{hlocb}) of $h(z,t)$ near the branch-point $z=b$, 
where the coefficients 
$C_0,C_1$ are given in Table \ref{expansions}, left column, to write
\be\label{hlocb^2}
h(z)=\frac{C_0}{(z+b)^\frac 32}(z^2-b^2)^\frac 32 + \frac{C_1}{(z+b)^\frac 52}(z^2-b^2)^\frac 52
 + \cdots~.
\ee
This expansion is a direct consequence of the modulation equation. Using (\ref{z_0}), 
we calculate $z^2-b^2=\frac{\sqrt{1+12t}}{-t}=\frac{\sqrt{12(t-t_0)}}{-t}$. Since for $t$ close to $t_0$ the saddle point $\hat z(t)$
(that satisfies $h_z(\hat z(t),t)\equiv 0$) is close to $b$,
we have
\be\label{brcurvelocz_0}
h(z(t),t)=\frac{C_0[12(t-t_0)]^\frac34}{(-2bt)^\frac 32}+\frac{C_1[12x(t-t_0)]^\frac54}{(-2bt)^\frac 52}+ O((t-t_0)^\frac 74)
=\frac{[12(t-t_0)]^\frac 54}{15(2b)^3(-t)^\frac 52}(10tb^2+4)+O((t-t_0)^\frac 74),
\ee
where we utilized the formulae for $C_0,C_1$. Now the requirement $\Re h=0$ yields $\frac 54 \arg(t-t_0)=\pm \frac \pi 2 + \pi k$, $k\in \Z$, so that
the breaking curve $\L$ near the point of gradient catastrophe $t_0$ is tangential to 
\be\label{asscript}
\arg(t-t_0)=\pm \frac{2\pi}{5}+\frac{4\pi}{5}k.
\ee
Note that there are various branches to keep track of: 
the principal branch of the radical $\sqrt {1+12 t}$ leads to the curve joining $t=-\frac 1{12}$ to $t=0$ ($u=0$ to $u=1$ correspondingly), 
light curve from $-\frac 1{12}$ to $0$ on Fig. \ref{allbreaks}; the secondary branch leads to the curve that joins $t=-\frac 1{12}$ to $t=\frac 1 {4}$ 
($u=0$ to $u=2$ correspondingly), on Fig. \ref{allbreaks}.

\subsection{Genus $0$ non-symmetric}\label{g0ns}
In this case we are looking for zeroes of $h'(z)$ satisfying $z^2 + az - b^2=0$, see (\ref{nsh'}).
They are given by  
\be\label{znonsym}
z=-\hf\pm\sqrt{\qt -\left(\frac ba\right)^2}.
\ee
Substituting (\ref{znonsym}) in $\Re h(z)=0$, where $h$ is defined by (\ref{nsh}), and repeating the previous arguments, we obtain
an implicit equation for the additional breaking curves. 
Leaving the lengthy but straightforward details aside, we obtain the curves on Fig. \ref{allbreaks} 
that join $t=\frac 1{15}$ to $t=0$ and $t=-\frac 1 {12}$ to $\frac 1{15}$ respectively.
\subsection{Genus $1$ symmetric}\label{g1}
\begin{figure}[h]
\begin{center}
\includegraphics[width=0.5\textwidth]{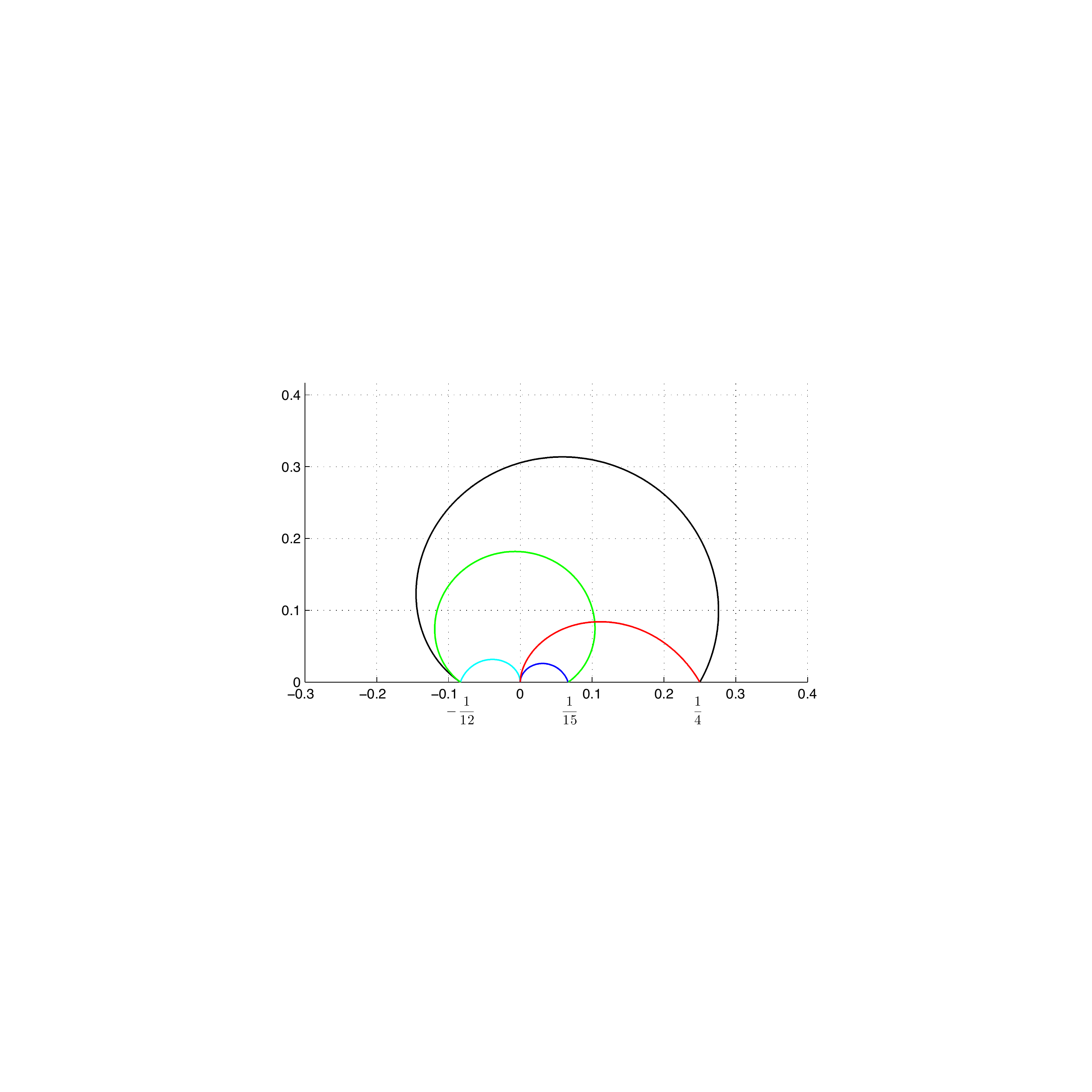}
\caption{ All breaking curves, summarized: they are symmetric about the real $t$--axis.  
Depending on the case under study, some of them may not be ``active'', namely they belong to different sheets of the phase portraits. We refer to Figs. \ref{Generic}, \ref{RealAxis}, \ref{Wedge}, \ref{Tri}, \ref{BiWedge}, \ref{BiWedgeSym} for the specifics.}
\label{allbreaks}
\end{center}
\end{figure}

For the case of genus $1$ there are $4$ branch-points; the only situation where we can have the saddle point $h'(z)=0$ on 
the zero-level set is when the saddle point is between to distinct connected components of the zero level-set of $\Re h(z)=0$. It is seen from the modulation equations that $h'(z)^2$ is always a polynomial of degree $6$; we look here for solutions where $(h'(z))^2$ is an {\bf even} polynomial.
 Since we are seeking a solution of genus $1$, there must be a single double root. By the symmetry this root must be at the origin; this allows us to write 
\be
h'(z;t) = -t z \sqrt{\le(z^2 + \frac {1+2\sqrt{t}}{t}\ri)\le(z^2 + \frac {1-2\sqrt{t}}{t}\ri)}
\ee
Indeed a simple Laurent expansion at $\infty$ yields $h'(z;t) = -t z^3 - z +\frac 2 z +\mathcal O(z^{-2})$, and evidently $h'(-z;t)=-h'(z;t)$.
Although the curve is of genus $1$, the integral of $h'(z)$ is elementary and a direct computation yields (recall that $h(z)$ vanishes at one of the branch-points)
\be
 \int_{\l_0}^0 h'(z;t){\rm d}z = h(0;t) = -\frac {\sqrt{1-4t}}{4t}  + \ln \le(
\frac {1 + \sqrt{1-4t}}{2\sqrt{t}}
\ri) 
\ee
We leave it to the reader to verify that $\Re h(0;t)$ is continuous at $t=\frac 14$ by using the identity $\frac {1 - \sqrt{1-4t}}{2\sqrt t} \frac {1 + \sqrt{1-4t}}{2\sqrt t}\equiv 1$.
The implicit equation of this breaking curve is then simply $\Re (h(0;t)) = 0$.
The curve is the one joining $t= \frac 14$ to the $t=0$ in Figs. \ref{allbreaks}, \ref{Generic}, \ref{RealAxis}, \ref{Tri}, \ref{BiWedge}, \ref{BiWedgeSym}.
\subsection{Phase portraits or distribution of genera in the complex $t$-plane}
\label{sectphase}
\begin{figure}
\begin{center}
\includegraphics[width=0.8\textwidth]{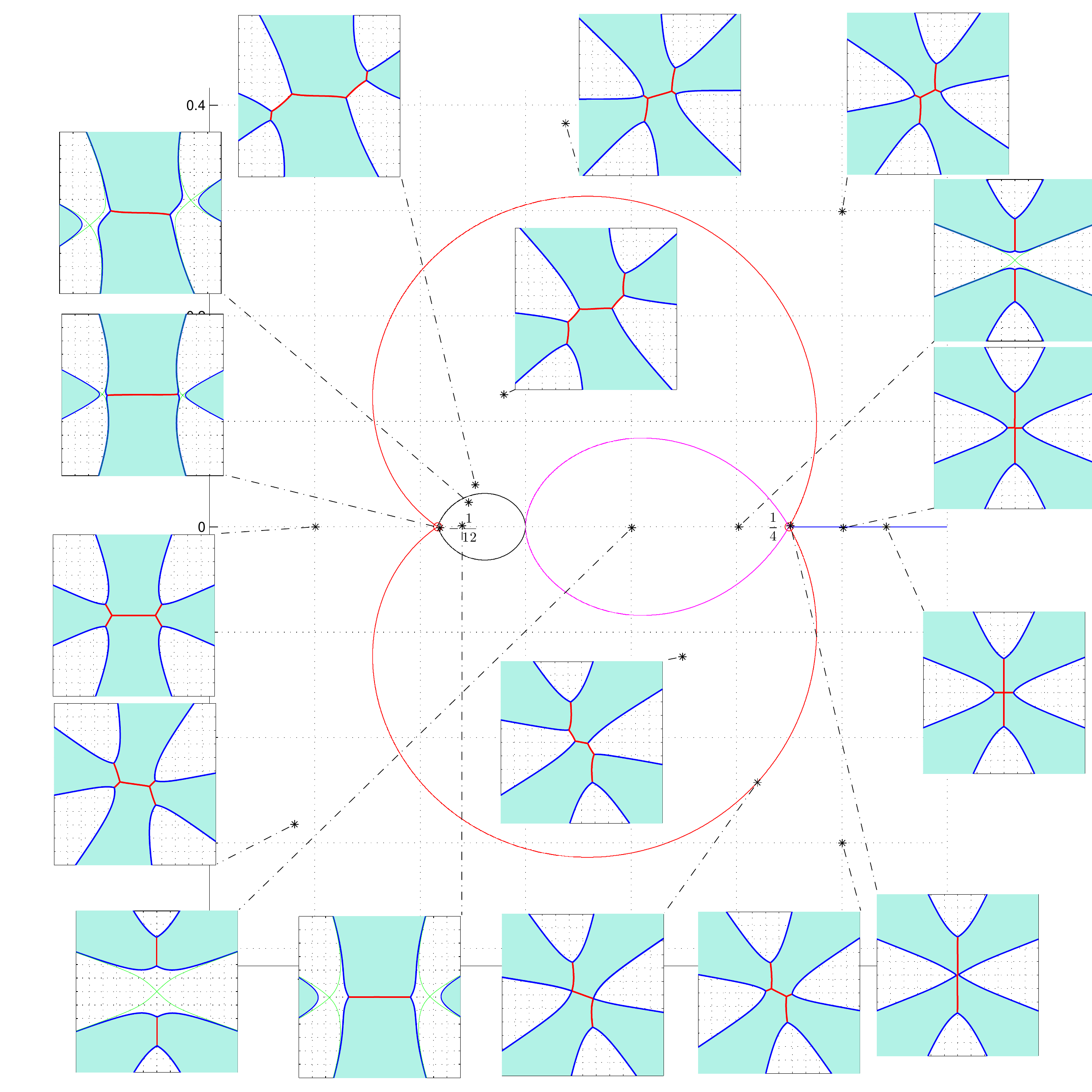}
\caption{{\bf Generic}. The region inside the curve joining $\frac 1 4$ to $0$ is of genus $1$; inside the curve that joins $t=-\frac 1 {12}$ and $t=0$ it is of genus 0 (symmetric about $z=0$). Everywhere else it is of genus $2$, except for the degeneration to genus $0$ occurring on the curve that joins $t=-\frac 1 {12}$ and $t=\frac 14$, and to genus $1$ on the ray $[\frac 1 4, \infty)$.  There is a Painlev\'e\ I transition at $t=-\frac 1 {12}$ and a Painlev\'e\ II transition at $t=\frac  14$.}
\label{Generic}
\end{center}
\end{figure}

\begin{figure}
\hspace{-0.1\textwidth}
\begin{minipage}{0.65\textwidth}
\includegraphics[width=1\textwidth]{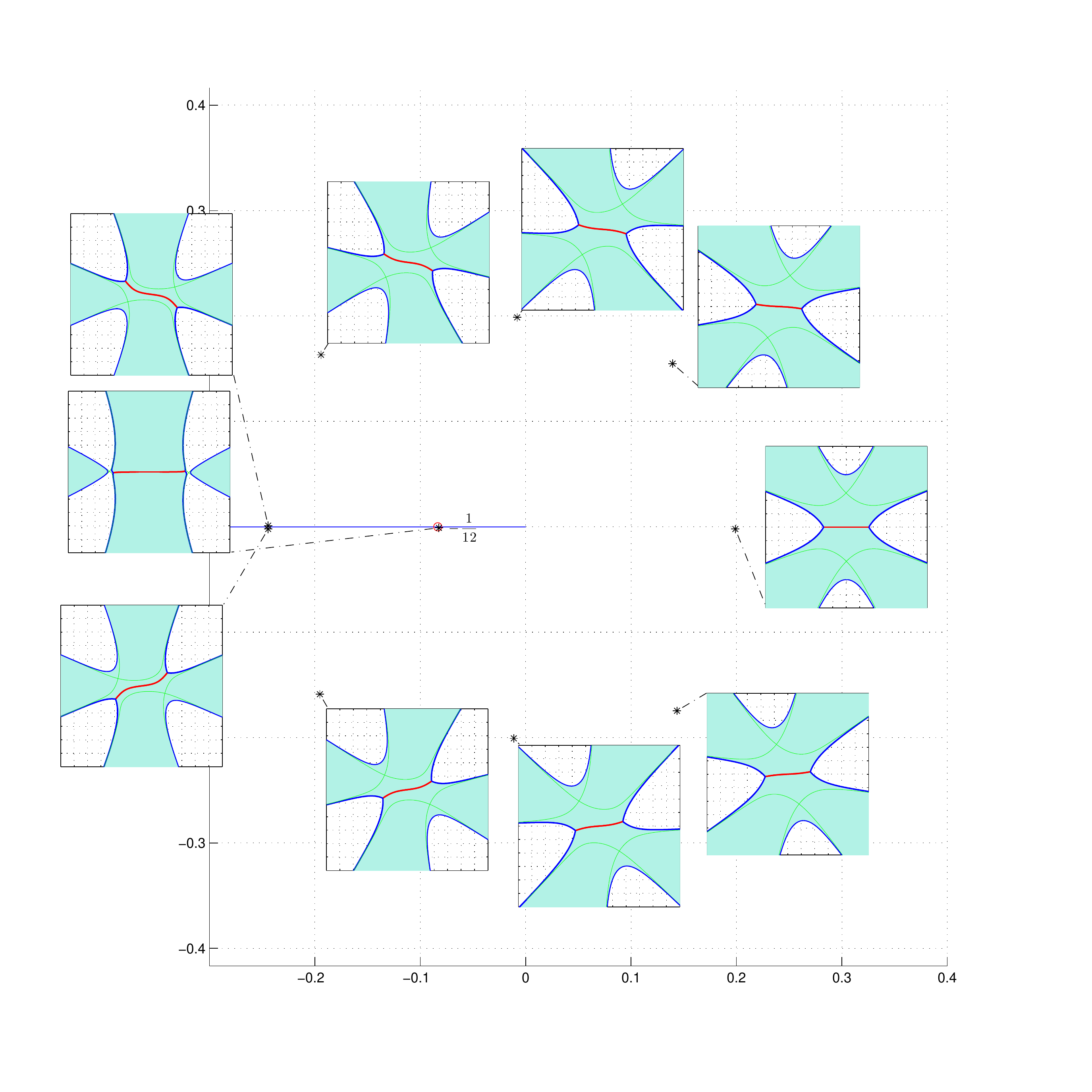}
\includegraphics[width=1\textwidth]{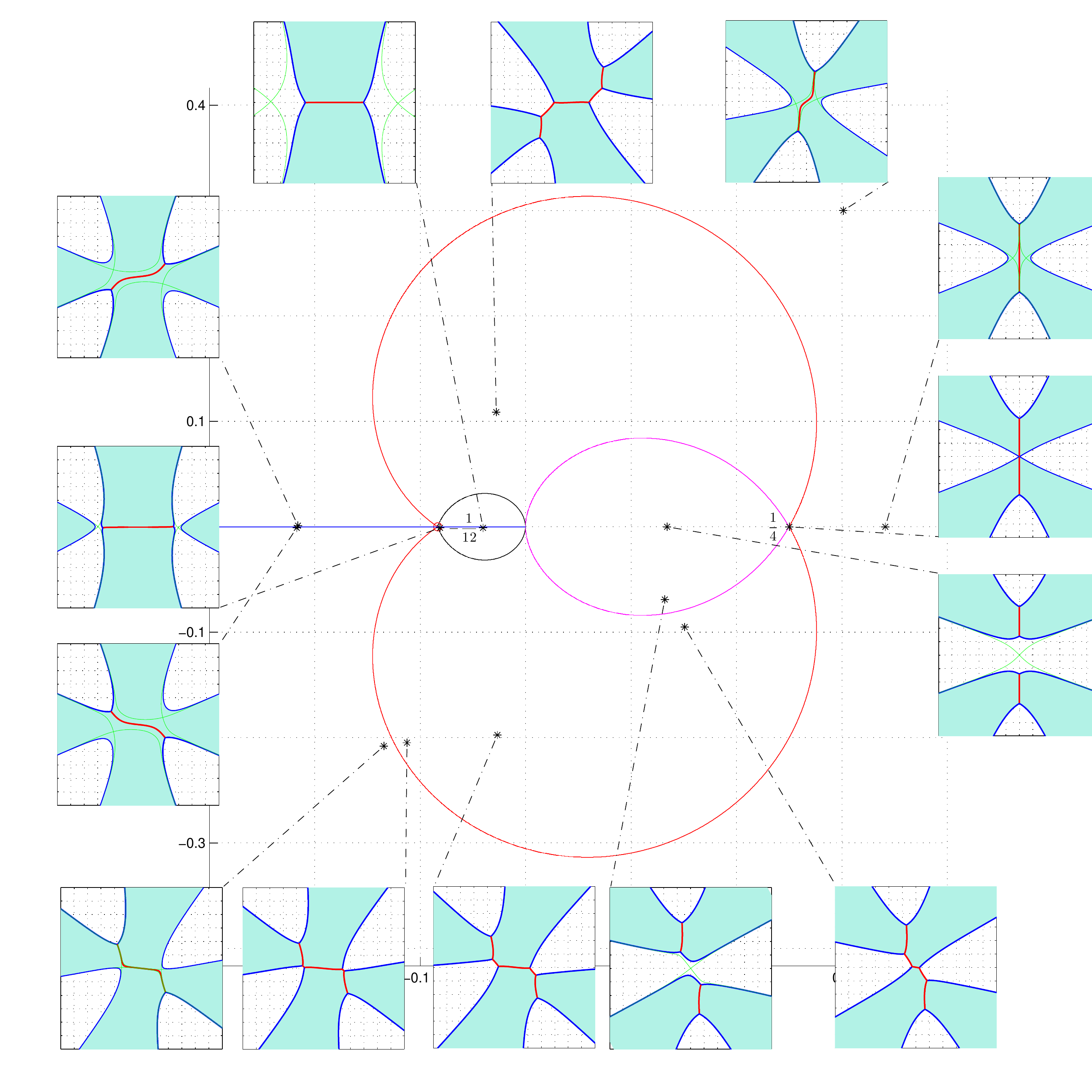}
\end{minipage}
\begin{minipage}[b]{0.35\textwidth}
\caption{{\bf Real Axis}.
There are  {\bf two sheets} glued along $t\in \R_-$; the level curves are always symmetric about $z=0$. On the first sheet the solution is always of genus $0$.
On the second sheet it is of genus $0$ outside of the curve connecting $t=-\frac 1{12}$ and $t=\frac 14$.
The region inside the curve joining $\frac 1 4$ to $0$ is of genus $1$; inside the curve that joins $t=-\frac 1 {12}$ and $t=0$ it is of genus 0 (symmetric about $z=0$). In the remaining part it is of genus $2$.}
\label{RealAxis}
\end{minipage}
\end{figure}

\begin{figure}\vspace{-0.1\textwidth}
\begin{minipage}{0.65\textwidth}
\includegraphics[width=1\textwidth]{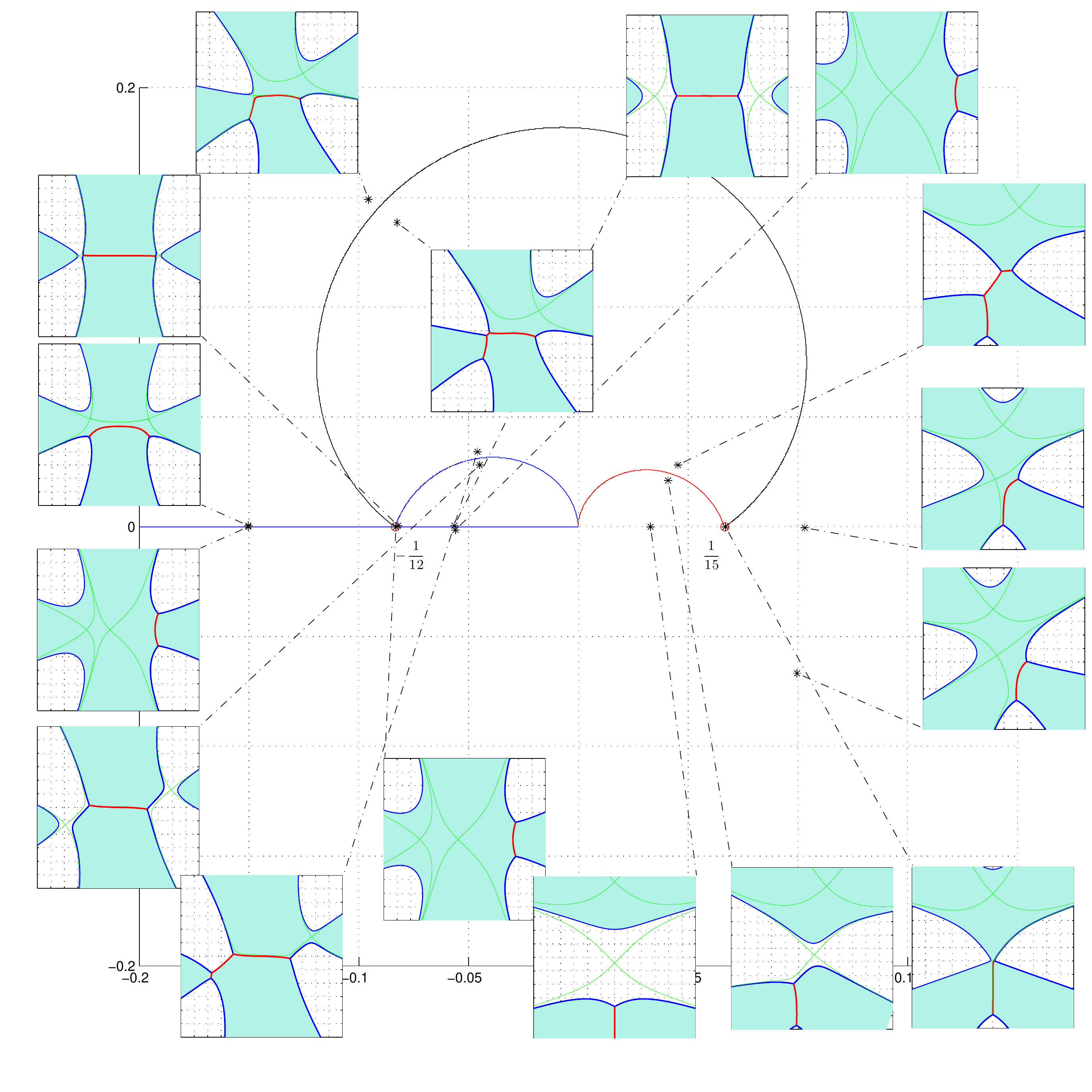}\\
\includegraphics[width=1\textwidth]{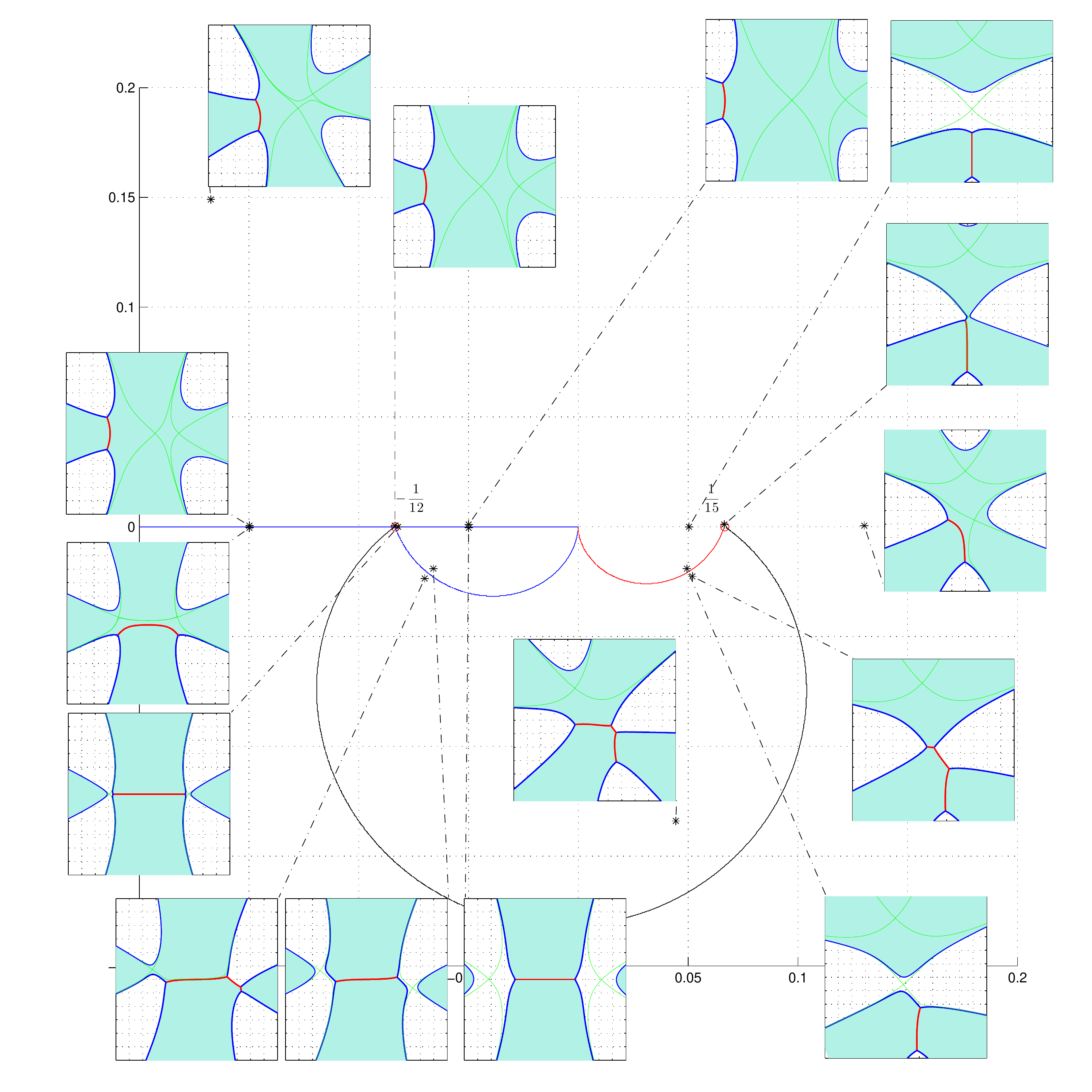}
\end{minipage}
\begin{minipage}[b]{0.35\textwidth}
\caption{
{\bf Single Wedge}.
There are {\bf four} sheets glued along $t\in\R_-$; shown here are only sheet $1$ and $2$, because the sheets $3,4$ are copies of sheet $1,2$ where the function $h(z)$ has undergone $z\mapsto -z$.
Note that there at the critical point $t=\frac 1{15}$ on all four sheets we have a transition of type Painlev\'e\ I (and also at $t=-\frac 1 {12}$).
}
\label{Wedge}
\end{minipage}
\end{figure}

\begin{figure}
\vspace{-0.1\textwidth}
\begin{minipage}{0.65\textwidth}
\includegraphics[width=1\textwidth]{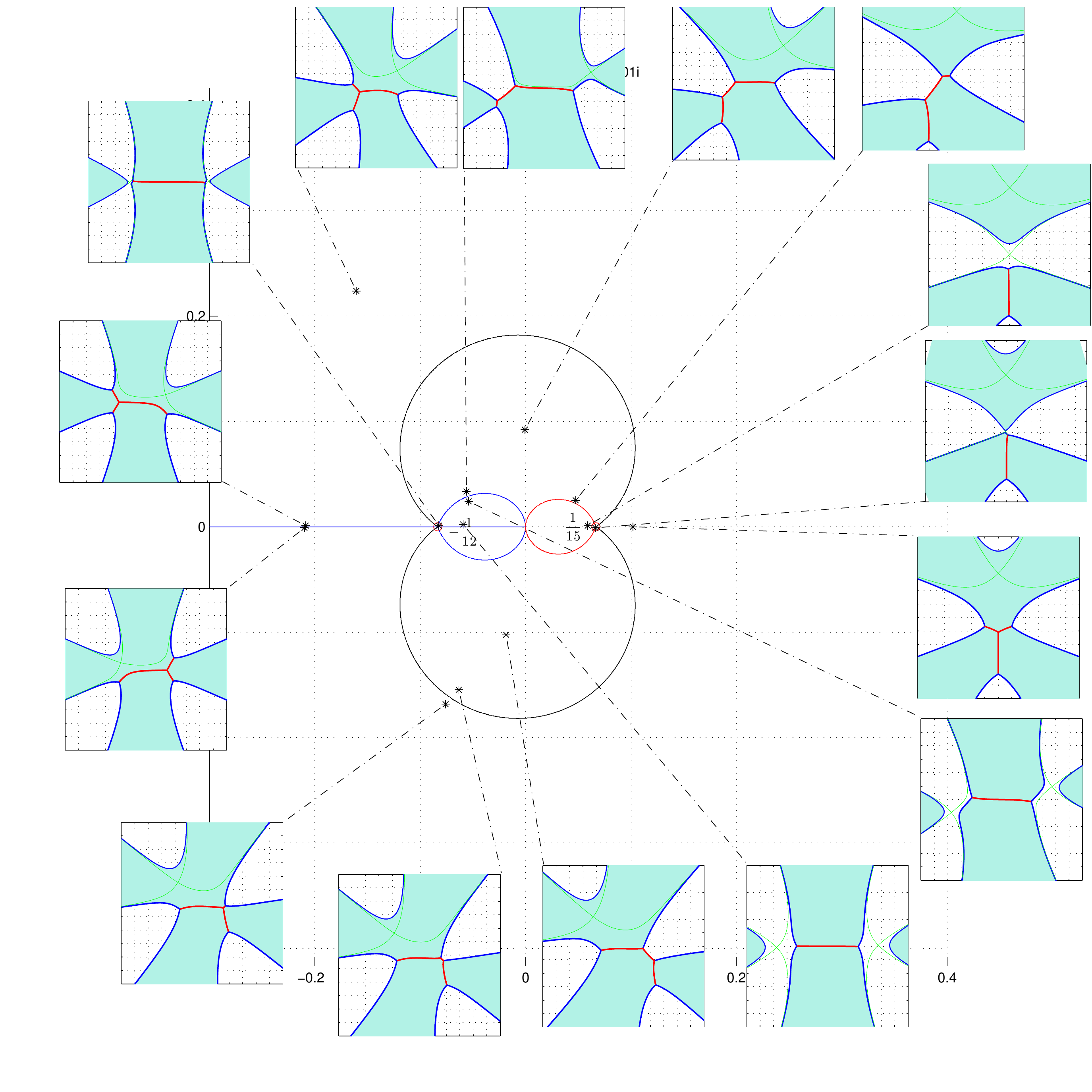}\\
\includegraphics[width=1\textwidth]{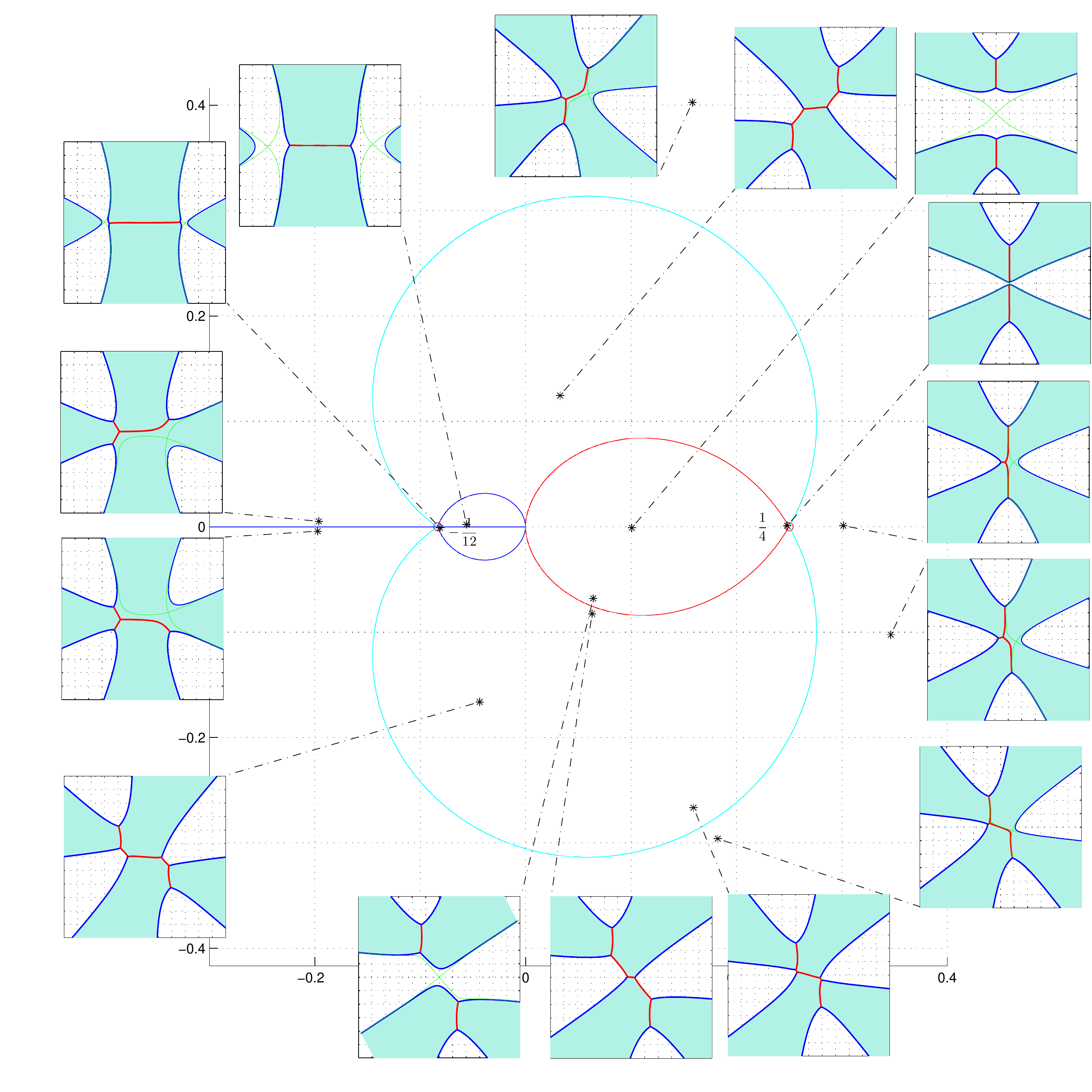}
\end{minipage}
\begin{minipage}[b]{0.35\textwidth}
\caption{{\bf Consecutive Wedges}.
There are {\bf four} sheets glued along $t\in \R_-$; shown here are only sheet $1$ and $2$, because the remaining two sheets are copies of sheet $1$ where the function $h(z)$ has undergone $z\mapsto -z$. Note the Painlev\'e\ I transition at both $t=-\frac 1 {12}$ and $t=\frac 1{15}$.
}
\label{Tri}
\end{minipage}
\end{figure}

\begin{figure}
\vspace{-0.1\textwidth}
\begin{minipage}{0.65\textwidth}
\includegraphics[width=1\textwidth]{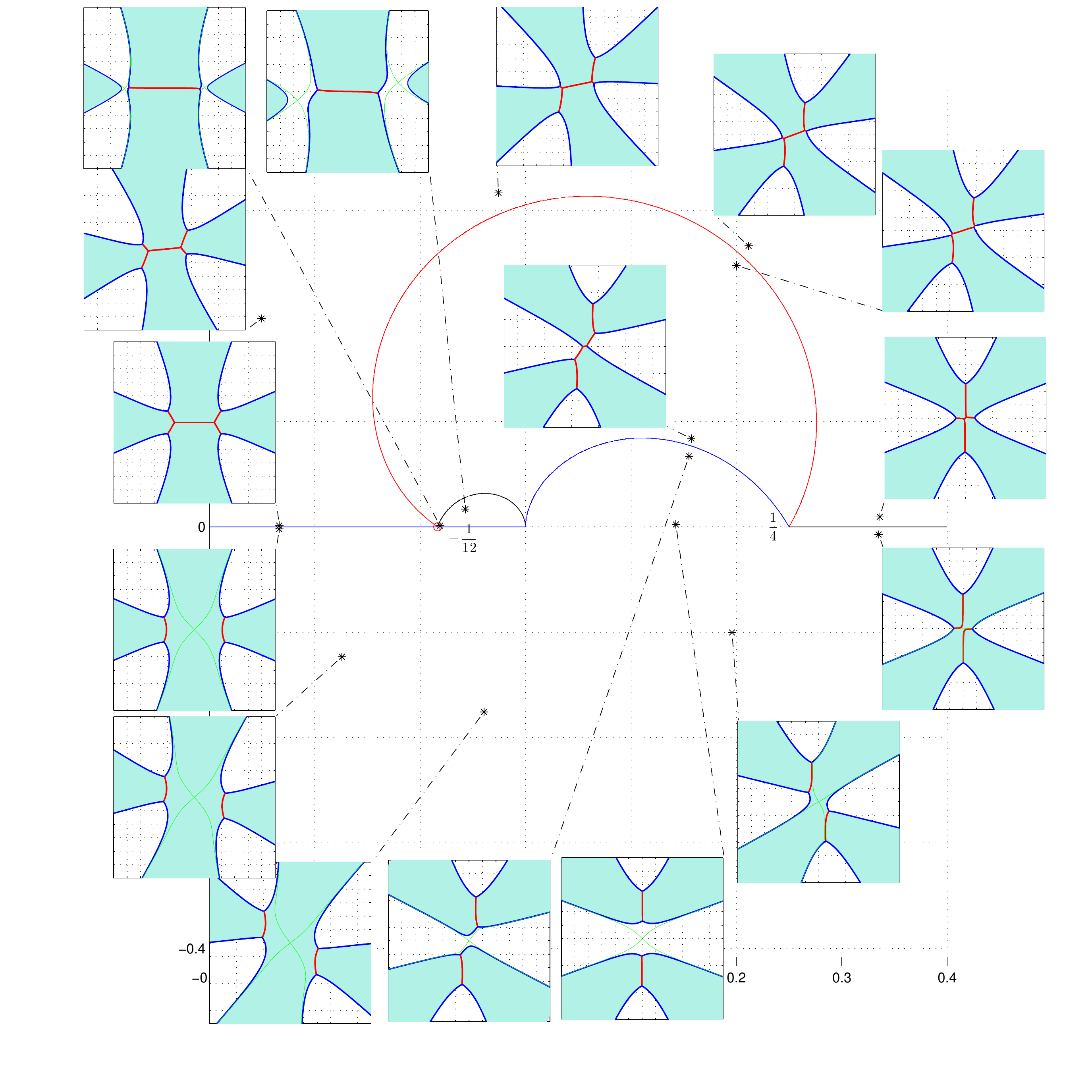}
\end{minipage}
\begin{minipage}[b]{0.35\textwidth}
\caption{{\bf Opposite Wedges, generic}.
There are {\bf two} sheets glued along $t\in \R_-$; shown here are only sheet $1$, because sheet $2$ is a copy 
of sheet $1$ where the function $h(z;t)$ has undergone $h(z,t)\mapsto \ov {h(\ov z;\ov t)}$. 
Note the Painlev\'e\ I transition at $t=-\frac 1 {12}$ and  Painlev\'e\ II transition at $t=\frac 1{4}$.
}
\label{BiWedge}
\end{minipage}
\end{figure}

\begin{figure}
\vspace{-0.1\textwidth}
\begin{minipage}{0.65\textwidth}
\includegraphics[width=1\textwidth]{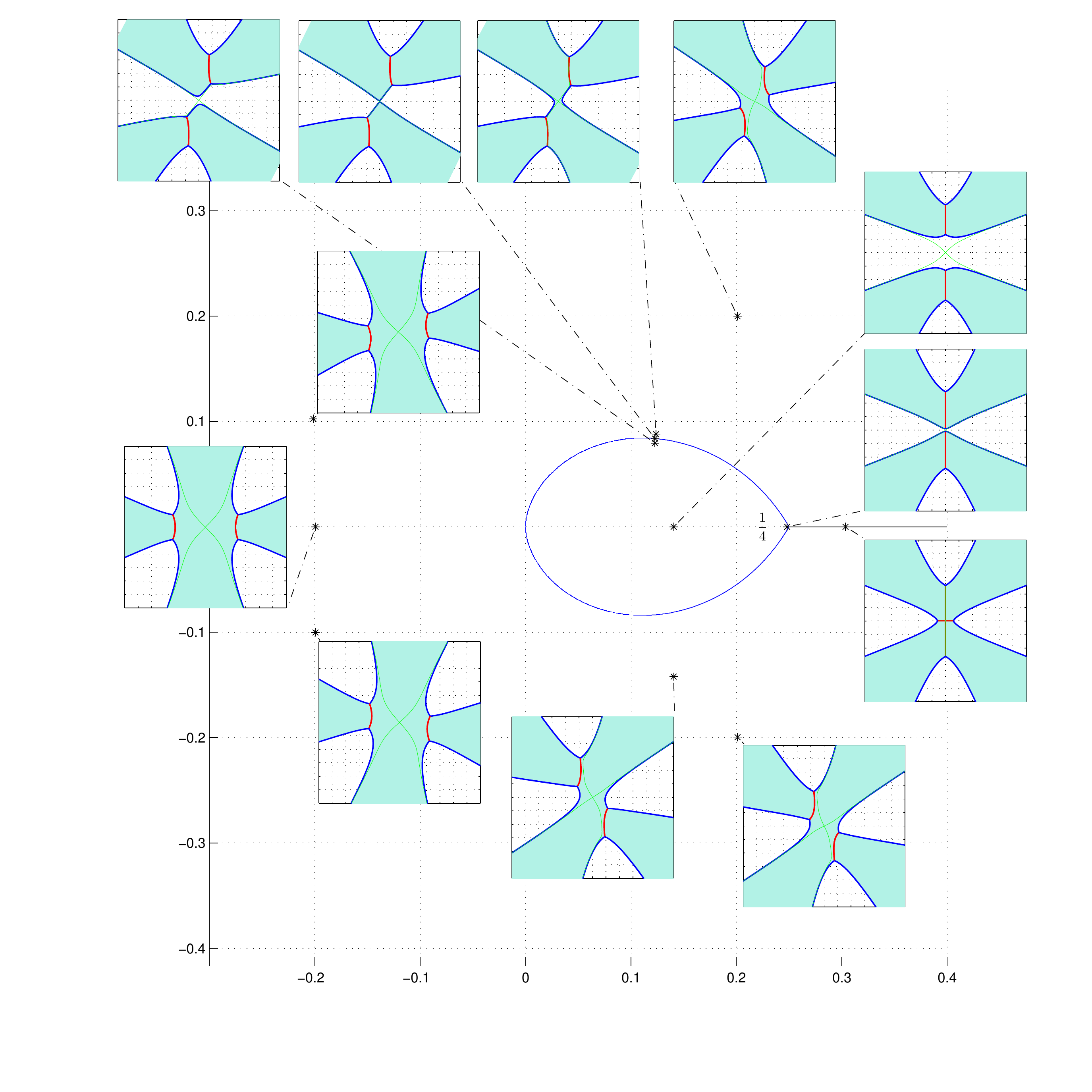}
\end{minipage}
\begin{minipage}[b]{0.35\textwidth}
\caption{{\bf Opposite Wedges, symmetric}.
There is only  {\bf one} sheet. Note the  Painlev\'e\ II transition at $t=\frac 1{4}$. The spectral curve is always 
of genus $1$ except at the point $t=\frac 14$.
}
\label{BiWedgeSym}
\end{minipage}
\end{figure}

There are six different situations depending on the values of $\nu_j$'s in the definition of the bilinear form eq. (\ref{orthogA}). 
Note that only their values up to common multiplication by nonzero constant is relevant, i.e., the orthogonal polynomials 
are parametrized by points  $[\nu_1:\nu_2:\nu_3]\in \C\mathbb P^2$. So, we have:
\begin{enumerate}
\item "Generic" case: $\nu_j\neq 0$, $\nu_1\neq \nu_2$, $\nu_2\neq \nu_3$;
\item "Real axis": $\nu_2=\nu_1$, $\nu_3=0$;
\item "Single Wedge": $\nu_1=0=\nu_2$\ ,$\nu_3\neq 0$.
\item "Consecutive Wedges": $\nu_3=0$, $\nu_2\neq \nu_1$, $\nu_2\neq 0$, $\nu_1\neq 0$;
\item "Opposite Wedges, generic": $\nu_2=0$, $\nu_3\neq \nu_1$, $\nu_1\nu_3\neq 0 $;
\item "Opposite Wedges, symmetric": $\nu_2=0$, $\nu_3 = \nu_1 \neq 0$.
\end{enumerate}
We provide the results of the 
computer-assisted investigation for all six cases in the tables that follow (Figures \ref{Generic}, \ref{RealAxis}, \ref{Wedge}, \ref{Tri}, 
\ref{BiWedge}, \ref{BiWedgeSym}). The common feature  is the following: as we move around the origin $t=0$ counterclockwise 
the asymptotic directions of the integration contours move {\em clockwise} by $\arg (t)/4$. Therefore, a counterclockwise loop around $t=0$ 
yields a new configuration of contours obtained by a clockwise rotation of  $\pi/2$ of the initial one. 

In general, thus, we can expect that our phase portraits to have {\bf four sheets}. In the "Generic" and "Opposite Wedges, 
symmetric" case, however, these four sheets are actually identical, and in the case of "Real Axis" two of them are equal. 

In the pictures that follow the cut (if necessary) is always along the negative $t$-axis and the gluing is the top of the 
negative axis of sheet $j$ is glued to the bottom of the negative axis of sheet $j+1$ (mod $4$). We hope that the pictorial 
representation will serve more than many pages of verbal explanation. 

We rather explain briefly the algorithm used to investigate the phase portraits; in \cite{BertoBoutroux} an algorithm to find 
numerically "Boutroux curves" was explained. The algorithm produces a solution of the "modulation equations" (for branch-points) in high genus, 
but will not enforce the sign distribution (sign conditions for $h(z)$) 
needed to have an appropriate $g$-function. In terms of the Remark \ref{eqmea}  it may yield a signed equilibrium measure. 
Plotting the level curves allows one to decide unambiguously whether the numerically produced $g$-function  satisfies the sign distribution. 

The pictures below are produced by some code written in Matlab which is available upon request; the code will allow 
"interactive exploration" of the $t$--plane and to produce the pictures interactively. 
\br[Zeroes of the orthogonal polynomials]
In all situations considered below, see Figures \ref{Generic}-\ref{BiWedgeSym}, the main arcs consist of all the red arcs that are 
surrounded by the shaded (light blue) regions on both sides. These arcs, as is well known (see for example \cite{BertolaMo, BertoBoutroux}),
 also represent the limiting arcs where the {\em roots} of the orthogonal polynomials accumulate, 
and the (weak) limit of their density can be recovered from the jump of $h'(z)$.
\er

\br
Although the results, presented in Figures \ref{Generic} - \ref{BiWedgeSym} are numerical, there is a straightforward way for 
their analytic justification. Consider, for example, the Generic case shown in Figures \ref{Generic}. According to Lemma \ref{lemma-signs},
the genus zero region contains the interval $(-\frac 1{12},0)$. Since a break can only occur at one of the curves defined
by (\ref{brcurveeq1}), the region contained inside the black curve on Figures \ref{Generic} is the  genus zero region.
According to the  continuation principle in the parameter space  (see \cite{BertolaMo} and \cite{TV1}), 
for any $t$ on the four-sheeted Riemann surface $\Xi$ (with branch-points $t=0$ and $t=\infty$)
there exists a contour $\O=\mathfrak M\cup\mathfrak C$ and $h_n(z)$, where $\mathfrak M$ is the union of all the branch-cuts of $h_n$,
such that $\Re h_n$ satisfies the sign conditions  on $\O$. Since there are 8 legs of zero level curves of $\Re h$, the genus
of the solution for any $t$ cannot be greater than two (as there can be no bounded closed loops of $\Re h=0$).
Let us take a point $t_*$, $\Im t_*>0$, on the main branch $\L$ of the breaking curve (\ref{brcurveeq1}), see Subsection \ref{g0s},
 that  contains genus zero region inside (the curve from $-\frac 1 {12}$ to $0$). Since $t_*$ is on the breaking curve, there exists a $z_*\in \C$,
such that the pairs
$(t_*,\pm z_*)$ satisfy (\ref{brcurveeqs}) (here we use the evenness of $\Re h(z)$). Choose $z_*$ so that $\Im z_*>0$. 
If we can show that 
\be\label{tran02}
\Re h_t(z_*;t_*)>0
\ee
as $t$ crosses $\L$ along $\Re t=\Re t_*$ going up, then we can prove that the genus of $h$ is changing from zero
to two as $t$ crosses $\L$.
According to the Cauchy-Riemann conditions, (\ref{tran02}) 
is equivalent to 
\be\label{tran02Im}
\Im h_t(z_*;t_*)<0,
\ee
where $\Im t=\Im t_*$. Using (\ref{hthx}) and the fact that $1+\frac{t_*b^2}{2}+t_*z_*^2=0$ (which follows from $h'(z_*;t_*)=0$),
we obtain $z_*^2=-\frac{1}{t}-\frac{b^2}{2}$, so that
\be\label{ht**}
h_t(z_*;t_*)=\frac{1}{4t_*}\sqrt{\left(-\frac 1{t_*}-\frac{b^2}{2}\right)\left(-\frac 1{t_*}-\frac{3b^2}{2}\right)}
\ee
To calculate the branch of the square root in (\ref{ht**}) we take $t_*\ra t_0$. As shown in subsection \ref{g0s}, in this limit 
$\arg(t_*-t_0)\ra\frac{2\pi}{5}$, so that, using (\ref{branchpeq}), we obtain 
\be\label{arght*rat0}
\arg h_t(z_*;t_*)\ra \frac{11\pi}{10}.
\ee
That proves inequality (\ref{tran02Im}) when $t_*$ is closed to $t_0$. Moreover, for any $t_*\in\L$ we obtain
\be\label{ht**fin}
h_t(z_*;t_*)=-\frac{1}{4\sqrt{3}t_*^2}\sqrt{2\sqrt{1+12t_*}-(1+12t_*)}~~~{\rm or}~~~h(u)=-12\frac{\sqrt{6u-u^2}}{(u^2-1)^2},
\ee
where $u=\sqrt{1+12t_*}$. It is easy to see that the upper halfplane part of the genus zero region (between $\L$ and $\R$)
is contained in the semistrip $0\leq \Re u\leq 1,~\Im u\geq 0$ of the $u$-plane. Direct calculations show
that: $\Im h(u)=0$ on $[0,1]$, and; $\Im h(u)<0$   on $i\R^+$, on $1+i\R^+$ and on any segment $\Im u=y,~\Re u\in[0,1]$,
where $y>0$ is sufficiently large. Thus, using the maximum principle for $\Im h(u)$ and the fact that $\frac{dh}{du}\neq 0$ on $[0,1]$, 
we conclude that $\Im h(u)<0$ inside the semistrip. So, we proved the transition from genus zero to genus two across $\L$. 
Similar considerations will lead to rigorous proofs of transitions through other level curves.
\er
\section{Double and multiple scaling analysis near the Painlev\'e\ I gradient catastrophe points}
\label{awayfromsect}
Having disposed of the global analysis of the problem in the complex $t$--plane we now focus on the so--called double (and multiple) scaling analysis near the two points of gradient catastrophe that are related to the Painlev\'e\ I transcendents. These are 
\be
t_0:= -\frac 1{12}\ ,\hbox { and} \  \ t_1:= \frac 1{15}\ .
\ee
There is another point of gradient catastrophe at $t_2:= \frac 1{4}$ which --however-- involves the Painlev\'e\ II transcendent and should be analyzed in a separate work. 

The two points $t_0, t_1$ can be analyzed much in a parallel fashion: they both necessitate of the same type local parametrix near one -or both- endpoints. The differences between the two cases appear by inspection of the figures: indeed 
\begin{itemize}
\item 
near $t=t_0$ the genus zero function $h(z;t) = h_0(z,t)$ always has symmetric level-curves and hence the Painlev\'e\ I parametrix  \insrt{(first introduced in \cite{FIK})} is needed near both endpoints $\lambda_0 = -b = -\lambda_1$ (see Figs. \ref{Generic}, \ref{RealAxis}, \ref{Wedge}, \ref{Tri}, \ref{BiWedge}, vignettes near $t_0$) ;
\item near $t=t_1$ the genus zero function $h(z;t) = h_0(z,t)$ does {\bf not} have any special symmetry and the PI parametrix is needed {\bf only near one endpoint} (see Figs. \ref{Generic}, \ref{RealAxis}, \ref{Wedge}, \ref{Tri}, \ref{BiWedge}, vignettes near $t_1$).
\end{itemize}
\subsection{Local analysis at the point of gradient catastrophe}\label{local}
Near an endpoint the genus-zero $h$--function has necessarily an expansion of the following form
\bea\label{hlocb}
\frac 1 2 h(z)&\&=C_0^{(j)} (z-\lambda_j)^\frac 32 + C_1^{(j)}(z-\lambda_j)^\frac 52 +C_2^{(j)}(z-\lambda_j)^\frac 72+ \cdots=\cr
&\&C_0^{(j)}(z-\lambda_j)^\frac 32\left(1+\frac{C_1^{(j)}}{C_0^{(j)}}(z-\lambda_j) +\frac{C_2^{(j)}}{C_0^{(j)}}(z-\lambda_j)^2+\cdots\right)\ ,\ \ j=0,1,
\eea
where the coefficients $C_k^{(j)} = C_k^{(j)}(t)$ depend on $t$. The gradient-catastrophe occurs when the 
leading coefficient $C_0^{(j)}(t)$ vanishes at one or both endpoints $\l_{0,1}$ of the main arc $\g_m$, 
while (in general) the next coefficient $C_1^{(j)}(t)$ does not. For our $f(z,t)$, the gradient catastrophe point  
is either $t_0$ or $t_1$. 
Elementary singularity theory\cite{ArnoldGZV-1} guarantees the validity of the following definition.

\bd[Scaling coordinate]
\label{defzetatau}
The {\bf scaling coordinate} $\zeta(z) =\zeta(z;t,N)$ and the {\bf exploration parameter} $\tau =\t(t,N)$ are  defined by 
\be
\frac N2 h(z;t) = \frac 45 \zeta^{\frac 5 2 } {(z;t,N)} + \tau {(t,N)}  \zeta^{\frac  3 2 } {(z;t,N)}\label{hzeta},
\ee
where $\zeta(b;t,N)\equiv 0$, $\zeta(z;t,N)$ is analytically invertible in $z$ in a 
fixed small neighborhood $\mathbb D_j$ of $z=\lambda_j$
and $\tau$ is analytic in $C_0^{(j)}$ at $C_0^{(j)}=0$, where $j=0,1$.
\ed
Let us consider the endpoint $\l_1$ near the point of gradient catastrophe $t_*$, where $t_*=t_0$ or  $t_*=t_0$.
The expression (\ref{hzeta}) is the {\bf normal form} of the singularity defined by $h(z;x,t)$ (in the sense of singularity theory \cite{ArnoldGZV-1}).
The local behaviour  (we suppress the superscripts)
\bea
\zeta&=&N^{\frac 2 5 } \le(\frac {5}4  C_1\ri)^\frac 25\le ( 1- \frac {6 C_0 C_2}{25 C_1^2} + \mathcal O(C_0^2)   \ri) (z-\lambda_1)(1+ \mathcal O(z-\lambda_1)), \label{zeta}\\
\tau&= &N^{ \frac 2 5}C_0  \le(\frac 4 {5C_1} \ri)^{\frac 3 5 }  \le(1 + \mathcal O(C_0)\ri)\label{tau}
\eea
was calculated in \cite{BT2}.
The determination of the root is fixed uniquely by the requirement that the image of the main arc $\g_m$, where $\Re h \equiv 0$, be mapped to 
the {\bf negative real} $\zeta$--axis. 
Following \cite{BT2}, we define: 
\bd
\label{painlevecoordinate}
The double scaling near $t=t_*$ shall be defined as the appropriate dependence of $t$ such that the variable 
\be
v = v(t,N):= \frac 3 8 \tau^2(t,N) = \frac 38 N^\frac 45 C_0^2 \le(\frac 5 4 C_1\ri)^{-\frac 6 5 } (1 + \mathcal O(C_0))
\ee
is kept within a disk of arbitrary but fixed (in $N$) radius around $v=0$. 
The variable $v$ shall be referred to as the {\bf Painlev\'e\ coordinate}.
\ed

\begin{table}[h]
\begin{center}
\begin{tabular}{c|c}
Near $t_0 = -\frac 1 {12}$, $\delta t := t-t_0$ & Near $t_1 = \frac 1{15}$ $\delta t := t-t_1$\\[10pt]
\hline\\
$\ds a=0$ &
 $\ds a = -3i \frac {\sqrt{ 1+ \frac {2i}3\sqrt{15 \d t}}}{\sqrt {1 + 15 \d t}} =
-3i +  \sqrt {15 \d t} + \mathcal O(\d t)$\\[10pt]
\hline\\
$\ds b =  \frac {\sqrt{8}}{\sqrt{1 + \sqrt{ 12\, \d t}}} = \sqrt{8} -\frac {\sqrt{8}}2\sqrt{12 \d t} + \mathcal O(\d t)$ & 
$ \ds b=\frac{2i}{ \sqrt{1 + i \sqrt{15 \d t}}} = 2i + \sqrt {15 \d t}  +  \mathcal O(\d t)$\\[10pt]
\hline\\
$\ds C_0=-\frac 1 6 \left(2+ 3 t b^2\right)\sqrt{2b}=-\frac {2\sqrt[4]{2} \sqrt{12\, \d t}}{ 3 \sqrt[4]{1+ \sqrt{12\, \d t}}} $ 
&
$\ds C_0 =  -\frac 13 t \sqrt{2 b} a(2a + 3b)= \frac 2 3  {\rm e}^{\frac{3i\pi} 4} \sqrt {15 \delta t} + \mathcal O(\delta t)
$
\\[10pt]
\hline \\
$ \ds C_1= -\frac {19 t b^2 + 2} {20\sqrt{2b}} 
=
 \frac {\sqrt[4]{ 1 + \sqrt {12\, \d t}} }{60 \sqrt[4]{2}} \le( 16 - 19 \sqrt{ 12\, \d t}\ri)$  
 &
$\ds  C_1 = -\frac 15 \frac {t(2a^2 + 15 ab + 8 b^2)}{2\sqrt{2b}}  = \frac {2 {\rm e}^{\frac {3i\pi}4}}{15} +  
\mathcal O(\sqrt{\d t})
$
 \\[10pt]
 \hline \\
$ \ds 
N^{-\frac 25} \z'(\l_1) = 
 3^{-\frac 25} 2^{-\frac 1{10}}+ \mathcal O(N^{-\frac 25})$
&
$\ds  N^{-\frac 25} \zeta'(\l_1) 
=
  6^{-\frac 25} {\rm e}^{\frac {3i\pi}{10}}\le(1 + \mathcal O( N^{-\frac 25}) \ri)$\\[10pt]
  \hline \\
 $\ds 
 \frac \tau{\zeta'(\l_1)} = \frac {4 C_0}{5 C_1} (1 + \mathcal O(C_0))  =  - \sqrt {8} \sqrt{12 \d t} + \mathcal O(\d t) 
$
 & $\ds 
 \frac \tau{\zeta'(\l_1)} = \frac {4 C_0}{5 C_1} (1 + \mathcal O(C_0))  = 4 \sqrt{15 \d t} + \mathcal O(\d t) 
$\\[10pt]
\hline\\
$\ds \ell = -\frac 32 + \ln 2 - 6 \delta t + \frac 2 3 (12 \delta t)^\frac 32 + \mathcal O(\delta t^2)$ 
&
$\ds \ell = 
\frac 94 +\ln(-1)- \frac {13}4 (15\d t) - \frac 2 3i (15 \d t)^\frac 32 + \mathcal O(\d t^2)
$
\end{tabular}
\end{center}
\caption{
The explicit expressions and relevant expansions of the indicated quantities: these are the result of straightforward algebraic manipulations using (\ref{branchpeq}, \ref{othersols}, \ref{h}, \ref{nsh}, \ref{l}, \ref{l2}, \ref{hlocb}, \ref{hzeta}, \ref{zeta}, \ref{tau}).
}
\label{expansions}
\end{table}
\bl
In the double scaling near $t=t_0$ for the symmetric genus zero case 
or near $t=t_1$ for the non-symmetric case,  the Painlev\'e\ coordinate $v$ has the following expansion
\bea\label{v(t)}
v(t)&\&=\frac 38 \t^2(t,N)=
 {- 3^\frac 65  2^{\frac {9}{5}}   \le(t + \frac 1 {12} \ri)N^{\frac 45} (1 + \mathcal O(\sqrt{t-t_0})) } 
,\  \ \ 
\\
v(t)&\& =\frac 38 \t^2(t,N)=3^\frac 65 \, 2^\frac 15 \,5\, {\rm e}^{\frac {3i\pi}{5}}  \le(t - \frac 1{15}\ri) N^{\frac 45} (1 + \mathcal O(\sqrt{t-t_1}))\ .
\label{v(t)2}
\eea
In either cases the function $v(t)$ is  a convergent series in $\sqrt{t-t_j}$; if $v$ is kept bounded as $N\to \infty$ then $t-t_j = \mathcal O(N^{-\frac 45})$. Therefore from (\ref{v(t)}, \ref{v(t)2}) it follows immediately that  with accuracy $\mathcal O(N^{-\frac 25})$ the map $v(t)$ is linear in $\frac {t-t_j}{N^{\frac 4 5}}$.
\el
 The proof is a direct computation with the help of Table \ref{expansions} and Def. \ref{defzetatau}.
%
%
\subsection{Asymptotics away from the poles}\label{assaway}
The asymptotic analysis now depends on the regions in the Painlev\'e\ variable $v$ (\ref{v(t)}) or (\ref{v(t)2}) that we are investigating. 
We will split this analysis into the following two cases.
\begin{itemize}
\item {\bf Away from the poles}: the variable $v$ is chosen within a fixed compact set $K$, that does not contain any pole of the relevant solutions to P1;
\item {\bf Near the poles}: the variable $v$ undergoes its own scaling limit and approaches a given pole at a certain rate.
\end{itemize}
Each of these two cases requires a slightly different analysis depending on the nature of the gradient catastrophe point, 
be it $t_0=-\frac 1 {12}$ or $t_1=\frac 1{15}$. 
In the former case the analysis was carried out in full in the regime "Away from the poles" by \cite{ArnoDu} and the 
relevant theorem is Thm. \ref{theo-ArnoDu}: we will not add anything to it. 

\begin{figure}[h]
\resizebox{0.4\textwidth}{!}{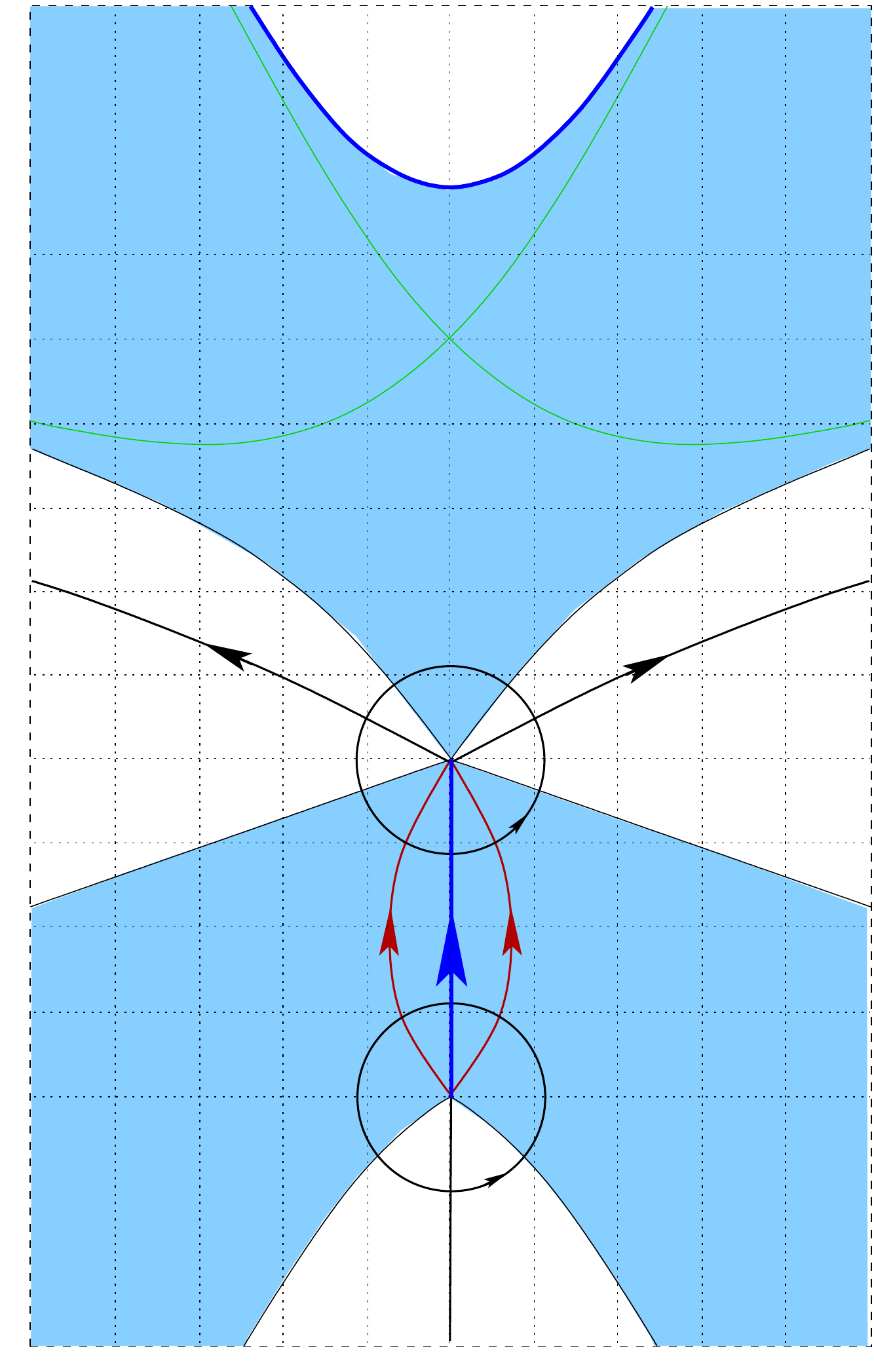}~~~~~~~~~~~~~~~~~~
\resizebox{0.45\textwidth}{!}{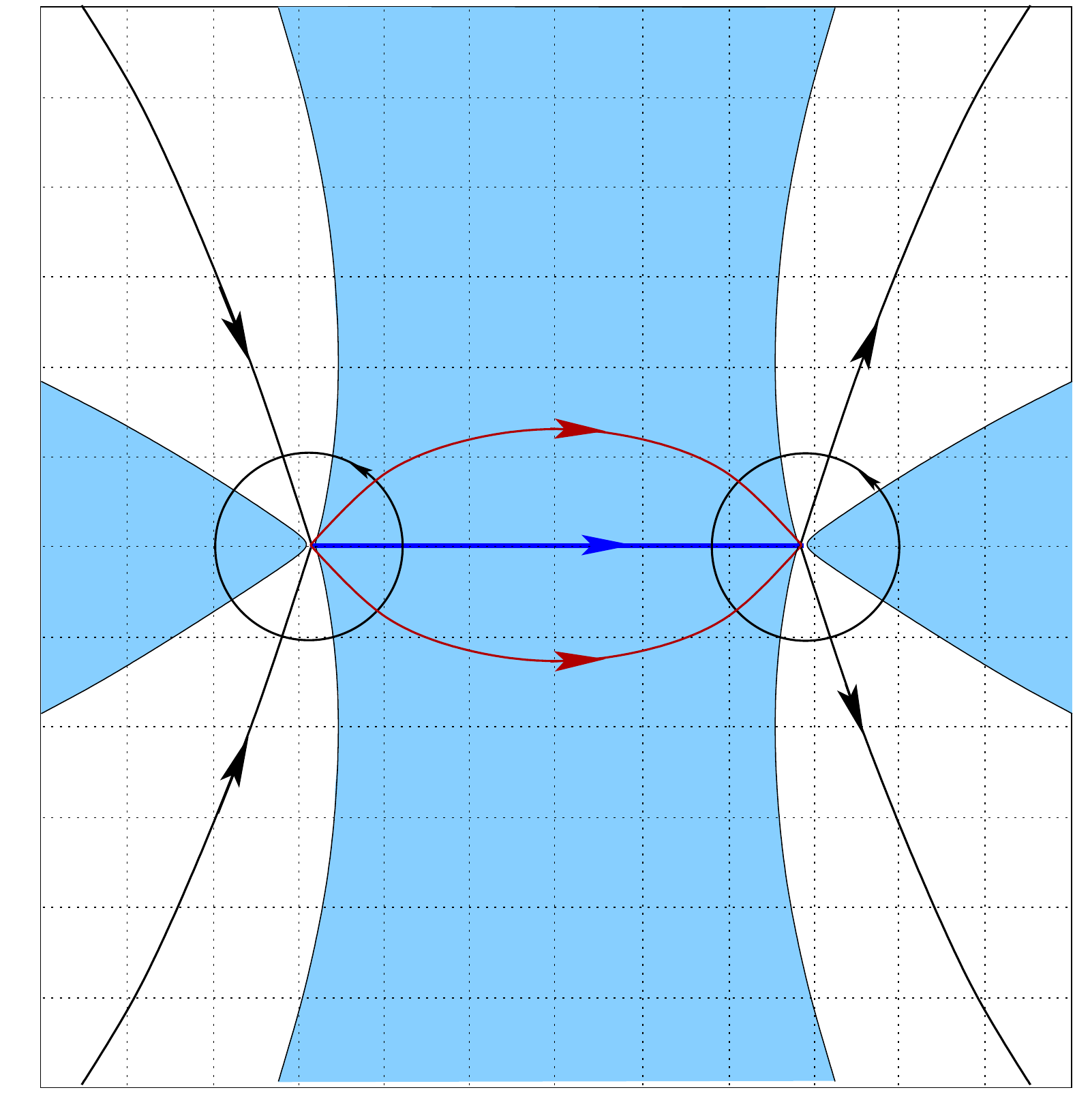}
\caption{The deformation of the contours and the partitioning in complementary (black) and main arcs (blue). Shown also are the lenses and the disks near the two branch-points $\l_1,\l_0$. The left frame refers to the case $t\sim \frac 1 {15}$, the right frame to $t\sim -\frac 1{12}$. The thin lines in the shaded area on the left frame are the level-curves passing through the saddle point. 
For the case on the left ($t=\frac  1{15}$) we have $\nu_2 = \nu_1\in \C$, and $\nu_3$ can be normalized to $\nu_3=1$; the deformed contour $\varpi_3$ consists of the complementary arc on the imaginary axis, the left one and the main arc (thick). The contour $\varpi_2+\varpi_1$ (homological sum) consists of the two left/right complementary arcs.
For the right frame, we have $t={\rm e}^{i\pi} \frac 1{12}$ and the weights are $\nu_2=1$ and $\nu_1,\nu_3\in \C$: the contour $\varpi_1$ is deformed to go through the two complementary arcs on the bottom and top right, and $\varpi_2$ runs along the top right, top left complementary arcs and the main arc, while $\varpi_3$ runs along the two complementary arcs on the left.
The weights of the various complementary arcs are indicated in the figure and determine the parameters of the Painlev\'e\ parametrices to be used according to Definition \ref{defyrl}. The level-curves in these pictures are numerically accurate.
}
\label{YRHP}
\end{figure}

The only case that is not covered by the mentioned theorem in the same regime  is when $t$ undergoes a double scaling limit near $t_1$ 
and a special Painlev\'e\ parametrix is needed only at one endpoint, say, at $\lambda_1$. Of course, one may still use the results of 
\cite{ArnoDu}  
with minor modifications to cover this new case, but since we will need some preparatory material, we briefly analyze this case below.
%
%
We shall construct an approximation to the matrix $\wh T(z; t,N)$ appearing in (\ref{hatT}) in the form 
\be
\wh T(z) = 
\le\{ 
\begin{array}{cc}
\mathcal E(z) \Psi_0(z) & \mbox { for $z$ {\bf outside} of the disks } \mathbb D_1, \mathbb D_{0},\\[4pt]
\mathcal E(z) \Psi_0(z)\mathcal P_{1}(z) & \mbox { for $z$ {\bf inside} of the disk } \mathbb D_1, \\[4pt]
\mathcal E(z) \Psi_0(z)\mathcal P_{0}(z) & \mbox { for $z$ {\bf inside} of the disk } {\mathbb D}_{0},\\[4pt]
\end{array}
\ri.
\label{leadingapprox}
\ee
where $\mathbb D_0,~\mathbb D_{1}$ are small fixed disks centered at $\lambda_0$ and $\lambda_1$ respectively, see Fig. \ref{YRHP}
and $\Psi_0$ as in (\ref{Psi0}).
Here $\mathcal E(z)$ is the so-called error matrix that will be shown to be close to the identity matrix $\1$ and
$\mathcal P_{0,1}(z)$ are local parametrices at $z=\lambda_0,\lambda_1$, respectively,  that will be constructed through the matrix $\Psi(\x,v)$ defined by
(\ref{PsiPain}). 
%
A local parametrix  $\mathcal P(z)$ (we drop the indices for convenience) must have a certain 
number of  properties (see Theorem \ref{thmlocalP}),
one of them being the restriction 
\be\label{param55}
\mathcal P(z)\bigg|_{z\in \pa\mathbb D} = \1 + o_\varepsilon (1)
\ee
on the boundary of the respective disk $\mathbb D$, 
where $o_\varepsilon (1)$ denotes some infinitesimal of $\varepsilon = \frac 1 N$, uniformly in $z\in \pa \mathbb D$ and 
in $t\in \hat K=v^{-1}(K)$. 

If the local parametrices $\mathcal P_{0,1}(z)$ satisfying (\ref{param55})  
can be found 
then the ``error matrix'' $\mathcal E(z)$ is 
seen to satisfy a {\em small--norms} RHP and, thus, be uniformly close to the identity. More precisely, the matrix $\mathcal E$ has jumps on:
\begin{enumerate}
\item[{\bf (a)}]the parts of the lenses and of the complementary arcs 
that lie outside of the disks $\mathbb D_0, \mathbb D_{1}$, and; 
\item[{\bf (b)}] on the boundaries of the two disks $\mathbb D_0, \mathbb D_{1}$.
\end{enumerate}
The jumps in {\bf (a)} are {\bf exponentially close} to the identity in any $L^p$ norm (including $L^{\infty}$) while {\bf (b)} on the 
boundary of the disks $\mathbb D_{0,1}$ we have 
\be
\mathcal E_+ (z) = \mathcal E_-(z) \Psi_0(z) \mathcal P^{-1}_{0,1}(z) \Psi_0^{-1}(z)\bigg|_{z\in \pa \mathbb D_{0,1}} = 
\mathcal E_- \le(\1 + o_{\varepsilon}(1)\ri).
\label{jumpErr}
\ee
From the analysis in \cite{DKMVZ} it follows that, for $|z|$ large enough,  $\|\mathcal E(z)-\1\| \leq \frac{ o_\varepsilon(1)}{|z|}$ (with the pointwise matrix norm)
and 
that the rate of convergence is estimated as the same as the $o_\varepsilon(1)$ that appears in (\ref{param55}) as $\e\ra 0$.

In the case at hand we keep in mind that near $t=t_1=\frac 1{15}$ the endpoint $\l_0$ requires the standard Airy parametrix 
and that the corresponding error term arising on the boundary of $\mathbb D_0$ is of order $\mathcal O(N^{-1})$. 


\begin{definition}[Local parametrix away from the poles]
\label{deflocalP}
Let $\zeta(z;\varepsilon)$ be the local conformal coordinate near $\l_1$  introduced in Def. \ref{defzetatau} so that 
\be
\frac N{2} h(z;x,t) = \theta(\zeta;\tau ) = \frac 4 5 \zeta^{\frac 5 2 } + \tau \zeta^{\frac 32}\ . 
\ee
Let $\Psi(\xi;v;\varkappa)$ denote the Psi--function of the Painlev\'e\ I Problem \ref{P1RHP} 
according to Def. \ref{defyrl}.
The parametrix $\mathcal P(z;\varkappa)$  is defined by  
\bea
\mathcal P(z;\varkappa) = \frac 1{\sqrt{2}} 
{\begin{bmatrix}
1 & 1 \\ i & -i
\end{bmatrix}}
\zeta^{-\frac {\s_3}4}
\Psi\le ( \zeta +  \frac \t 2; \frac 3 8 \t^2;\varkappa\ri)   {\rm e}^{-\theta(\zeta;\t) \s_3}\ ,\ \ \  \zeta:= \zeta(z). \label{localP}
\eea
\end{definition}

\begin{theorem}\label{thmlocalP}
The matrix $\mathcal P_1(z):=\mathcal P(z;\nu_1) $ 
satisfies: 
\begin{enumerate}
\item Within $\mathbb D_1$, the matrix  $\mathcal P_1(z)$ solves the exact jump conditions on the lenses and on the complementary arc;
\item On the main arc (cut) $\mathcal P_1(z)$ satisfies 
\be
\mathcal P_{1,+}(z) = \sigma_2 \mathcal P_{1,-}(z) \s_2\ ,\label{oddjump0}
\ee
so that $\Psi_0 \mathcal P_1$ within $\mathbb D_1$ solves the exact jumps on all arcs contained therein  (the left-multiplier in the jump (\ref{oddjump0}) 
cancels against the jump of $\Psi_0$);
\item The product $\Psi_0 (z) \mathcal P_1(z)$ (and its inverse) are --as  functions of $z$-- bounded within $\mathbb D_1$, namely the 
matrix $\mathcal P_1(z)$ cancels the growth of $\Psi_0$ at $z=\lambda_1$;
\item The restriction of $\mathcal P_1(z)$ on the boundary of $\mathbb D_1$ is 
\be\label{Papproxaway}
\mathcal P_1(z)\bigg|_{z\in \pa\mathbb D_1} = \1 - \le(H_I + \frac {\t^3}{16}\ri)\frac {\s_3}{\sqrt{\zeta}} 
+ \frac 1 {2\zeta} \le[\le( H_I  + \frac {\tau^3 }{16} \ri)^2\1  + 
{\le(y+\frac \tau 4\ri)} \s_2\ri] + 
\mathcal O(\zeta^{-\frac 3 2}),
\ee
where $v = \frac 3 8 \tau^2$, $y(v)=y^{(1)}(v)$ and
$H_I = \frac 1 2 (y')^2 + y v - 2 y^3 = \int y (s) {\rm d} s$.
\end{enumerate}
\end{theorem}

The proof of Theorem \ref{thmlocalP} can be found in Theorem 5.1 of \cite{BT2} (although it was for the 
tritronqu\'ee solution, the proof is identical for the general case).\footnote{
The parametrix $\mathcal P_1$ coincides with the parametrix considered in \cite{BT2}  up to conjugation by $\s_2$.}  
\subsection{Computation of the correction near $t_1$: proof of Theorem \ref{t1away}}
{\bf Proof of Theorem \ref{t1away}.} 
According to (\ref{leadingapprox}), we have 
\be\label{jumpE}
\ds 
\mathcal E_+ = \mathcal E_- \Psi_0 \mathcal P_{0,1}^{-1} \Psi_0^{-1} = \mathcal E_- \le(\1 +\Delta M(z)\ri)~~{\rm on}~~ \partial \mathbb D_{0,1}.
\ee
In particular, according to
(\ref{Papproxaway}),
\bea
\Psi_0 \mathcal P_1^{-1} \Psi_0^{-1} =&\&  \1 + \le(H_I + \frac {\tau^3}{16} \ri) \le( \frac  {\s_3 -i \s_1}{2\sqrt{\zeta p}}
 + \sqrt{\frac p \zeta}\frac {\s_3 + i \s_1}{2}\ri)
+\cr
&\& +
  \frac 1 {2\zeta} \le( \le(H_I + \frac {\tau^3}{16} \ri)^2 \1 - \le(y + \frac \tau 4\ri) \s_2 \ri) + \mathcal O(N^{-\frac 35})\ ,\ \ \  
p:= \frac {z-\l_1}{z-\l_0}.
\eea
Using  $\1 +\Delta M(z)$ to denote the jump-matrix of  $\mathcal E$ on {\bf all} the contours (see below), 
we can rewrite (\ref{jumpE}) as the integral equation
\be\label{intE}
\mathcal E(z) = \1 + \frac 1{2i\pi} \int \frac {\mathcal E_-(s) \Delta M(s) {\rm d}s}{s-z}, 
\ee
where the integral is taken along all the jumps of $\mathcal E$,
that is, along the parts of the lenses and the 
complementary arcs that lie outside  $\mathbb D_0\cup\mathbb D_1$ as well as along the boundaries of $\mathbb D_0$, $\mathbb D_1$.
However, the contribution to $\mathcal E$ coming from the integrals along all these contours, except for $\partial\mathbb D_1$, 
are of order not exceeding $O(N^{-1})$
(note that the parametrix in $\mathbb D_0$ is the standard Airy parametrix). Therefore, to obtain
the leading order solution,  we consider (\ref{intE}) with the contour  $\partial\mathbb D_1$. This integral 
equation will be solved by iterations.
The first iteration yields
\bea
\mathcal E^{(1)}(z)  = \1  +   \frac 1{\l_1-z}\le\{
\le(H_I + \frac {\tau^3}{16} \ri) \le( \frac  {\s_3 - i \s_1}{2\sqrt{\zeta'(\l_1)/(\l_1-\l_0)}} \ri) + \frac 1 {2\zeta'(\l_1)} \le( \le(H_I + \frac {\tau^3}{16} \ri)^2 \1 - \le(y + \frac \tau 4\ri) \s_2 \ri) 
\ri\}. 
\eea
 Retaining only the terms up to order $\mathcal O(N^{-\frac 25})$ in the second iteration, we obtain
\bea
\mathcal E^{(2)}_-(z)  =\mathcal E^{(1)}_-(z) 
 +\res{s=\l_1} \le(H_I + \frac {\tau^3}{16} \ri)^2
  \le( \frac  {\s_3 - i \s_1}{2\sqrt{\zeta'(\l_1)/(\l_1-\l_0)}} \ri)  \frac 1{(\l_1-s)(s-z)} 
\le(  \sqrt{\frac {p(s)}{ \zeta(s)}}\frac {\s_3 + i \s_1}{2}\ri){\rm d}s=\cr
 = \1 + \frac {\mathcal E^{(1)}_1} {\l_1-z} -  \le(H_I + \frac {\tau^3}{16} \ri)^2
  \le( \frac  {1}{\zeta'(\l_1)} \ri)  
  \frac 1{\l_1-z} \le(  \frac {\1 - \s_2}{2}\ri) =\cr
  = \1  +   \frac 1{\l_1-z}\le\{
\le(H_I + \frac {\tau^3}{16} \ri) \le( \frac  {\s_3 -i \s_1}{2\sqrt{\zeta'(\l_1)/(\l_1-\l_0)}} \ri) - \frac 1 {2\zeta'(\l_1)}  
 \le(y + \frac \tau 4 - \le(H_I + \frac {\tau^3}{16} \ri)^2 \ri) \s_2 
\ri\}.
\eea
Therefore, using the fact that  $\l_1 = a+b, \ \l_0 = a-b$, we have 
\bea
T(z) = \le( \1 + \frac {E_1}{\l_1-z}  + \mathcal O(N^{-\frac 3 5}) \ri) \le( \1 - \frac {(\l_1-\l_0) \s_2}{4z}  + 
\frac{ (\l_1-\l_0)^2  -4(\l_1^2 - \l_0^2)\s_2 }  {32 z^2} \ri) = \cr
= \1 - \frac {2 E_1 + b\s_2}{2z} - \frac {(a+b) E_1}{z^2} + \frac {b E_1 \s_2}{2z^2} + \frac{ b^2  -4ab\s_2 }  {8 z^2}. 
\eea
From this we can read off the relevant matrix entries:
\bea
(T_1)_{22}&\& =
 \frac  {\le(H_I + \frac {\tau^3}{16} \ri)}{\sqrt{2\zeta'(\l_1)/b}}  =: {\bf G}, \label{526}\\
 (T_1)_{12} &\& = i{\bf G} + \frac {ib}2
-
\frac i {2\zeta'(\l_1)}   \le(y + \frac \tau 4\ri)  + \frac i b {\bf G}^2,  \\
(T_1)_{21} &\& = i{\bf G}
-\frac {ib}2 
+
\frac i {2\zeta'(\l_1)}   \le(y + \frac \tau 4\ri) -\frac i b {\bf G}^2, \\
(T_2)_{12} &\& = \frac {iab}{2} 
-
(a+b) \le[ i{\bf G}
+ 
\frac i {2\zeta'(\l_1)}   \le(y + \frac \tau 4\ri) -\frac i b {\bf G}^2\ri] 
-\frac {ib{\bf G}}2,
\eea
where all the terms have accuracy $\mathcal O(N^{-\frac 35})$.
Direct computation using (\ref{Talphabeta}) shows
\bea
\a_n 
=\frac {b^2}4  - \frac b{2\zeta'(\l_1)}  \le(y + \frac \tau 4  \ri) + \mathcal O(N^{-\frac 35})\ ,\ \ 
\ \b_n = a - \frac {y + \frac \tau 4}{\zeta'(\l_1)} + \mathcal O(N^{-\frac 35}).
\eea
Using Table \ref{expansions}, we see that   
\be
a = a_0 +  \frac {\tau}{4 \zeta'(\l_1)} + \mathcal O(N^{-\frac 45 })\ ,\ \ \ b= b_0 + \frac {\tau}{4 \zeta'(\l_1)} + \mathcal O(N^{-\frac 45 }),
\label{abdelta}
\ee
where $a_0=-3i$, $b_0=2i$, and, thus 
\bea
\a_n = \frac {b_0^2}4  - \frac {b_0}{2\zeta'(\l_1)} y + \mathcal O(N^{-\frac 35}) = 
-1  + \frac{i 6^{\frac 25} {\rm e}^{-\frac {3i\pi}{10}}}{N^{\frac 25} } y(v) + \mathcal O(N^{-\frac 35}),
\\
 \b_n = a_0 - \frac {y}{\zeta'(\l_1)} + \mathcal O(N^{-\frac 35}) = 
 -3i -  \frac {6^\frac 25 {\rm e}^{-\frac {3i\pi}{10}}}{N^{\frac 25}} y(v) + \mathcal O(N^{-\frac 35}).
\eea

To compute $\h_n$ we use (\ref{Talphabeta}). Noticing that ${\bf G} = \mathcal O(N^{-\frac 1  5})$, we can rearrange $(T_1)_{12}$ as follows: 
\bea
(T_1)_{12} = \frac {ib_0}2 - \frac {i y}{2\z'(\l_1)} + i{\bf G} + \frac i {b_0} {\bf G}^2 + \mathcal O(N^{-\frac 35})  
=\frac {ib_0}2 \le(
1 - \frac {y}{b_0 \z'(\l_1)} + \frac {2{\bf G}}{b_0} + \frac {2 {\bf G}^2 }{b_0^2} + \mathcal O(N^{-\frac 35})
\ri)\cr
=\frac {ib_0}2 \le(
1 - \frac {y}{b_0 \z'(\l_1)}\ri){\rm e}^{\frac {2{\bf G}}{b_0}} (1 + \mathcal O(N^{-\frac 35})).
\eea
Therefore 
\bea
\h_n = -2i\pi (T_1)_{12} {\rm e}^{N\ell} =\pi b_0 \le(
1 - \frac {y}{b_0 \z'(\l_1)}\ri){\rm e}^{N \ell + \frac {2{\bf G}}{b_0}} (1 + \mathcal O(N^{-\frac 35})).
\eea
Utilizing (\ref{526}) and the values in Table \ref{expansions}  we find:
\bea
&\&\h_n =  2i\pi \le(
1 - \frac {y}{2i \z'(\l_1)}\ri)\exp\le[ N\ell  
-i \le(
\frac {H_I}{\sqrt{-i\z'(\l_1)}} + \frac {\t^3}{16 \z'(\l_1)^3 \sqrt{-i} } (\z'(\l_1)^5)^\frac 1 2 
\ri)
\ri] (1 + \mathcal O(N^{-\frac 35}))\cr
&\&=  2i\pi \le(
1 - \frac {y}{2i \z'(\l_1)}\ri)\exp\le[ N\ell 
-i\le(
\frac {H_I}{\sqrt{-i\z'(\l_1)}} - 
 \frac {4 (15 \d t)^\frac 32}{\sqrt{-i} } \frac N 6 {\rm e}^{-\frac{i\pi }4}
\ri)
\ri] (1 + \mathcal O(N^{-\frac 35}))\cr
&\&=  2i\pi (-1)^N \le(
1 -\frac{{\rm e}^{-\frac {4}{5}i\pi} }{3^{-\frac 25 }2 ^{\frac 35} } \frac {y(v) }{N^{\frac 25 }}\ri)\exp\le[
	\frac {9N}4 
	- 
 	\frac {13N }4 (15\d t) 
	-
	N \frac 2 3i (15 \d t)^\frac 32 
-i
\le({\rm e}^{\frac 1{10} i\pi} \frac { 6^{\frac 1 5} H_I}{N^{\frac 15} } -  
\frac {2N}3  (15 \d t)^\frac 32 
\ri)
\ri]\cr
&\& (1 + \mathcal O(N^{-\frac 35}))
\eea
\bea
=  2i\pi  (-1)^N\le(
1 -\frac{3^\frac 25}{2 ^{\frac 35} }{\rm e}^{-\frac {4}{5}i\pi} \frac {y(v) }{N^{\frac 25 }}\ri)\exp\le[ \frac {9N}4 
-
 \frac {195N }4 \d t
  + {\rm e}^{-\frac 25 i\pi} \frac{6^\frac 15}{N^{\frac 1 5}} H_I 
\ri] (1 + \mathcal O(N^{-\frac 35})).
\eea
\QED
\section{Analysis near the poles: triple scaling limit}\label{Spikessect}

The analysis in \cite{ArnoDu} was carried through under the assumption that --in the double scaling limit-- the Painlev\'e\ coordinate is 
chosen in an arbitrary compact set that does not contain any of the poles of the functions $y^{(0)}, y^{(1)}$ (see Theorem \ref{theo-ArnoDu}). 
Our special interest now is the analysis in the vicinity of {anyone of such poles}. 

To set the stage in general terms,  we shall consider the case where the Painlev\'e\ variable $v$ {\em  undergoes its own scaling}.
If $v_p$ is the pole under scrutiny, we shall consider the following {\bf triple scaling limit}, whereby,
in addition to $N\ra\infty$ and $N^\frac 45(t-t_*)$ being bounded, we also impose 
\be
v- v_p = \mathcal O\le(N^{-\frac 1 5 - \rho}\ri), \label{triplescale}
\ee
where $\rho\geq 0$ (depending on the situation, it may be bounded above). 

There are two distinct scenarios depending on whether the coalescence of the 
saddle points (zeroes of $h'(z)$)  with the branch-points $\l_{0,1}$
occurs at both branch-points or only at one, say, at $\l_1$.
These scenarios
 corresponds to the analysis near the critical points $t_0=-\frac 1{12}$, and $t_1 = \frac 1{15}$ respectively.
We recall that $t_0=\frac 1{12}$ is  a point of gradient catastrophe  in all the situations discussed in 
Sect. \ref{sectphase}, with the exception of situation "Opposite Wedges, symmetric" (Fig. \ref{BiWedgeSym}).
Viceversa, the gradient catastrophe point $t_1 =  \frac 1{15}$ occurs only in "Single Wedge" (Fig. \ref{Wedge}) 
and  "Consecutive Wedges" (Fig. \ref{Tri}).

%
%

\subsection{The asymmetric case}\label{yrneqyl}
Under this title we treat both the case where $t$ is near $t_1$ 
(which requires a  special parametrix  only near one endpoint, say $\l_1$, and the standard Airy parametrix near the other) 
{\bf and} the case of $t$ near $t_0$ {\bf but with $\nu_1\neq {1-\nu_3}$}
The latter case  requires some special parametrix at both endpoints;
 but a given value of $v$, generically, can be  near the pole of only one of the two special 
solution $y^{(0)}(v),~y^{(1)}(v)$ of the Painlev\'e\ I equation that enter in Theorem \ref{theo-ArnoDu}.
Below, we assume that $v$ is close to the pole $v_p$ of $y^{(1)}(v)$.
The case when a pole $v_p$ of $y^{(1)}(v)$ is simultaneously a pole of $y^{(0)}(v)$ even though $\nu_1\neq {1-\nu_3}$ and, thus,
$y^{(0)}\not\equiv y^{(1)}$, could be treated as the symmetric case (Subsection \ref{yryl}) with minor modifications (but we shall not consider it here for simplicity).

We define the approximate solution
to the RHP (\ref{RHPT}) with the jump matrix (\ref{RHPlips}) as
\be
\Phi(z) = 
\le\{ 
\begin{array}{cc}
 E(z) \Psi_0(z) & \mbox { for $z$ {\bf outside} of the disks } \mathbb D_{0}, \mathbb D_{1}\\[4pt]
E(z) \Psi_0(z)\mathcal P_{0}(z) & \mbox { for $z$ {\bf inside} of the disk } \mathbb D_{0}, \\[4pt]
 E(z) \Psi_0(z)\hat\mathcal P_{1}(z) & \mbox { for $z$ {\bf inside} of the disk } {\mathbb D}_{1}.\\[4pt]
\end{array}
\ri.
\label{leadingapproxPoles2}
\ee
where the matrix $E(z)$, discussed below, is needed to ``adjust'' the situation due to the pole $v_p$.
Here the parametrix $\mathcal P_0(z)$ is the Airy parametrix if we are near $t_{1}$. If we are near
$t_{0}$, the parametrix $\mathcal P_0(z)$
is given by 
\be
\mathcal P_0(z):= \s_3 \mathcal P(-z;1-\nu_3)\s_3,
\ee
where $\mathcal P(z;\varkappa)$ was introduced in Definition \ref{deflocalP}. 
To introduce the parametrix $\hat \mathcal P_1(z)$, we first  define $\wh \Psi$ by the Masoero factorization (\cite{Masoero})
\be\label{Masfactor}
 \Psi(\xi;v;\varkappa  )= (\xi-y)^{ - \sigma_3/2} \le[ 
\begin{array}{cc}
\frac 1 2\le(y' + \frac 1{2(\xi-y)}\ri) & 1\\
1& 0 
\end{array}\ri]\wh \Psi(\xi;v;\varkappa)
\ee
with $\Psi$ as in Def. \ref{deflocalP} and $y= y(v;\varkappa)$ (prime denotes derivative in $v$). 
\begin{definition}[Local parametrix near the poles.]
\label{deflocalPspike}
The parametrix $\wh {\mathcal P}_{1}(z)$ is defined in $\mathbb D_{1}$ as
\bea\label{localPspike}
\wh {\mathcal P}_{1}(z) = \wh {\mathcal P}_{1}(z;\nu_1)=
\frac 1{\sqrt{2}}  \begin{bmatrix}
i&i\\
-1 & 1
\end{bmatrix}
\zeta^{\frac 3 4 {\s_3}} 
\wh \Psi\le ( \zeta +  \frac \t 2; \frac 3 8 \t^2;\nu_1\ri) {\rm e}^{-\theta(\zeta;\t) \s_3}, 
\eea
\end{definition}
where  $\zeta(z;\varepsilon)$ is the local conformal coordinate in  $\mathbb D_{1}$, see Definition \ref{deflocalP}. 
We can then formulate the statement corresponding to Theorem \ref{thmlocalP} for the new local parametrix.
\bt[Theorem 6.1 in \cite{BT2}]
\label{thmlocalP1}
The matrix $\wh{\mathcal P}_1$ satisfies: 
\begin{enumerate}
\item Within $\mathbb D_1$, the matrix  $\wh {\mathcal P}_1(z)$ solves the exact jump conditions on the lenses and on the complementary arcs;
\item On the main arc (cut) $\wh{\mathcal P}_1(z)$ satisfies 
\be
\wh {\mathcal P}_{1+}(z) = \sigma_2 \wh{\mathcal P}_{1-}(z) \s_2\ ,\label{oddjump}
\ee
so that $\Psi_0 \wh {\mathcal P}_1$ within $\mathbb D_1$ solves the exact jumps on all arcs contained therein  (the left-multiplier in the jump (\ref{oddjump}) 
cancels against the jump of $\Psi_0$);
\item The product $\Psi_0 (z) (z-\l_1)^{-\s_2 }  \wh {\mathcal P}_1(z)$ (and its inverse) are --as  functions of $z$-- bounded within $\mathbb D_1$, namely the 
matrix $\wh {\mathcal P}_1(z)$ cancels the growth of $\Psi_0(z) (z-\l_1)^{-\s_2 }$ at $z=\lambda_1$; 
\item The restriction of $\wh{\mathcal P}_1(z)$ on the boundary of $\mathbb D_1$ is 
\be
\wh {\mathcal P}_1(z)\bigg|_{z\in \pa\mathbb D_\a} = \le(\1 + 
\mathcal O(\zeta^{-\frac 1 2})\ri) \le( \frac{ \sqrt{1-\zeta/y}}{ 1+ \sqrt{\zeta/y}} \ri)^{-\s_3},
\label{611} 
\ee
where $\mathcal O(\zeta^{-\frac 12})$ is uniform w.r.t. $v$ in a small, compact neighborhood of a pole $\pole$ that does  not contain any zero of $y(v)$.
\end{enumerate}
\et
The statements in \cite{BT2} were tailored to the case of the tritronqu\'ee solution and there was a slightly different normalization,
 but the proof goes through in identical fashion. Also note that the parametrix in \cite{BT2} differs from $\wh {\mathcal P}_1$ 
by a conjugation by $\s_2$.

\subsubsection{Triple scaling: proof of Theorem \ref{theor-nonsym}}
%
%
%
%
%
%
%
%
%

Before delving into the proof we make some preparatory remarks: first off, recall that we are choosing $v$ so that $v-v_p=\mathcal O(N^{-\frac 1 5 - \rho})$, $\rho\geq 0$; this means that $y (v)= \frac 1{(v-v_p)^2} + \mathcal O(v-v_p)^2$ also grows at a rate $y(v) = \mathcal O(N^{\frac 25+2\rho})$. 
Recall also that for $z\in \pa\mathbb D_1$ we have $\zeta(z) = \mathcal O(N^{\frac 25})$; {\em therefore} 
\be
\frac {\zeta(z)}{y(v)} = \mathcal O (N^{-2\rho}), \ \ \ \ z\in \pa\mathbb D_1,\ \rho\geq 0.
\ee
In the case $\rho=0$ the disk around $\lambda_1$ shall be chosen sufficiently small so that $|\zeta/y|<1-\d$ for some $\d>0$; 
this means that the rightmost factor in (\ref{611}) is a uniformly {\bf smooth and  bounded} matrix on $\pa \mathbb D_1$. In fact it 
also tends to the identity if $\rho>0$, but in  general it does so very slowly (in $N$) or not at all (if $\rho=0$, which is the most interesting case).
Therefore we can move the rightmost factor in (\ref{611}) to the left at ``no cost". So, we can write
\be\label{611_1}
\hat\mathcal P_1(z)\bigg|_{z\in \pa\mathbb D_1} = \le(\1 + 
\mathcal O(\zeta^{-\frac 1 2})\ri) \le( \frac{ \sqrt{1-\zeta/y}}{ 1+ \sqrt{\zeta/y}} \ri)^{-\s_3}
= \le( \frac{ \sqrt{1-\zeta/y}}{ 1+ \sqrt{\zeta/y}} \ri)^{-\s_3} \le(\1 + 
\mathcal O(\zeta^{-\frac 1 2})\ri). 
\ee
If $\rho=0$, the above mentioned factor  does not tend to identity.

We require that the approximate solution $\Phi(z)$ from (\ref{leadingapproxPoles2}) satisfies
\be
\Phi(z)=\le\{ 
\begin{array}{cc}
 \Phi_+(z)=\Phi_-(z)(\1 +o(1)) & \mbox { uniformly in $z$ on  } \partial\mathbb D_{0}\cup\partial\mathbb D_{1}\\[4pt]
\Phi(z) & \mbox {is bounded for $z$ { inside}  the disks  $\mathbb D_{0}\cup\mathbb D_{1}$}, \\[4pt]
\end{array}
\ri.
\label{Condforapprox}
\ee
In particular, in view of point 3 in Theorem \ref{thmlocalP1}, the requirements of (\ref{Condforapprox}) will become true if  
the matrix $E(z)$, introduced in (\ref{leadingapproxPoles2}), would satisfy the following RHP problem for $E(z)$.
\begin{problem}
\be\label{RHPE}
\le\{ 
\begin{array}{cc}
 E_+(z)=E_-(z) \Psi_0(z) \le( \frac{ \sqrt{1-\zeta/y}}{ 1+ \sqrt{\zeta/y}} \ri)^{\s_3} \Psi^{-1}_0(z) & \mbox { on } \partial\mathbb D_{1},\\[4pt]
E(z) =O(1)(z-\l_1)^{-\s_3}(\s_2+\s_3) & \mbox { as } z\ra \l_1, \\[4pt]
E(z)=\1+O(\frac{1}{z}) & \mbox { as } z\ra\infty,
\end{array}
\ri.
\ee
where $O(1)$ means an invertible matrix analytic at $z=\lambda_1$,  bounded together with its inverse, and the circle $\partial\mathbb D_{1}$ 
has positive orientation.
\end{problem}

Note  that the second condition
of (\ref{RHPE}) is equivalent to 
\be\label{Eloc}
E(z)\Psi_0(z) \begin{bmatrix}
    i & i \\ -1 & 1
    \end{bmatrix}\z(z)^{\frac 34 \s_3}  =O(1)\qquad \mbox { as } z\ra \l_1, 
\ee
given that $\zeta(z) = \mathcal O(z-\l_1)$. 
Equation (\ref{Eloc}) together with Theorem \ref{thmlocalP1}, item 3, guarantee the boundedness of  
\be
\Phi (z) = E(z) \Psi_0 (z)\hat\mathcal P_{1}(z) =\underbrace{E(z) \Psi_0(z)
\frac 1{\sqrt{2}} 
{ \begin{bmatrix}
i&i\\
-1 & 1
\end{bmatrix}}
\zeta^{\frac 3 4 {\s_3}} }_{=\mathcal O(1)}
\overbrace{\wh \Psi\le ( \zeta +  \frac \t 2; \frac 3 8 \t^2\ri)   {\rm e}^{-\theta(\zeta;\t) \s_3}}^{=\mathcal O(1)}, 
\ee
within the disc $\mathbb D_1$.

\paragraph{Proof of solution of the Problem \ref{RHPE}} Let $\wh E(z)=\hf(\s_2+\s_3)E(z)(\s_2+\s_3)$. Then
\be
\le\{ 
\begin{array}{cc}
\wh E_+(z)=\wh E_-(z)M(z) & \mbox { on } \partial\mathbb D_{1},\\[4pt]
\wh E(z) =O(1)(z-\l_1)^{-\s_3} & \mbox { as } z\ra \l_1, \\[4pt]
\wh E(z)=\1+O(\frac{1}{z}) & \mbox { as } z\ra\infty,\\[4pt]
\end{array}
\ri.
\label{RHPhatE}
\ee
where 
\be\label{Mgen}
M = \frac 12 (\s_2 + \s_3) \Psi_0 \le(\frac {\sqrt{ 1 - \z/y}}{1 + \sqrt{\z/y}}\ri)^{\s_3} \Psi_0^{-1}(\s_2 + \s_3).
\ee
Using (\ref{Psi0}) and the fact that
\be\label{S^s2}
S^{\s_2}=e^{\ln S \s_2}=\cosh(\ln S \s_2)+\sinh (\ln S \s_2)= \hf(S+S^{-1})\1+\hf(S-S^{-1})\s_2,
\ee
we calculate 
\bea
 M(z)=[A(z),B(z)]=
  \frac 1{\sqrt{1-\frac \zeta y}} \le(\1 + i  \sqrt{\frac {\z p}{y}} \s_+   - i \sqrt{ \frac {\z}{py}} \s_-\ri),\  
{\rm where}~~ p:= \frac {z-\l_1}{z-\l_0}
\eea
and $\s_\pm=\hf(\s_1\pm i\s_2)$.
We make the Ansatz that $\wh E_-(z) = \1 + \frac L{z-\l_1}$: then the (constant in $z$) matrix $L$ must be chosen so that 
$\wh E_+(z)= \wh E_-(z) M(z)$  satisfies
\be\label{eqL}
\wh E_+(z)(z-\l_1)^{\s_3}=\left(\1+\frac{L}{z-\l_1}\right)[A,B](z-\l_1)^{\s_3}=\mathcal O(1),~~~z\in\mathbb D_{1} .
\ee
In light of (\ref{zeta}) we see that $A(\l_1)=\mathcal O (1)$, and thus  we need to consider only the second column of (\ref{eqL}):
\be
\left(\1+\frac{L}{z-\l_1}\right)\frac{B(z)}{z-\l_1}=\frac{B(z)}{z-\l_1}+\frac{LB(z)}{(z-\l_1)^2}=\mathcal O(1)
\ee
or 
\be
\le\{ 
\begin{array}{cc}
LB(\l_1)=0 & \\[4pt]
B(\l_1)+LB'(\l_1)=0. & \\[4pt]
\end{array}
\ri.
\label{Bsyst}
\ee
Calculating 
\be 
B(\l_1)=(0,1)^T\ ,\qquad 
B'(\l_1)=  \le(i\sqrt{ \frac{\z'(\l_1)}{2b y} } , \frac {\z'(\l_1)}{2y}\ri)^T ,   
\ee
we see that
\be\label{L}
L=i\sqrt{\frac{(\l_1-\l_0) y}{\z'(\l_1)}}\s_-
\ee
solves the system (\ref{Bsyst}). Thus
\be
\le\{ 
\begin{array}{cl}
\ds \wh E_-(z)&=\ds \1+  i\frac{\sqrt{\frac{y(\l_1-\l_0)}{\z'(\l_1)}}\s_-}{z-\l_1}, \\[9pt]
\ds \wh E_+(z)& = \left(\1+i \frac{\sqrt{\frac{y(\l_1-\l_0)}{\z'(\l_1)}}\s_-}{z-\l_1}\right)  M(z) 
\end{array}
\ri.
\label{solRHPhatE}
\ee
solves the RHP (\ref{RHPhatE}).
 Since $\frac i 2(\s_2+\s_3)\s_-(\s_2+\s_3)=\frac 12(\s_3 -i \s_1)$, we obtain that
\be
\le\{ 
\begin{array}{cl}
\ds  E_-(z)& \ds =\1 {+}\frac{\sqrt{\frac{y(\l_1-\l_0)}{\z'(\l_1)}}(\s_3-i\s_1)}{2(z-\l_1)},  \\[9pt]
\ds  E_+(z)& \ds = \left(\1   {+}  \frac{\sqrt{\frac{y(\l_1-\l_0)}{\z'(\l_1)}}(\s_3-i\s_1)}{2(z-\l_1)}\right) 
\Psi_0 \le(\frac {\sqrt{ 1 - \z/y}}{1 + \sqrt{\z/y}}\ri)^{\s_3} \Psi_0^{-1}
\end{array}
\ri.
\label{solRHPE}
\ee
solves the RHP (\ref{RHPE}). \QED

\paragraph{Error analysis.} The error matrix $\mathcal E(z)=T(z)\Phi^{-1}(z)$ has jumps on the lenses and on the complementary arcs
outside the disks $\mathbb D_{0}, \mathbb D_{1}$, as well as on the boundary of these disks. 
The jump matrices on the lenses and on the complementary arcs approach $\1$ exponentially fast in $N^{-1}$ and uniformly in $z$.
It is also clear that  $\mathcal E\ra \1$ as $z\ra\infty$ since both $T$ and $\Phi=E\Psi_0$ do so. So, it remains only to prove the uniform convergence to $\1$ of the jump
matrix on  $\partial\mathbb D_{1}$ (convergence on $\partial\mathbb D_{0}$ was established in Subsection \ref{assaway}). 
Indeed, using (\ref{leadingapproxPoles2}), (\ref{RHPE}), (\ref{611}), (\ref{611_1}) and (\ref{solRHPE}), we have
\bea\label{errSpikes2ttmp}
\mathcal E_+&\&=T\Phi_+^{-1}=T\hat\mathcal P^{-1}_{1}\Psi^{-1}_0E^{-1}_+=T\Phi^{-1}_{-}E_-\Psi_0 \hat\mathcal P^{-1}_{1}\Psi^{-1}_0
\Psi_0 \le( \frac{ \sqrt{1-\zeta/y}}{ 1+ \sqrt{\zeta/y}} \ri)^{-\s_3} \Psi^{-1}_0E^{-1}_- \cr
&\& =\mathcal E_-E_-\Psi_0(\1+\mathcal O(\z^{-1/2}))\Psi^{-1}_0E^{-1}_-=\mathcal E_-E_-(\1+\mathcal O(\z^{-1/2}))E^{-1}_-.
\eea
On the boundary $z\in \pa\mathbb D_1$ we have $\zeta = \mathcal O(N^{\frac 25})$ and in our triple scaling 
$y = \mathcal O(N^{\frac 25  + 2 \rho})$ with $\rho\geq 0$. 
Then $L$ is of the order
$\mathcal O(N^{-\frac 15} \sqrt y) = \mathcal O(N^{\rho})$.
Thus, 
\bea
E_-\mathcal O(\z^{-1/2})E^{-1}_- = \mathcal O(\z^{-1/2}) + \frac{ [L, \mathcal O(\z^{-1/2})]}{z-\l_1} - \frac {L\mathcal O(\z^{-1/2})L}{(z-\l_1)^2} = 
\mathcal O(N^{-\frac 1 5}) + \mathcal O(N^{-\frac 1 5+ \rho}) + \mathcal O(N^{-\frac 1 5+ 2\rho}). 
\eea
So, it is the last term that contributes the slowest decay. Therefore, we obtain 
\bea\label{errSpikes2}
\mathcal E_+=\mathcal E_-(\1+\mathcal O(N^{-\frac35}y)),~~~~~z\in \partial \mathbb D_{1}.
\eea
The latter estimate shows that we can control the error provided 
\be\label{estspike}
y=y^{(1)}=O(N^{\frac 25 + \rho}),~~~{\rm or,~equivalently,~}~~ v-v_p=O(N^{-\frac 1{5}-\rho}),
\ee 
where $0\leq \rho <  \frac 1 5$.

\paragraph{Computation of the recurrence coefficients:}
We need to use  (\ref{expressalpbet}) and  (\ref{errSpikes2}).
Using  (\ref{leadingapproxPoles2}),
 (\ref{solRHPE}) and the expansion of $\Psi_0$ (\ref{Psi0ass}) 
we obtain
\bea\label{residuespike2}
\Phi(z)&\&=\ds E_-(z)\Psi_0(z)=
\left(\1{+} \frac 1  2 \frac{k {(\s_3-i\s_1)}}{z-\l_1}\right)
\left(  \1-\frac{b}{2z}\s_2+\frac{b^2}{8z^2}\1 -  \frac {ab\s_2}{2z^2}+O(z^{-3})  \right)\cr
&\&=
\left(\1 +  \frac 1 2 k {(\s_3-i\s_1)} \le( \frac 1 z + \frac {\l_1}{z^2}\ri)\right)
\left(  \1-\frac{b}{2z}\s_2+\frac{b^2}{8z^2}\1 - \frac {ab\s_2}{2z^2}+O(z^{-3})  \right)
\\
&\&=
\1 + \frac 1 {z} \le[  \frac k 2  \le(\s_3-i\s_1\ri) - \frac b 2 \s_2\ri] 
+ \frac 1{z^2} \le[
\frac {b^2\1 }8 - \frac {ab }2  \s_2 + \frac k 2\le(a + \frac {1} 2 b\ri) \le(\s_3-i\s_1\ri) 
\ri] \\
&\&\ds k:=\sqrt{\frac{(\l_1-\l_0) y}{\z'(\l_1)}}=\sqrt{\frac{2b y}{\z'(\l_1)}}
\eea
We introduce 
\be
s:= \sqrt{\frac {\zeta'(\l_1)}{2 b y(v)}}= \sqrt{\frac {\zeta'(\l_1)}{2 b_0 }} (v-v_p) + \mathcal O\le( N^{\frac 1 5} (v-v_p)^5\ri),
\ee
where the latter expression follows from (\ref{TaylorP1}). 
Here and henceforth $a_0,b_0$  denote the values of $a,b$ calculated exactly at one of the critical points $t_0$ or $t_1$.
Assuming in $\rho=0$ in (\ref{estspike}), we obtain
$ s = \mathcal O(1)$ as $N\ra\infty$. 
On the other hand, if $\rho\in(0,\frac 15)$, then, consequently,  
$s$  scales as $\mathcal O( N^{-\rho})$.
Thus, we have the {\bf triple scaling limit}
\be
t - t_j =
\frac {v}{\kappa N^{\frac 45}} = 
\frac {v_p}{\kappa N^{\frac 45}}    + \frac {\sqrt{\frac {2b_0}{\z'(\l_1)}} }{\kappa N^{\frac 45}} s,
 \ee 
where $\kappa$ is the constant in front of $\delta t=t-t_j$ appearing in formul\ae\ (\ref{v(t)}, \ref{v(t)2}). Explicitly, using Table \ref{expansions},
we obtain 
\be\label{deltat's}
t+\frac 1 {12} =  -\frac{  v_p}{3^\frac 6 5 2^{\frac 9 5} N^{\frac 45} } -\frac {s}{3\sqrt 2 N}\ ,\qquad
t-\frac 1{15} = \frac {v_p {\rm e}^{-\frac {3i\pi}5}} {3^\frac 6 5 2^\frac 1 {5}  5 N^{\frac 45} } -i \frac {2 s} 
{15 N}.
\ee
Now, according to (\ref{Talphabeta}), (\ref{l}), (\ref{l2}) and (\ref{errSpikes2}), we obtain: 
\bea\label{Yfinalspike2}
\a_n &\& = (T_1)_{12}(T_1)_{21}=\frac{b^2}{4}{-} \frac{by(v)}{2N^{\frac 25}C} + \mathcal O(N^{-\frac 3 5 } y) = 
\frac {b_0^2}4 - \frac 1{4s^2} + \mathcal O(N^{-\frac 3 5 } y, N^{-\frac 25}),
\\
\label{betaspike}
\b_n &\& = \frac{\left(T_2\right)_{12}}{\left(T_1\right)_{12}}-\left(T_1\right)_{22}=
 a_0+\frac 1{2s(1-b_0s) +\mathcal O(N^{-\frac 32}y)} + \mathcal O(N^{-\frac 32}y,N^{-\frac 25}),
 \\
 \h_n&\&= -2i\pi (T_1)_{12}{\rm e}^{N\ell}
=\left[\pi \le(b_0-\frac 1 s \ri)+\mathcal O(N^{-\frac 32}y, N^{-\frac 25})\right]{\rm e}^{\le[
N \ln \frac {b^2}4 -N\frac {2a^2 + b^2}8 - \frac N 2
\ri]}
.
\eea
Here $\mathcal O(N^{-\frac 25})$ error term comes from replacing $a,b$ with their respective values $a_0, b_0$
considered at the  critical point $t_{0}$ or $t_1$. Note, however, that in the regime (\ref{triplescale}), the 
$\mathcal O(N^{-\frac 25})$ term is of a smaller order than the $\mathcal O(N^{-\frac 32}y)$ term. Therefore, 
in all these expressions, in the regime (\ref{triplescale}),  the error is at best $\mathcal O(N^{-\frac 15})$ (recall that $y = \mathcal O(N^{\frac 2 5 +2\rho})$ $\rho\in[0, \frac 1 5)$). 
Thus in the exponent ${\rm e}^{N\ell}$ we can use the expansion in Table \ref{expansions} up to order $\delta t$ included. So,
\be
\h_n = \pi \le(\frac {b_0^2}4\ri)^N \exp \le[
 -N\frac {2a_0^2 + b_0^2+4}8 + c N\delta t  
\ri]\le(b_0-\frac 1 s  +  \mathcal O(N^{-\frac 32}y)\ri),
\ee
where  $c=-6$ for the case $t\sim t_0$ and $c=-\frac {13\cdot 15}4 $ for the case $t\sim t_1$. 
One has then to replace $\delta t$ by the expressions in (\ref{deltat's}). So, in the leading order, 
\bea
\h_n &\&=\pi\le(\sqrt{8} - \frac 1 s \ri) 2^N \exp \le[-\frac {3 N}2   + \frac{  N^{\frac 1 5} v_p}{3^{\frac 1 5 } 2^{\frac 4 5}} +\sqrt{2} \,{s}     \ri]
\ ,\ \ \ t\sim -\frac 1{12},  \\
\h_n &\& = \pi \le(2i - \frac 1 s\ri) (-1)^N \exp\le[\frac {9N}4  -   \frac {13 }4\frac {N^\frac 1 5v_p {\rm e}^{-\frac {3i\pi}5}} 
{3^\frac 1 5 2^\frac 1 {5}    }  + i \frac {13}{2}  s \ri] \ ,\ \ \ t\sim \frac 1 {15}.
\eea

It is remarkable to note that the genus zero leading order asymptotics $\a_n(t)\sim \frac{b^2}{4}$ and $\b_n(t)\sim a$ are valid as long
$y=o(N^{\frac 25})$ with the accuracy $O(\frac{y}{N^{\frac 25}})$. However, when $y=O(N^{\frac 25})$, both terms
in (\ref{Yfinalspike2}), (\ref{betaspike}), contribute to the leading order, whereas, when $y=O(N^{\frac 25+2\rho})$ with $\rho\in[0,\frac 1{5})$,
the asymptotics are determined by the latter terms of  (\ref{Yfinalspike2}), (\ref{betaspike}). 
In this case, both $\a_n$ and $\b_n$ are unbounded as $N\ra\infty$. 

So, the proof of Theorem \ref{theor-nonsym} is completed.  \QED


%
%
%
%
\subsection{The  symmetric case: proof of Theorem \ref{theor-sym} }\label{yryl}
We are now in the symmetric situation and hence the critical point to consider can  only 
be $t_0=-\frac 1 {12}$, where $\l_1 = b$, $\l_0=-b$ and $a\equiv 0$. 
This case is significantly different from the previous inasmuch as the two Painlev\'e\ parametrices in $\mathbb D_{0,1}$ are identical: 
in particular $y^{(1)} = y^{(0)}=y$. Thus, if the double scaling is such that we are close to a pole $v_p$ of $y(v)$, 
this will {\em simultaneously} affect the both parametrices  and, as we shall see, will have  a significant effect on the asymptotics
of $\a_n$. On the other hand, due to the exact symmetry of the bilinear pairing, the orthogonal polynomials have the same parity of their degree and thus automatically $\b_n\equiv 0$.

It will be advantageous for us  to use a {\em different} solution to the model problem (\ref{RHPPsi_0}),
 which has a different growth rate near the branch-points: such modification 
 (see  \cite{JMU2}) 
is called a {\em discrete Schlesinger transformation}. In terms of the RHP (\ref{RHPPsi_0}), 
this amounts to replacing the solution $\Psi_0$ (\ref{Psi0}) with 
\be
\Psi_1(z):= \frac 1  2 
(\s_3+\s_2)
\le(\frac{z-b}{z+b}\ri)^{-\frac 34 \sigma_3} 
(\s_3+\s_2)=\le(\frac{z-b}{z+b}\ri)^{-\frac 34 \sigma_2} .
\label{Psi1}
\ee
This matrix satisfies all the conditions of the RHP (\ref{RHPPsi_0}) except the last one, as it   
clearly has a different growth behaviour near the endpoints $\pm b$. 
We then  shall construct an approximate solution 
\be
\Phi(z) = 
\le\{ 
\begin{array}{cc}
 E(z) \Psi_1(z) & \mbox { for $z$ {\bf outside} of the disks } \mathbb D_{0}, \mathbb D_{1}\\[4pt]
E(z) \Psi_1(z)\wh {\mathcal P}_{0}(z) & \mbox { for $z$ {\bf inside} of the disk } \mathbb D_{0}, \\[4pt]
 E(z) \Psi_1(z)\hat\mathcal P_{1}(z) & \mbox { for $z$ {\bf inside} of the disk } {\mathbb D}_{1},\\[4pt]
\end{array}
\ri.
\label{leadingapproxPolessym}
\ee 
where $\hat\mathcal P_{1}(z)$ is defined by (\ref{localPspike}) and
\be
\wh {\mathcal P}_{0}(z)  = \s_3 \wh {\mathcal P}_{1}(-z)\s_3 \ .
\ee
Due to the fact that we are using $\Psi_1$ instead of $\Psi_0$, 
the boundedness of 
the product $\Psi_1 \wh{\mathcal P}_{0,1}$ at $\l_0,\l_1$ 
follows immediately 
(see also Theorem \ref{thmlocalP1}, item 3).
 Hence, the requirements on the left multiplier $E(z)$ are now different 
compare with the asymmetric case studied above (we reuse the same symbol $E$ with a new meaning relative to the previous section).
\begin{problem} 
\label{problem72} Find
the matrix $E(z)$  is analytic (together with its inverse) 
on $\C\setminus (\pa\mathbb D_0 \cup \pa\mathbb D_1)$ and satisfies
\be\label{RHPE2}
\le\{ 
\begin{array}{cc}
E_+(z)=E_-(z) \Psi_1(z) \le( \frac{ \sqrt{1-\zeta/y}}{ 1+ \sqrt{\zeta/y}} \ri)^{\s_3} \Psi^{-1}_1(z) & \mbox { on } \partial\mathbb D_{0,1},\\[4pt]
E(z)=\1+\mathcal O(\frac{1}{z}) & \mbox { as } z\ra\infty,
\end{array}
\ri.
\ee
where the contours $\partial\mathbb D_{0,1}$ have positive orientation.
\end{problem} 
{\bf Proof of solution of Problem \ref{problem72}}. Note that any solution to this RHP has unit determinant and hence its inverse is also analytic and bounded. As before, we find it more convenient to solve the RHP for $\wh E(z)= F E(z)F$
instead of the RHP  (\ref{RHPE2}). Here $F= \frac {\s_2 + \s_3}{\sqrt{2}}$.
The jump matrix for the new RHP is    
\be 
M(z)=F^{-1}\Psi_1(z)Q^{-1}(z)\Psi_1^{-1}(z)F,\label{defM}
\ee 
where 
\be\label{Q}
Q:=\le(\frac {\sqrt{1-\zeta/y}}{1+\sqrt{\zeta/y}}\ri)^{-\s_3}, 
\ee
$y=y(v)$,
$v$ was defined by (\ref{v(t)}) and the local scaling coordinate $\zeta = \zeta(z,N)$ 
near $z=b$ was introduced in (\ref{zeta}).

Direct calculations yield
\be\label{M}
M (z)=\frac{1}{\sqrt{1-\z(z)/y}}
\begin{bmatrix}
1 & i \sqrt{\frac {\zeta(z) (z+b)^3}{ y (z-b)^3}}\\
-i \sqrt{\frac {\zeta(z) (z- b)^3}{ y (z+b)^3}} & 1
\end{bmatrix}.
\ee

Similarly, near $z=-b$ we obtain
\be\label{tildM}
\tilde M(z)=\frac{1}{\sqrt{1-\tilde\z(z)/ y}}
\begin{bmatrix}
1 & i \sqrt{\frac {\tilde\zeta(z) (z+b)^3}{  y (z-b)^3}}\\
-i \sqrt{\frac {\tilde\zeta(z) (z- b)^3}{ y (z+b)^3}} & 1
\end{bmatrix},
\ee
\be\label{tildezeta}
{\rm where}~~~~~\tilde \z=\tilde \z(z)=\zeta(-z)\ ,\ \ z\in \mathbb D_{0}.
\ee
 Note that the orthogonal polynomials in this case are even/odd and the symmetry of the RHP implies (which can be verified directly from the above formul\ae\ and also as a consequence of (\ref{defM}))
\be
\Psi_1(z) = \s_3 \Psi_1(-z) \s_3 \ \ \Rightarrow \ \ \ \wt M(z) = \s_2 M(-z) \s_2
\ee

Using (\ref{M}), (\ref{tildM}), (\ref{zeta}), (\ref{tildezeta}), we obtain 
\be\label{Mstr}
M(z)-\1=\frac{\eta\s_+}{z-b} +O_1(z),~~~~\tilde M(z)-\1=\frac{ \eta\s_-}{z+b} +O_0(z),
\ee
where
\be\label{eta}
\eta = iN^\frac 15\sqrt{\frac{(2b)^3C}{y}}.
\ee
Here $C= N^{-\frac 2 5} \tilde \z'(b)$ and 
 $\tilde C = N^{-\frac 2 5} \tilde \z'(-b) = -C$ by the symmetry of the problem.
Note that the $O_{0,1}$ terms in (\ref{Mstr}) are analytic at $z=\l_{0,1}$ and  when evaluated  at $z=\l_{0,1}=\pm b$  are proportional 
to $\s_\pm$  (respectively). 
The matrix $\wh E(z)$ satisfies 
\bea
\wh E(z) = \1 + \oint_{|s-b|=r} \!\!\!\! \frac {\wh E_-(s) (M(s)-\1)}{s-z} \frac{{\rm d}s}{2i\pi}+ \oint_{|s+b|=r} \!\!\!\! 
\frac {\wh E_-(s) (\tilde M(s)-\1)}{s-z} \frac{{\rm d}s}{2i\pi}. \label{ERHP}
\eea
We pose the Ansatz
\be
\wh E_-(z)=\1+\frac{A}{z-b}+\frac{\tilde A}{z+b},
\ee
and  obtain 
\be\label{Eterms}
\frac{A}{z-b} + \frac {\tilde A}{z+b}  = 
- \frac {A\eta \s_+ }{(z-b)^2} 
- \frac {\eta\s_+}{z-b} 
- \frac {\tilde A\eta \s_+}{2b(z-b)} 
- \frac {\tilde A\eta \s_- }{(z+b)^2}  
- \frac {A O_1(b)} {z-b}   
- \frac {\tilde A O_0(-b)} {z+b}
 - \frac {\eta\s_-}{z+b} 
 - \frac { A \eta \s_- }{2b(z+b)}.
\ee
That leads to
the following system for the unknown $A,\tilde A$ (recall that  $O_1(b)\propto \s_+, \ O_0(-b)\propto \s_-$):
\bea\label{etaeq}
A \s_+ =0,  &&
\tilde A \s_-  = 0,\cr
A  + \frac {\eta}{ 2b} \tilde A \s_+ = -\eta \s_+, & & 
\tilde A + \frac {\eta}{2b} A \s_-  = -\eta \s_-.
\eea
This system has the solution
\bea
A  = \frac {1}{
1 +  \frac {\eta^2}{(2b)^2}
}
\begin{bmatrix}
     0& -\eta  \\  0 & \frac{\eta^2}{{2b}}
    \end{bmatrix},~~~~~~~  
\tilde A   =  
-\frac {1}{
1 +  \frac {\eta^2}{(2b)^2}
}
\begin{bmatrix}
    \frac{\eta^2}{{2b}} & 0 \\ \eta & 0
    \end{bmatrix} = -\s_2 A \s_2
\label{AAhat}
\eea
So, we found $\wh E(z)$ and, thus,  $E(z)$.
Note that the function $E(z)$ in the region {\bf outside} of the disks  is a rational function with poles at $\pm b$, while, inside the disks, it is analytic and given by formula (\ref{ERHP}). \QED

\paragraph{Error analysis.} The error matrix $\mathcal E(z)=T(z)\Phi^{-1}(z)$ has jumps on the lenses and on the complementary arcs
outside the disks $\mathbb D_{0}, \mathbb D_{1}$, as well as on the boundary of these disks. 
The jump matrices on the lenses and on the complementary arcs approach $\1$ exponentially fast in $N$  and uniformly in $z$.
It is also clear that  $\mathcal E\ra \1$ as $z\ra\infty$. So, it remains only to prove the uniform convergence to $\1$ of the jump
matrix on  $\partial\mathbb D_{0,1}$: the computations are absolutely parallel and we report only the one for $\pa \mathbb D_1$.
Using (\ref{leadingapproxPoles2}), the solution to Problem \ref{problem72} and eq. (\ref{611}), we have
\bea\label{errSpikes2tmp}
\mathcal E_+&\&=T\Phi_+^{-1}=T\hat\mathcal P^{-1}_{1}\Psi^{-1}_1 E^{-1}_+=T\Phi^{-1}_{-}E_-\Psi_1 \hat\mathcal P^{-1}_{1}\Psi^{-1}_1
\Psi_1 \le( \frac{ \sqrt{1-\zeta/y}}{ 1+ \sqrt{\zeta/y}} \ri)^{-\s_3} \Psi^{-1}_1 E^{-1}_- \cr
&\& =\mathcal E_-E_-\Psi_1(\1+\mathcal O(\z^{-1/2}))\Psi^{-1}_1 E^{-1}_-=\mathcal E_-E_-(\1+\mathcal O(\z^{-1/2}))E^{-1}_-.
\eea
On the boundary $z\in \pa\mathbb D_1$ we have $\zeta = \mathcal O(N^{\frac 25})$ and in our double scaling $y = \mathcal O(N^{\frac 25 + 2\rho})$, where $\rho\in[0,\frac 15)$. 
Moreover, $E_- = \1 + \frac {FAF}{z-b} + \frac {F\tilde AF}{z+b}, $ where  $A, \tilde A$ are of the same order. 
That creates 
 the situation that is drastically different from the previous: for example, the matrices $A, \tilde A$ (\ref{AAhat}) 
remain bounded no matter how fast $y$ 
grows (and hence $\eta \to 0$ (\ref{eta})). The only unboundedness occurs when the denominators in (\ref{AAhat})
vanish, which means that $\eta$ has a {\em finite} 
value $ \eta^2 =- 4b^2$ or, equivalently,
\be
N^{-\frac 25} y(v) = 2 C b. \label{blowup}
\ee
Condition (\ref{blowup}) identifies two points near the pole $v=v_p$ at a distance of order $\mathcal O(N^{-\frac 15})$.
Thus,  in (\ref{errSpikes2tmp}) we have
\bea\label{errE2}
E_-\mathcal O(\z^{-1/2})E^{-1}_- = \mathcal O(\z^{-1/2}) +\mathcal O \le(\z^{-\frac 12 } \le( 1 + \frac {\eta^2}{4b^2}\ri)^{-1}\ri) =  \mathcal O(N^{-\frac 15}) 
+\mathcal O \le( N^{-\frac 15 }\le( 1 + \frac {\eta^2}{4b^2}\ri )^{-1}\ri) 
\eea
The very last contribution to the error term comes from the denominators of the matrices $A, \tilde A$ (\ref{AAhat}) and prevents us from getting close ``too fast'' to the points where they vanish.
\paragraph{Computation of the recurrence coefficients:}
Following \cite{BT2}, we  find the expansion  of the matrix $\Phi(z)=E(z) \Psi_1(z)$
at $z=\infty$:
\be\label{leadinfF}
E\Psi_1 = F\hat E(z) \left(\frac{z-b}{z+b}\right)^{-\frac 34 \s_3}F^{-1}=\1+\frac{F(A+\tilde A)F^{-1}+ \frac 32 b\s_2}{z}+
\frac{\frac b2 F(\tilde A-A)F^{-1}+\frac 98 b^2\1}{z^2}+O(z^{-3}).
\ee
Using (\ref{AAhat}), we obtain
\bea\label{ApmAhat}
 \tilde A+A=-\frac{1}{1 +  \frac {\eta^2}{(2b)^2}}\left(
 \frac{\eta^2}{2b}\s_3+
\eta \s_1
 \right), ~~~~~~
 \tilde A-A=-\frac{1}{1 +  \frac {\eta^2}{(2b)^2}}\left(\frac{
 \eta^2}{2b}\1
 - i \eta  \s_2 
  \right),\\ 
\eea
so that
\bea
&\& F (\tilde A+A)F^{-1}=\frac{1}{1 + \frac {\eta^2}{(2b)^2}}\left[
-\frac{\eta^2}{2b}\s_2 + \eta \s_1 \right]\ ,
\qquad F (\tilde A-A)F^{-1}=\frac{1}{1+  \frac {\eta^2}{(2b)^2}}\left[- \frac{\eta^2}{2b}\1
+i \eta  \s_3\right].\label{FApmAhatF}
\eea
It follows from (\ref{leadinfF}) and (\ref{FApmAhatF}) that the residue of $\Phi$ at infinity, which we denote by 
$\Phi_1$, is
\bea\label{Y_1}
\Phi_1=F(A +\wt A) F^{-1} + \frac 32 b\s_2
=
b\frac 1{1+ \frac {\eta^2}{4b^2}} \le(\le(-\frac {\eta^2}{2b^2}\s_2 + \frac \eta b \s_1\ri) + \frac 3 2 \le(1+ \frac {\eta^2}{4b^2}\ri)   \s_2\ri)=\cr
=b\frac 1{1+ \frac {\eta^2}{4b^2}} \le(\le(-\frac {\eta^2}{8b^2}\s_2 + \frac \eta b \s_1\ri) + \frac 3 2   \s_2\ri)
\eea
We note in passing that $\Phi_1$ is off-diagonal and $\Phi_2$ is diagonal (which implies $\b_n=0$, which -of course- is  identity and not just an approximation due to the special symmetry of this case).
Then 
\bea\label{Y_1ent}
(\Phi_1)_{12}=\frac b 2\frac{-3i+\frac{i\eta^2}{4b^2} + 2\frac \eta b }{1 +  \frac {\eta^2}{(2b)^2}} = -i \frac b 2 
\frac{3 + \frac {i \eta}{2b} }{ 1  - \frac {i\eta}{2b}}
\\
(\Phi_1)_{21}=\frac b 2 \frac{3i - \frac{i\eta^2}{4b^2} +2 \frac \eta b}{1 + \frac {\eta^2}{(2b)^2}} = 
 i \frac  b 2  \frac { 3 - \frac {i\eta}{2b}  }{1 + \frac {i\eta}{2b} }
\eea
Using (\ref{errE2}), we can now calculate the (leading order)   final expressions
\bea\label{Y1221}
&&\a_n = (T_1)_{12}(T_1)_{21}=\frac{b^2}{4}\frac{9 + \frac{\eta^2}{4b^2}}{1 +  \frac{\eta^2}{4b^2}}\!\!\!,
\quad
\b_n=0
\eea
where it is understood that both expressions (also in the denominators) are affected  by an error of the order indicated in (\ref{errE2}).
Introducing
\be\label{s}
s = -\frac {i \eta}{2b} =  N^\frac 1 5 \sqrt{ \frac {8b^3 C}{y}},
\ee
we note that $\mathcal O(N^{-\frac 1 5} (1 + \eta^2/(4b^2))^{-1}) = \mathcal O (N^{-\frac 1 5} (s^2 - 1)^{-1})$ and 
we find finally (using  Table \ref{expansions} for the symmetric case)\footnote{The error terms can have been collected in a more elegant form as indicated.}
\bea
\label{Yfinal}
\a_n&\& =\frac{b^2}{4}\frac{9-s^2+ \mathcal O(N^{-\frac 1 5})}{1- s^2 +\mathcal O(N^{-\frac 1 5})}
,\qquad
\b_n=0 
, \cr
\h_n  &\& = 2^N\pi  \sqrt{8}
 \exp\le[- \frac {3N}2  - 
 6 N^{\frac 1 5} \frac {\d t}{N^{-\frac 45}}  \ri]\le(
 \frac {3-s}{1+s } +  \mathcal O\le(\frac {N^{-\frac 1 5}}{1-s^2}\ri) 
 \ri).
\label{bfinalsym}
\eea
 %
Using  (\ref{v(t)}) and Table \ref{expansions} to relate $s$ and $t$, we can write (\ref{bfinalsym}) as
\be
\h_n  = \pi  \sqrt{8}\,2^N
 \exp\le[-\frac {3N}2  +
    \frac {N^\frac 1 5 \,v_p}{ 3^\frac 15 2^\frac 45} - \frac {s}{4 } 
   \ri]\le(
  \frac {3-s }{1+s }+  \mathcal O\le(\frac {N^{-\frac 1 5}}{1-s^2}\ri)
 \ri).
\ee
 \QED
%
%

\end{document}